\documentclass[twocolumn]{aastex6}

\usepackage{natbib,graphicx} 
\fullcollaborationName{The KINGFISH Collaboration}

\begin{document}


\title{The Radio Spectral Energy Distribution and Star Formation Rate Calibration in Galaxies}


\author{F.~S. Tabatabaei\altaffilmark{1}}
\affil{Instituto de Astrof\'{i}sica de Canarias, V\'{i}a L\'{a}ctea S/N, E-38205 La Laguna, Spain}
\affil{Departamento de Astrof\'{i}sica, Universidad de La Laguna, E-38206 La Laguna, Spain}
\affil{Max-Planck-Institut f\"ur Astronomie, K\"onigstuhl 17, 69117 Heidelberg, Germany}
\author{E. Schinnerer}
\affil{Max-Planck-Institut f\"ur Astronomie, K\"onigstuhl 17, 69117 Heidelberg, Germany}
\author{M. Krause}
\affil{Max-Planck Institut f\"ur Radioastronomie, Auf dem H\"ugel 69, 53121 Bonn, Germany}
\author{G. Dumas \& S. Meidt}
\affil{Max-Planck-Institut f\"ur Astronomie, K\"onigstuhl 17, 69117 Heidelberg, Germany}
\author{A. Damas-Segovia \& R. Beck}
\affil{Max-Planck Institut f\"ur Radioastronomie, Auf dem H\"ugel 69, 53121 Bonn, Germany}
\author{E.\,J.\,Murphy}
\affil{National Radio Astronomy Observatory, 520 Edgemont Road, Charlottesville, VA 22903, US}
\author{D.\,D. Mulcahy}
\affil{Jodrell Bank Centre for Astrophysics, Alan Turing Building, School of Physics and Astronomy, The University of Manchester, Oxford Road, Manchester, M13 9PL, U.K}
\author{B. Groves}
\affil{Research School of Astronomy and Astrophysics, Australian National University, Canberra, ACT 2611, Australia}
\author{A. Bolatto}
\affil{Department of Astronomy and Joint Space Institute, University of Maryland, MD 20642, US}
\author{D. Dale}
\affil{Department of Physics \& Astronomy, University of Wyoming, Laramie, WY 82071, USA}
\author{M. Galametz}
\affil{European Southern Observatory, Karl-Schwarzschild-Str. 2, 85748 Garching, Germany}
\author{K. Sandstrom}
\affil{Center for Astrophysics and Space Sciences, Department of Physics, University of California, San Diego, 9500 Gilman Drive, La Jolla, CA 92093, USA}
\author{M. Boquien}
\affil{Unidad de Astronom\'ia, Fac. Cs. B\'asicas, Universidad de Antofagasta, Avda. U. de Antofagasta 02800, Antofagasta, Chile}
\author{D. Calzetti}
\affil{Department of Astronomy, University of Massachusetts–Amherst, Amherst, MA 01003, USA}
\author{R.\,C. Kennicutt}
\affil{Institute of Astronomy, University of Cambridge, Madingley Road, Cambridge CB3 0HA, UK}
\author{L.\,K. Hunt}
\affil{INAF-Osservatorio Astrofisico di Arcetri, Largo E. Fermi 5, 50125, Firenze, Italy}
\author{I.~De~Looze}
\affil{Department of Physics \& Astronomy, University College London, Gower Place, London WC1E 6BT, UK}
\and
\author{E. W. Pellegrini}
\affil{Zentrum f\"ur Astronomie der Universit\"at Heidelberg, Institut f\"ur Theoretische Astrophysik, Albert-Ueberle-Str. 2, 69120 Heidelberg, Germany}

%
%
\altaffiltext{1}{ftaba@iac.es}

\begin{abstract}

We study the spectral energy distribution (SED) of the radio continuum emission from the KINGFISH sample of nearby galaxies to understand the energetics and origin of this emission.  Effelsberg multi-wavelength observations at 1.4\,GHz, 4.8\,GHz, 8.5\,GHz, and 10.5\,GHz combined with archive data allow us, for the first time, to determine the mid-radio continuum (1-10\,GHz, MRC)  bolometric luminosities and further present calibration relations vs. the monochromatic radio luminosities. The 1-10\,GHz radio SED is fitted using a Bayesian Markov Chain Monte Carlo (MCMC) technique leading to measurements for the nonthermal spectral index ($S_{\nu} \sim \nu^{-\alpha_{\rm nt}}$)  and the thermal fraction ($f_{\rm th}$) {with mean values of $\alpha_{\rm nt}$=\,0.97$\pm$0.16 (0.79$\pm$0.15 for the total spectral index) and $f_{\rm th}$=(10$\pm$9)\% at 1.4\,GHz. The MRC luminosity changes over $\sim$3 orders of magnitude in the sample, $4.3\times\,10^2\,L_{\sun}<$\,MRC\,$<\, 3.9\times\,10^5\,L_{\sun}$. The thermal emission is responsible for $\sim$23\% of the MRC on average.  We also compare the extinction-corrected diagnostics of star formation rate  with the thermal and nonthermal radio tracers and derive the first star formation calibration relations using the MRC radio luminosity. The nonthermal spectral index flattens with increasing star formation rate surface density, indicating the effect of the star formation feedback on the cosmic ray electron population in galaxies.}
Comparing the radio and IR SEDs, we find  that the FIR-to-MRC ratio could decrease with star formation rate, due to the amplification of the magnetic fields in star forming regions. This particularly implies a decrease in the ratio at high redshifts, where mostly luminous/star forming galaxies are detected. 

\end{abstract}

\keywords{galaxies:star formation -- galaxies: ISM --- catalogs --- surveys}


%
\section{Introduction} \label{sec:intro}
The use of the radio continuum (RC) emission as an extinction-free tracer of star formation in galaxies was first suggested by the tight empirical radio--infrared (IR) correlation, extending to more than 4 orders of magnitude in luminosity \citep[see ][and references therein]{Condon_2}.  However, some authors have raised the possibility of conspiracy of several factors as the cause of the radio-IR correlation \citep{Bell,Lacki_10}. More direct studies of the radio emission properties at several  frequencies are needed to understand the origins, energetics,  and the thermal and nonthermal processes producing the RC emission observed in galaxies.  
Star forming regions as the most powerful source of the RC emission are directly evident in the resolved maps of not only the thermal free-free emission but also the nonthermal synchrotron emission in nearby galaxies \citep{Tabatabaei_3_07,Taba_13,Taba_13_b,Srivastava,Heesen_14}. This is understandable as massive star formation activities like supernova explosions, their shocks, and remnants increase the number density of high-energy cosmic ray electrons (CREs) and/or accelerate them, on one hand, and amplify the turbulent magnetic field strength, on the other hand. The net effect of these processes is a strong nonthermal emission in or around star forming regions. These maps also show that extended structures in non star forming regions emit RC, as well, but at lower intensities than in star forming regions. How these various sources/emission shape the RC spectrum globally and locally is a pressing question today.  

Studying the spectral energy distribution (SED) provides significant information on the origin, energetics, and physics of the electromagnetic radiation in general. The shapes of the SEDs usually reflect the radiation laws and their parameters such as power-law energy index or emissivity index as well as physical phenomenon affecting those parameters like cooling/heating mechanisms in the interstellar medium.  Integrating the SEDs determines the total energy output of a source over a certain frequency range which is a useful parameter to compare the energetics from different regimes of the electromagnetic radiation. 
Comparing the SEDs at different regimes (like in the radio and infrared) provides key insights on the origin/nature of the emission and general factors setting their energy balance. To date, the infrared (IR) SEDs of various astrophysical objects have been dissected thanks to the coherent and simultaneous observations at several bands/frequencies with space telescopes like IRAS, ISO, Spitzer, and Herschel. In radio, however, most of the surveys have targeted a single radio frequency/band (mainly 1.4\,GHz) with different sensitivities/resolutions/observational instruments  prohibiting a coherent (i.e., consistent in terms of performance/observations, targets and selection limits) radio-SED analysis for galaxy samples. This has been mainly because of a simple assumption under which the nonthermal radio spectrum has a fixed power law index of $\alpha_{\rm nt}\sim$\,0.8 (for $S\sim \nu^{-\alpha_{\rm nt}}$). 
However, this assumption cannot explain either the resolved spectra of galaxies \citep[e.g.][]{Tabatabaei_3_07, Taba_13} or the integrated spectra \citep{Duric_88,Marvil}. 

The radio SED of galaxies can be divided into 2 main domains: the nonthermal domain at $\nu\lesssim\,10$\,GHz and the thermal domain at frequencies $10-20\,{\rm GHz}\,<\nu<\,100$\,GHz. The aging of cosmic ray electrons (CREs) and the thermal free-free absorption could cause curvature or flattening of the nonthermal SEDs. Such a flattening and curvature mostly occurs at low frequencies $\nu<\,1$\,GHz in galaxies \citep[e.g.][]{Condon_92_1,Adebahr,Mulcahy,Marvil}. In the mid-frequency range of $1\,<\nu<\,10$\,GHz, the synchrotron power-law index faces minimal variations with frequency, on one hand, and the radio continuum has the least contribution from  spinning dust, on the other hand. Hence, the power-law SED \citep[which is expected if the cooling and aging of CREs occur in a clumpy ISM,][]{Basu_15} could be optimally constrained in this frequency range.
Extrapolating the 1-10\,GHz SEDs  toward lower frequencies would then provide a basis to obtain the amplitude of the various effects causing possible flattening or curvature of the nonthermal spectrum. Toward higher frequencies, the extrapolations would help uncover potential contribution from anomalous dust emission.        

This paper presents a coherent  multi-band survey of the 1-10\,GHz SEDs in a statistically meaningful nearby galaxy sample, the KINGFISH \citep[Key Insights on Nearby Galaxies; a Far-Infrared Survey with Herschel,][]{Kennicutt_11}  sample, providing a wide range in star formation rate, morphology, and  mass with the 100-m Effelsberg telescope.  The KINGFISH sample is ideally suited to characterize the radio SEDs with respect to their IR SEDs that have been presented in \citet{Dale_12}. Without any pre-assumption about $\alpha_{\rm nt}$, the true range of radio SED parameters are searched by means of the Bayesian MCMC technique. The dependence of the radio SED parameters on the star formation rate (SFR) are then studied using the measurements already available for the KINGFISH sample \citep{Kennicutt_11}.
The thermal and nonthermal radio fluxes separated using the SED modeling allow us to estimate the SFR using the basic thermal/nonthermal radio SFR calibration relations presented in \citep{Murphy_11} and to compare the radio SFRs with other extinction-free SFR tracers. 

The paper is organized as follows. After presenting the observations and the data (Sect.~2), we describe the SED modeling and present the results (Sect.~3). In Sect.~4, we introduce the MRC bolometric SED and determine the contribution of the standard radio bands. The calibration relations based on the radio emission are presented in Sect.~5. The decomposed nonthermal emission allows estimation of the equipartition magnetic field strength for the sample (Sect.~6). We then discuss the results (Sect.~7) and summarize our findings (Sect.~8). 
\begin{table*}
\begin{center}
\caption{Basic properties of the galaxy sample.}
\begin{tabular}{ l l l l l l l l l l} 
\hline
Galaxy  &  R.A.   & Dec. & Hubble& Size$^a$ & Inclination$^b$& Distance$^c$& Nuclear & log(TIR)$^d$ & SFR$^c$ \\ 
Name & (J2000) & (J2000) & Type$^a$ & [$\arcmin\times \arcmin$] & [degrees] & [Mpc] &Type$^c$  & [L$_{\odot}$] & [M$_{\sun}$\,yr$^{-1}$] \\ \hline \hline
DDO053      &  08 34 07.2 & +66 10 54 &  Im       &    1.5$\times$1.3& 31  &3.61 &... &7.0& 0.006 \\
DDO154      &  12 54 05.2 & +27 08 55 &  IBm       &   3.0$\times$2.2 & 66  & 4.3&... &6.9$^c$& 0.002 \\
DDO165      &  13 06 24.8 & +67 42 25 & Im             &3.5$\times$1.9& 61&4.57&...&7.0$^c$& 0.002 \\
HoI         &  09 40 32.3 & +71 10 56 &  IABm    &    3.6$\times$3.0 & 12 & 3.9 &...&7.1& 0.004 \\
IC0342      &  03 46 48.5  &  +68 05 46  &  SABcd &  21.4$\times$20.9& 31  & 3.28 & SF&10.1 &1.87 \\
IC2574      &  10 28 21.2  &  +68 24 43  &  SABm  &  13.2$\times$5.4  & 53  &3.79  &  SF& 8.3 & 0.057\\
M81DwB	    &  10 05 30.6 & +70 21 52 &  Im        &    0.9$\times$0.6 & 48 &3.6 & ...&6.5 & 0.001 \\
NGC~0337     &  00 59 50.0  &  -07 34 41  &  SBd    &   2.9$\times$01.8  &52 & 19.3 & SF &10.1& 1.30\\
NGC~0584     &  01 31 20.7 & -06 52 04 &  E4        &    4.2$\times$2.3 &  58& 20.8 &... & 8.8& ...\\
NGC~0628     &  01 36 41.7  &  +15 47 01  &  SAc   &  10.5$\times$09.5  &  25&7.2 &...&9.9 & 0.68\\
NGC~0855     &  02 14 03.6 & +27 52 39 &  E          &    2.6$\times$1.0 &  70& 9.73 &SF & 8.6 &... \\
NGC~0925     &  02 27 17.1  &  +33 34 45  &  SABd   &  10.5$\times$05.9  & 66 & 9.12&SF& 9.7 & 0.54\\
NGC~1266     &  03 16 00.7  &  -02 25 38  &  SB0    &   1.5$\times$01.0  &  32&30.6 & AGN& 10.4& ...\\
NGC~1377     &  03 36 39.1 & -20 54 08 &  S0        &    1.8$\times$0.9 & 62 &24.6 &...& 10.1 & 1.86 \\
NGC~1482  &  03 54 38.9  &  -20 30 08  &  SA0     &   2.5$\times$01.4  &57 &  22.6  & SF & 10.6 & 3.57\\
NGC~2146  &  06 18 37.7  &  +78 21 25  &  Sbab    &   6.0$\times$03.4  & 57&17.2  & SF& 11.0 & 7.94 \\
NGC~2798  &  09 17 22.9  &  +42 00 00  &  SBa    &   2.6$\times$01.0  & 68 &25.8  & SF/AGN& 10.6 & 3.38\\
NGC~2841  &  09 22 02.6  &  +50 58 35  &  SAb    &   8.1$\times$3.5  & 74 &14.1   &AGN& 10.1 & 2.45\\
NGC~2976  &  09 47 15.3  &  +67 55 00  &  SAc    &   5.9$\times$2.7  & 65 &3.55   &  SF& 8.9 & 0.082\\
NGC~3049  &  09 54 49.6  &  +09 16 17  &  SBab  &   2.2$\times$1.4  & 61 & 19.2   & SF& 9.5 & 0.61\\
NGC~3077  &  10 03 19.1  &  +68 44 02  &  I0pec  &   5.4$\times$4.5  & 33& 3.83    & SF&8.9& 0.094\\
NGC~3184  &  10 18 16.9  &  +41 25 28  &  SABcd  &   7.4$\times$6.9  & 16& 11.7   & SF&10.0 & 0.66\\
NGC~3190  &  10 18 05.6 & +21 49 56 &  SAap    &    4.4$\times$1.5   & 73& 19.3      & AGN& 9.9 & 0.38\\ 
NGC~3198  &  10 19 54.9  &  +45 32 59  &  SBc    &   8.5$\times$3.3  & 72 &14.1  & SF &10.0 & 1.01\\
NGC~3265  &  10 31 06.7  &  +28 47 48  &  E        &   1.3$\times$1.0  & 46 &19.6 & SF&9.4 & 0.38\\
NGC~3351  &  10 43 57.7  &  +11 42 13  &  SBb    &   7.4$\times$5.0  &  41&9.93   & SF&9.9 & 0.58\\
NGC~3521  &  10 05 48.6 &  -00 02 09  &  SABbc  &   11.0$\times$5.1  & 73 &11.2   &   SF/AGN&10.5 & 1.95\\
NGC~3627  & 11 20 14.9  & +12 59 30    & SABb    & 9.1$\times$4.2    & 62& 9.38   & AGN &10.4& 1.70\\
NGC~3773  &  11 38 13.0 & +12 06 44 &  SA0      &    1.2$\times$1.0 & 34& 12.4& SF       &8.8 & 0.16\\
NGC~3938  &  11 52 49.4 &  +44 07 15  &  SAc    &  5.4$\times$4.9   &25&17.9     & SF &10.3 & 1.77\\
NGC~4236  &12 16 42.1   & +69 27 45    & SBdm    & 21.9$\times$7.2  &72 &4.45       & SF&8.7 & 0.13\\
NGC~4254  & 12 18 49.6  & +14 24 59    & SAc    & 5.4$\times$4.7  & 29 & 14.4       & SF/AGN&10.6 & 3.92\\
NGC~4321  & 12 22 54.8  & +15 49 19  & SABbc   & 7.4$\times$6.3    & 32&   14.3 & AGN &10.5 & 2.61\\
NGC~4536  &  12 34 27.0 & +02 11 17 & SABbc    & 7.6$\times$3.2    &67 &14.5   & SF/AGN&10.3 & 2.17\\
NGC~4559  &  12 35 57.7 &  +27 57 36 &  SABcd  &  10.7$\times$4.4  &66 &6.98      & SF & 9.5 & 0.37\\
NGC~4569  & 12 36 49.8  & +13 09 47 & SABab &9.5$\times$4.4       &64 &9.86   & AGN& 9.7 & 0.29\\
NGC~4579  & 12 37 43.5  & +11 49 05 & SABb  &5.9$\times$4.7       & 38&16.4   &AGN& 10.1 & 1.10\\
NGC~4594  & 12 39 59.4  & -11 37 23 & SAa   & 8.7$\times$3.5       &69 & 9.08   & AGN& 9.6 & 0.18\\
NGC~4625   &  12 41 52.6 & +41 16 26 &  SABmp &    2.2$\times$1.9 &    30&9.3 &SF &8.8 & 0.052\\
\hline \hline
\label{tab:list}
\end{tabular}
\tablecomments{$a$- NASA Extragalactic Database, $b$- \citet{Hunt_15} and references therein, $c$- \citet{Kennicutt_11} and references therein, $d$- \citet{Dale_12}}
\end{center}
\end{table*}
\begin{table*}
\begin{center}
\caption{Table\,1  continued.}
\begin{tabular}{ l l l l l l l l l l} 
\hline
Galaxy  &  R.A.   & Dec. & Hubble& Size$^a$ &Inclination$^b$ &Distance$^c$& Nuclear & log(TIR)$^d$ & SFR$^c$ \\ 
Name & (J2000) & (J2000) & Type$^a$ & [$\arcmin\times \arcmin$] & [degrees]&[Mpc] &Type$^c$  & [L$_{\odot}$] & [M$_{\sun}$\,yr$^{-1}$] \\ \hline \hline
NGC~4631  & 12 42 08.0  & +32 32 29 & SBd   & 15.5$\times$2.7 &  83 & 7.62           & SF&10.4 & 1.70\\
NGC~4725  & 12 50 26.6  & +25 30 03  &  SABab  &  10.7$\times$7.6& 45 &  11.9     & AGN& 9.9 & 0.44\\
NGC~4736  &  12 50 53.1 &  +41 07 13  &  SAab  &  11.2$\times$9.1  &41 & 4.66     &   AGN&9.8 & 0.38\\
NGC~4826  &  12 56 43.7  &  +21 41 00  &  SAab   &  10.0$\times$5.4 & 65& 5.27    & AGN &9.6 & 0.26\\
NGC~5055  &  13 15 49.3  &  +42 01 46  &  SAbc   &  12.6$\times$7.2  & 59& 7.94    & AGN &10.3& 1.04\\
NGC~5457  &  14 03 12.6  &  +54 20 57  &  SABcd &  28.8$\times$26.9  & 18&6.7     &   SF &10.4 & 2.33\\
NGC~5474  &  14 05 01.5  & +53 39 45   &  SAcd  &   4.8$\times$4.3   & 26 &6.8     & SF &8.8 & 0.091\\
NGC~5713  &  14 40 11.5  &  -00 17 20  &  SABbcp &   2.8$\times$2.5  & 33&21.4    & SF &10.5& 2.52 \\ 
NGC~5866  &  15 06 29.5  &  +55 45 48  &  S0     &   4.7$\times$01.9  & 68 &15.3  & AGN& 9.8 & 0.26 \\
NGC~6946  &  20 34 52.3  & +60 09 14   & SABcd   &   11.5$\times$9.8  & 33&  6.8    & SF& 10.5 & 7.12\\
NGC~7331  &  22 37 04.1  & +34 24 56  &  SAb     &  10.5$\times$03.7  &76 &14.5   &  AGN&10.7 & 2.74 \\
M51     & 13 29 56.2   & +47 13 50  & SAbc     &  11.2$\times$6.9    & 22&7.6$^e$& AGN$^f$&10.5$^g$ & 5.0\\
\hline \hline
\label{tab:list2}
\end{tabular}
\tablecomments{$a$- NASA Extragalactic Database, $b$- \citet{Hunt_15} and references therein, $c$- \citet{Kennicutt_11} and references therein, $d$- \citet{Dale_12}, $e$-~\citet{Ciardullo}, $f$- \citet{Matsushita}, $g$- \citet{Rujopakarn}}
\end{center}
\end{table*}
\section{Data}
\subsection{Radio Observations and Data Reduction}
The KINGFISH sample consists of 61 nearby galaxies of different morphological types. From this sample, we selected all galaxies with declinations $\geq$ 21\degr\ so that they can be observed with the Effelsberg 100-m single dish telescope to obtain global measurements of the radio continuum at 20\,cm, 6\,cm and 3.6\,cm\footnote{Based on observations with the 100-m telescope of the Max-Planck-Institut f\"ur Radioastronomie at Effelsberg}. About 50  galaxies fulfill this criterion. The non-KINGFISH galaxy, M51, is also included in this study. We observed 35 of these galaxies at 6\,cm, 10 galaxies at 20\,cm and 7 at 3.6\,cm to complete already existing archival data during 4 observation runs listed in Table~\ref{tab:date}.
\begin{table}
\begin{center}
\caption{Effelsberg projects}
\begin{tabular}{ll} 
\hline
Project Code&  Observation Date \\ \hline \hline
78--08& December 2008 \\
10--09&  December 2009\\
20--10& April 2010 \\
72--10& December 2010 \& March 2012 \\
\hline \hline
\label{tab:date}
\end{tabular}
\end{center}
\end{table}
Tables~\ref{tab:list} and \ref{tab:list2}  summarize some KINGFISH sample properties, and Table~\ref{tab:obs} the new Effelsberg observations. 
\subsubsection{The 6\,cm observations}
At 6\,cm, the beam size of the Effelsberg telescope is 2\farcm5~ which is comparable to the optical sizes of some of our targets.  
Two modes of observation were used, depending on the size of the target. The 19 smaller and fainter galaxies were observed in the cross-scan mode (point source observations). In this mode, the objects were observed in 20\arcmin\ long scans in azimuth and in elevation  with a velocity of 30\arcmin /min. For galaxies with 20\,cm flux densities lower than $\sim$10 mJy and those not detected in NVSS (11 galaxies), 30 cross-scans were used leading to an on-source time of 30\,min per target. For the other five bright compact galaxies, only 10 cross-scans ($\sim$10\,min per target) were used.  The remaining 16 galaxies  were observed in the mapping mode. The Effelsberg maps at 6\,cm are scanned in the azimuthal direction with a two-horn secondary-focus system, using software beam-switching \citep{Emerson}, corrected for baselevel, and transformed into the RA, DEC coordinate system. We obtained maps of 18\arcmin$\times$10\arcmin\ (grid size of 60\arcsec) for the 5 sources with optical sizes of $D<7\arcmin$, and 26\arcmin$\times$18\arcmin\ maps  for the remaining 10  galaxies. A map size of 28\arcmin$\times$20\arcmin\ was used for NGC~5055.  With 20 coverages per target, we achieved a 0.3\,mJy/beam rms noise. The total on-source observing times are 200min (=20$\times$10min) for the  18\arcmin$\times$10\arcmin\ maps,  320min (=20$\times$16min) for the  26\arcmin$\times$18\arcmin\ maps, and  480min (=20$\times$24min) for NGC~5055. 

The pointed observations were reduced using the program package {\it Toolbox}\footnote{https://eff100mwiki.mpifr-bonn.mpg.de/}. The resulting fluxes were then corrected for opacity and pointing offsets. After correcting for various effects including the gain curve, the conversion from Kelvin to Jansky was applied. The errors reported in Table~\ref{tab:flux} are uncertainties in fitting the cross-scan profiles.
\begin{table}
\begin{center}
\caption{Observing modes and covering areas of the galaxies observed with the 100-m telescope at the three wavelengths.}
\begin{tabular}{ l l l l} 
\hline
Galaxy  & 3.6\,cm & 6\,cm & 20\,cm \\ 
\hline
DDO053   &      ...          & pointed&...\\ 
DDO154   &      ...          & pointed  &... \\ 
DDO165   &       ...          & pointed & ...\\ 
HoI      &       ...          & pointed & ...\\ 
IC2574   &   $21\arcmin \times 14\arcmin$ & ...&... \\
M81DwB   &        ...        & pointed& ...\\ 
NGC~0337  &   $10\arcmin \times 10\arcmin $& pointed  & ...\\
NGC~0584  &            ...     & pointed  &... \\ 
NGC~0628  &   $21\arcmin \times 21\arcmin $  &$26\arcmin \times 18\arcmin$ & $51\arcmin  \times 51\arcmin $ \\
NGC~0855  &             ...    & pointed & ...\\ 
NGC~0925  &             ...    &$26\arcmin  \times 18\arcmin $ & $51\arcmin  \times 51\arcmin $ \\
NGC~1266  &   $10\arcmin  \times 10$\arcmin & pointed & ...\\
NGC~1377  &           ...      & pointed& ... \\ 
NGC~1482  &  $10\arcmin  \times 10\arcmin $ & pointed&... \\
NGC~2146  &           ...     & $18\arcmin  \times 10\arcmin  $ & ...\\
NGC~2798  &  $10\arcmin  \times 10\arcmin $ &pointed  &... \\
NGC~2841  &        ...        & $26\arcmin  \times 18\arcmin $ &... \\
NGC~2976  &       ...         & $26\arcmin  \times 18\arcmin $ &... \\
NGC~3049  &       ...         &pointed & ...\\
NGC~3077  &       ...         & $18\arcmin  \times 10\arcmin $ & ...\\
NGC~3184  &        ...        & $26\arcmin  \times 18\arcmin $ & ...\\
NGC~3190  &        ...      & pointed& ...\\ 
NGC~3198  &        ...    & $26\arcmin  \times 18\arcmin $ & ...\\
NGC~3265  &        ...    & pointed & ...\\
NGC~3351  &        ...     & $26\arcmin  \times 18\arcmin $ & ...\\
NGC~3521  &        ...      &      ...         & $ 51\arcmin  \times 51\arcmin $\\
NGC~3773  &        ...      & pointed&... \\
NGC~3938  &         ...     & $26\arcmin  \times 18\arcmin $ & ...\\
NGC~4559  &        ...      & $26\arcmin  \times 18\arcmin $ & $ 51\arcmin  \times 51\arcmin $  \\
NGC~4625  &         ...     & pointed&\\ 
NGC~4725  &$42\arcmin  \times 28\arcmin $ & $26\arcmin  \times 18\arcmin $ & $ 51\arcmin  \times 51\arcmin $ \\
NGC~4736  &       ...       &                   & $ 51\arcmin  \times 51\arcmin $ \\
NGC~4826  &        ...      & $26\arcmin  \times 18\arcmin $ & $ 51\arcmin  \times 51\arcmin $ \\
NGC~5055  &        ...      & $28\arcmin  \times 20\arcmin $ & $ 51\arcmin  \times 51\arcmin $ \\
NGC~5457  &        ...       &      ...         & $ 90\arcmin  \times 90\arcmin $ \\
NGC~5474  &        ...      & pointed        &  ...                     \\
NGC~5713  &        ...       & pointed        &       ...                  \\ 
NGC~5866  &   $25\arcmin \times 25\arcmin $ &$ 18\arcmin  \times 10\arcmin $ &... \\
NGC~7331  &         ...         & $26\arcmin  \times 18\arcmin $ & $ 51\arcmin  \times 51\arcmin $ \\
\hline \hline 
\label{tab:obs}
\end{tabular}
\tablecomments{The sizes refer to the areas of the observations. The center of the areas are the galaxy centers. Pointed means cross-scan observing mode. See text for details. }
\end{center}
\end{table}
\subsubsection{The 20\,cm observations}
No archival 20\,cm data existed for 10 large galaxies ($>10\arcmin$ in extent). Hence, they were observed in our last run of observations (obs. code 20--10). The Effelsberg maps at 20\,cm (and 3.6\,cm, see below) were scanned alternating in RA and DEC with one-horn systems and combined using the spatial-frequency weighting method by \citet{Emerson_etal_88}. We obtained maps of $51\arcmin \times 51\arcmin $ for all these galaxies but NGC~5457 (M~101) for which a map of $90\arcmin \times 90\arcmin$ was obtained due to its large size.  The beam size at 20\,cm is $9\farcm15$ and we used a sampling of 3\arcmin\, and a scanning velocity of 3\,deg/min. In order to reach the rms noise of about 6\,mJy/beam, we used 4 coverages of 12\,min exposure time for each galaxy (4$\times$26\,min for M\,101).
\subsubsection{The 3.6\,cm observations}
At 3.6\,cm, we observed 7 galaxies with a grid size of 30\arcsec\, and a scanning velocity of 20\arcmin/min. With 13 coverages, we reached an rms noise of 0.5\,mJy/beam.
The beam size at 3.6\,cm is 1.5\arcmin. The map sizes are provided in Table~\ref{tab:obs}.

The data reduction was performed using the NOD2 (and NOD3, M\"uller et al. in prep.) data reduction system ~\citep{Haslam}. The maps were reduced using the program package {\it Ozmapax}. In order to remove scanning effects due to ground radiation, weather condition, and receiver instabilities, we applied the scanning removal program, $\it{Presse}$, of \citet{Sofue_etal_79} in the mapping mode. 

Throughout our observations, the quasars 3C48, 3C138 , 3C147 and 3C286 were used as pointing, focus,  and flux calibrators. 
\subsection{Other data}
The Effelsberg observations complement the already available radio data sets for the KINGFISH sample, which were mainly picked from the NVSS \citep[][]{Condon_98} at 20\,cm and the Atlas of Shapley-Ames Galaxies at 2.8\,cm \citep{Niklas_97_1}. Depending on the galaxy/wavelength, we also used the archival Effelsberg   radio data (see Table~\ref{tab:flux}).   

Herschel data were used to compare the radio and IR SEDs. The sample was observed with the Herschel Space Observatory as part of the KINGFISH project \citep[][]{Kennicutt_11} as described in detail in \citet{Dale_12}, \citet{Aniano_12}. Although we used the  calibrations by \citet[][]{Dale_12}, the newer calibrations reported by \citet{Hunt_15}
change the luminosities by no more than 10-15\%, within the 20\% uncertainties quoted here. Table~\ref{tab:flux} lists the total IR luminosities (TIR) based on Herschel PACS \citep{Poglitsch} and SPIRE \citep{Griffin} data.

We also used the Spitzer MIPS 24\,$\mu$m and the H$\alpha$ data from  SINGS \citep{Dale_07,Kennicutt_03},  and the FUV  data from GALEX \citep{GildePaz_07} as star formation tracers.  
\section{Radio spectral energy distributions}
Table~\ref{tab:flux} lists the integrated radio flux densities at various frequencies. 
The integration was performed up to the optical radius in order to be consistent with the measurements in the IR \citep{Dale_12} (see Sect.~7.3). The background estimate was determined far beyond the optical radius. 
In those cases where particularly bright background radio sources were present in the
field, such sources were first interactively blanked from the image,
before integration. {The integrated radio flux densities can be uncertain in different ways via the calibration uncertainty, map fluctuations, and the baselevel uncertainty of the single-dish observations. The calibration error ($\delta_{\rm cal}$) of the Effelsberg observations is $\simeq$5\% at 3.6\,cm and 6\,cm, and $\simeq$2\% at 20\,cm   \citep[the error in the absolute scale of the radio flux densities is similar , $\simeq$5\%, at different wavelengths,][]{Baars_etal_77}. The error due to the map fluctuations is given by
\begin{equation}
\delta_{\rm rms}=  \sigma_{\rm rms} \sqrt{\rm N_{\rm beam}}= \sigma_{\rm rms}\,\, \frac{a}{\theta} \, \sqrt{\frac{\rm N}{1.133}},
\end{equation}
where $\sigma_{\rm rms}$ is the rms noise level, ${\rm N_{\rm beam}}$ the number of beams, $\theta$ the angular resolution, N the number of pixels, and $a$ the pixel size.  The error due to the baselevel uncertainty  is $\delta_{\rm base}~=~\sigma_{0}\,\, {\rm N_{\rm beam}}$ with $\sigma_{0}$ the zero level uncertainty ($\sigma_{0}= 0.2\,\, \sigma_{\rm rms}$ for the Effelsberg measurements). The total error in the integrated flux densities is hence $\delta= \sqrt{\delta_{\rm cal}^2 + \delta_{\rm rms}^2 + \delta_{\rm base}^2}$, which is  $\simeq$7\% at 3.6, $\simeq$6\% at 6\,cm, and $\simeq$4\% at 20\,cm averaged over the observed galaxy sample.  } 
\begin{figure*}
\resizebox{\hsize}{!}{\includegraphics*{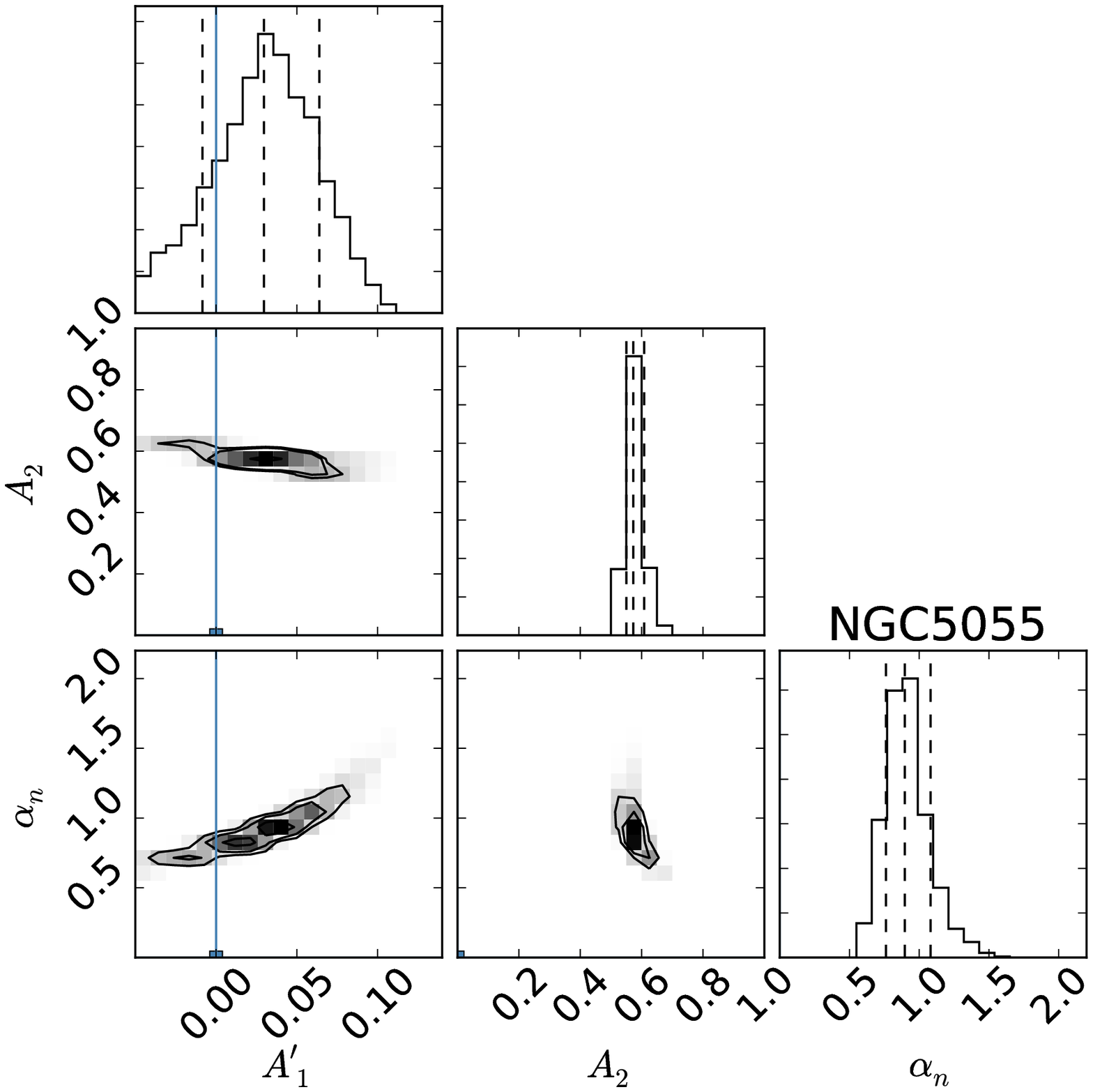}\includegraphics*{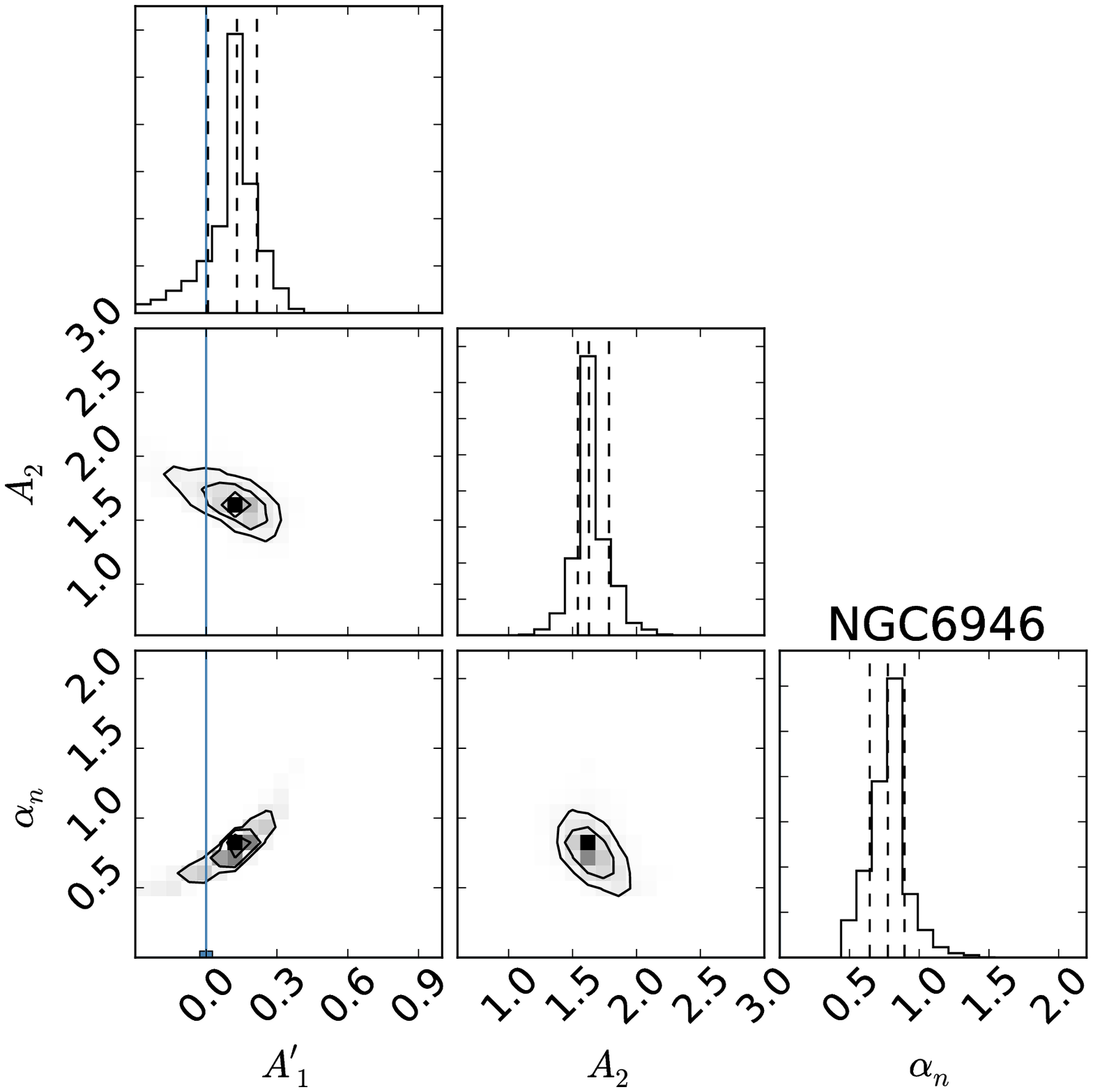}\includegraphics*{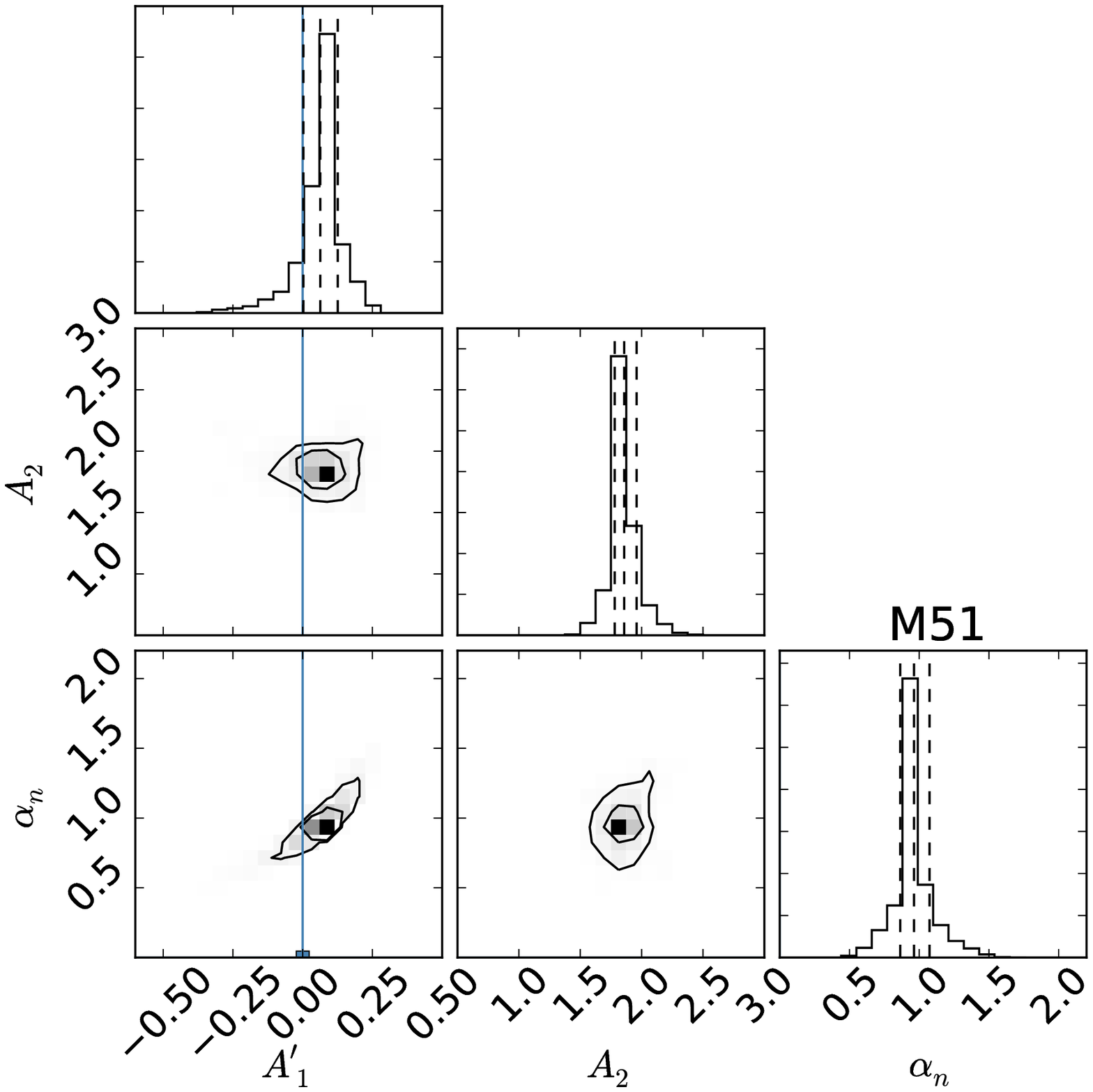}}
\resizebox{\hsize}{!}{\includegraphics*{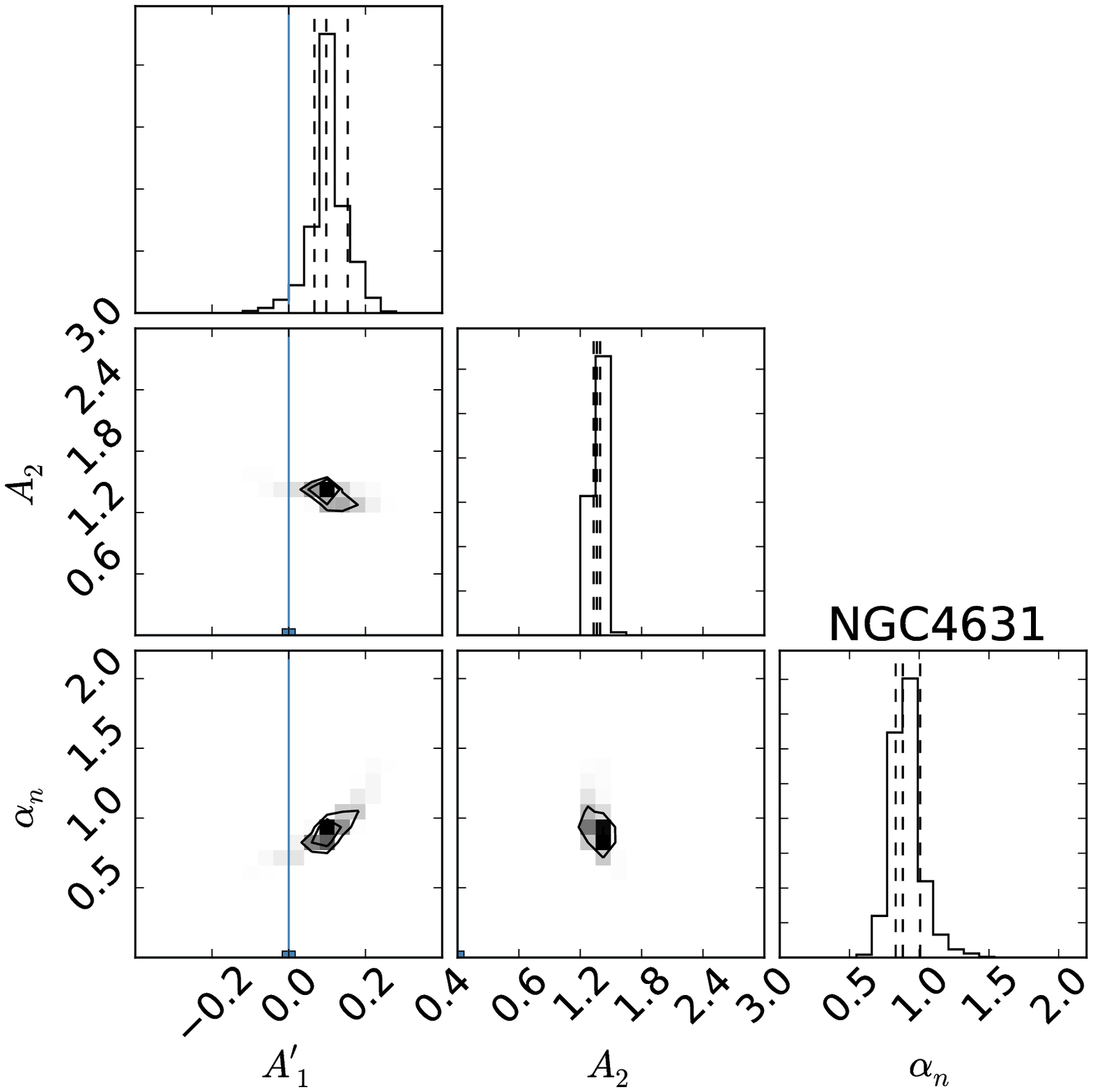}\includegraphics*{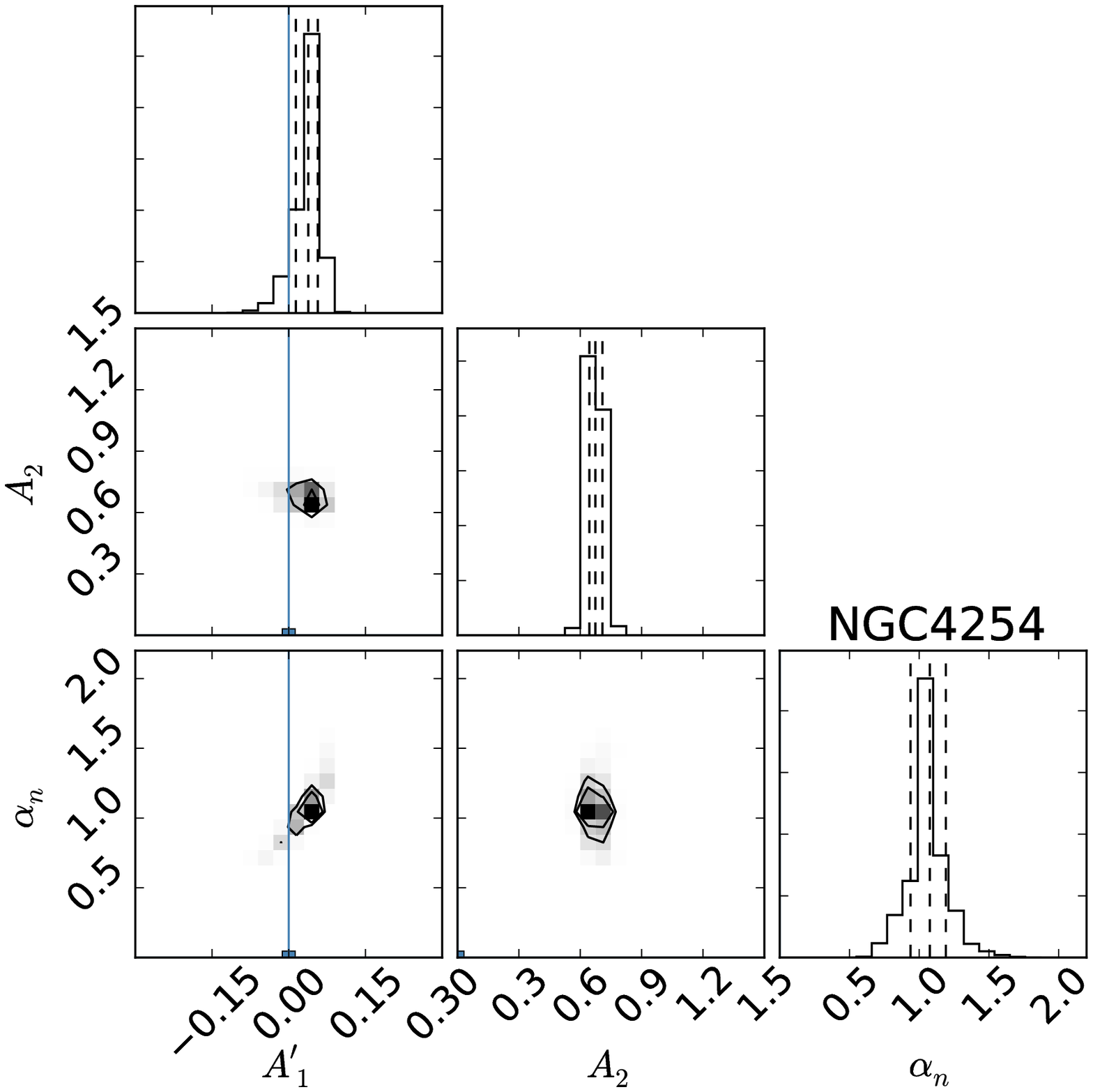}\includegraphics*{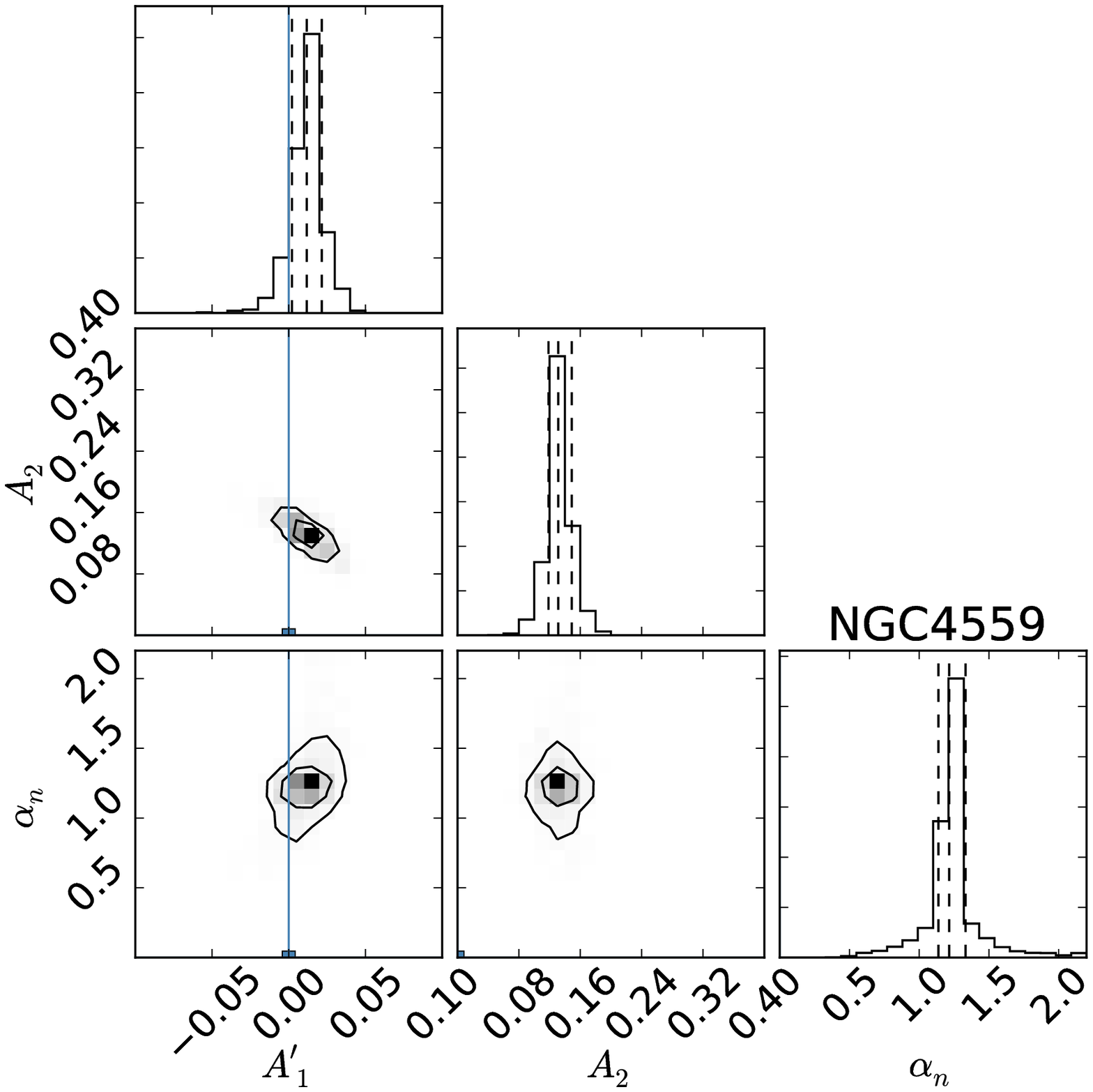}}
\resizebox{\hsize}{!}{\includegraphics*{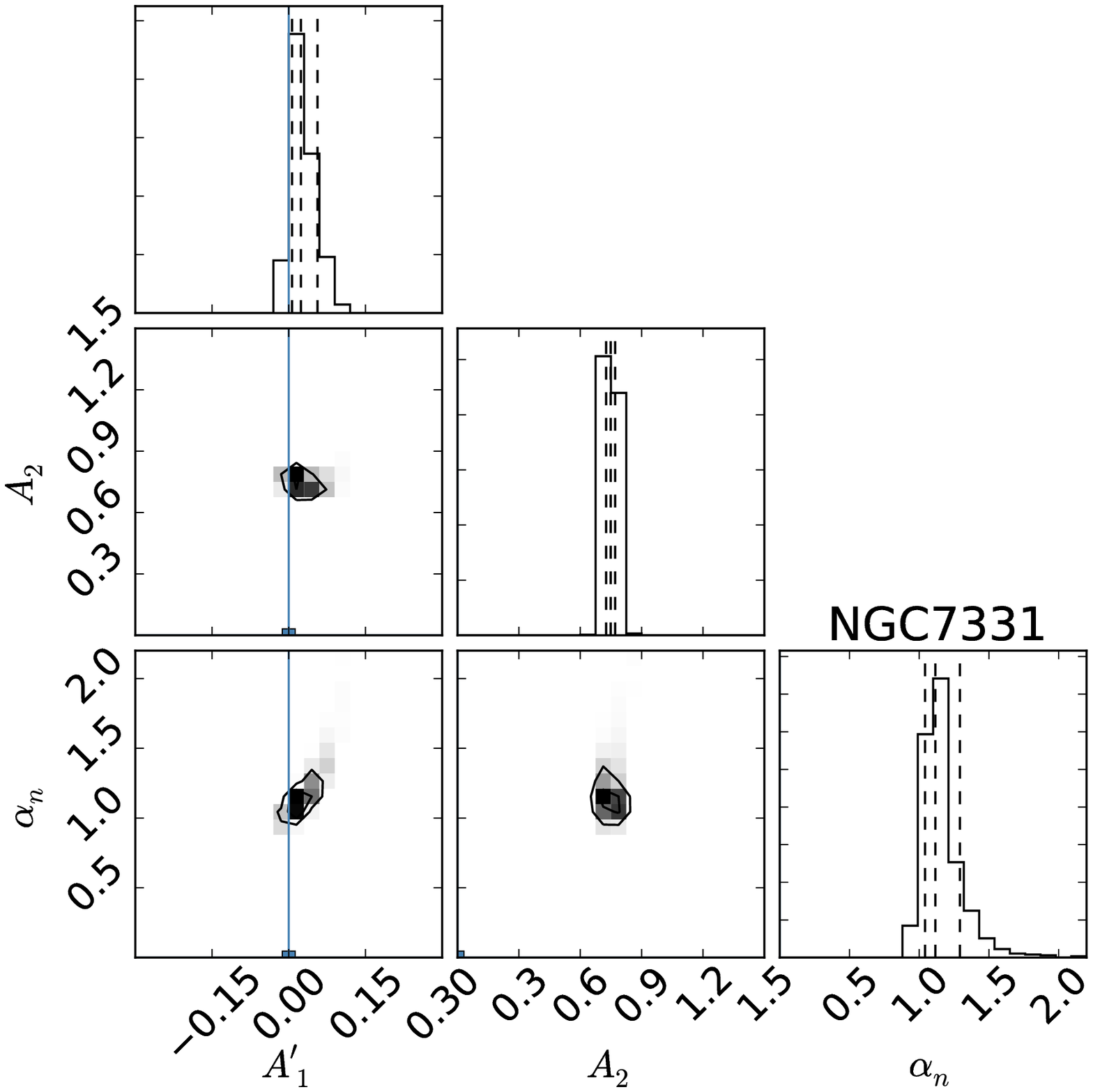}\includegraphics*{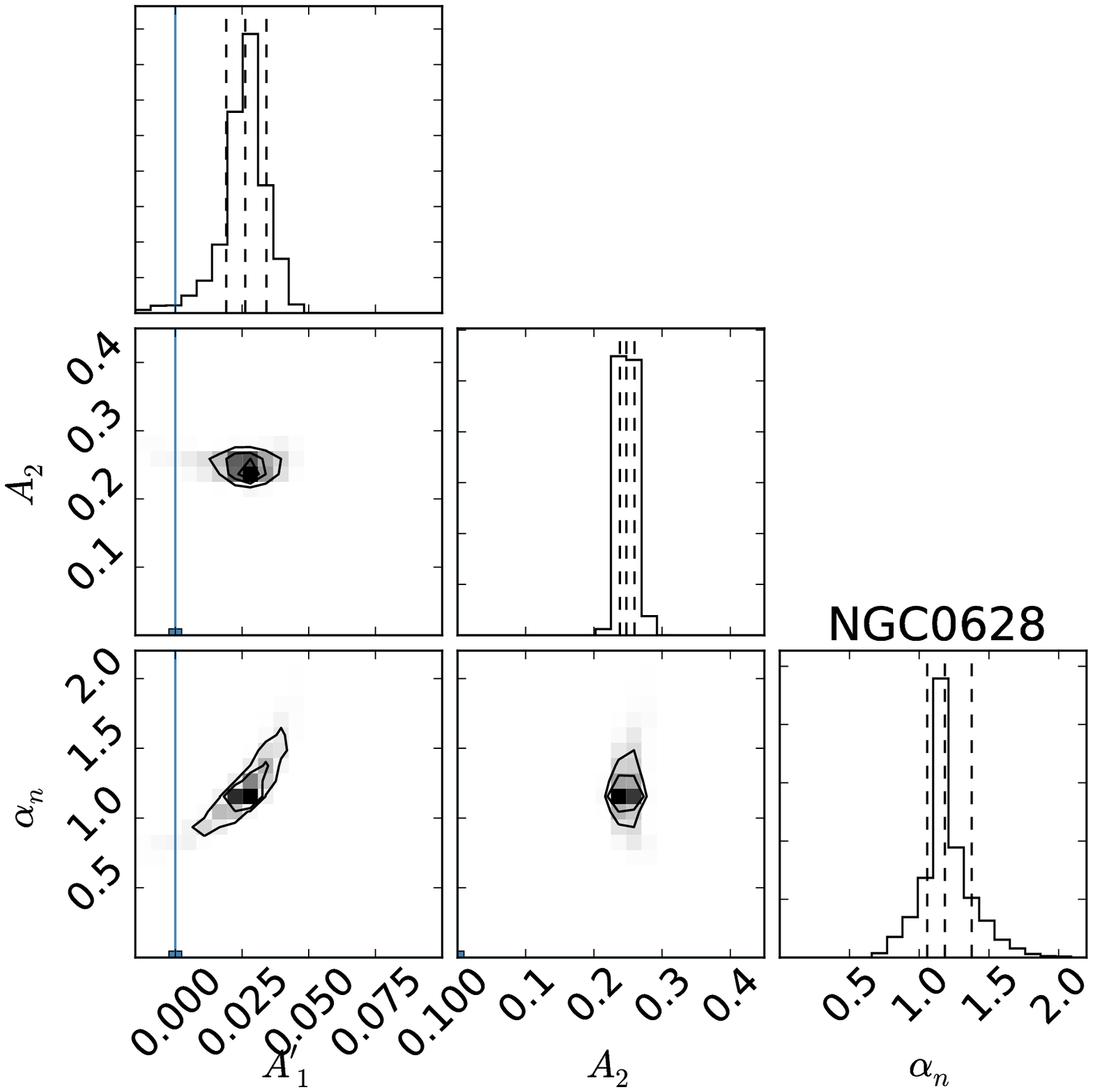}\includegraphics*{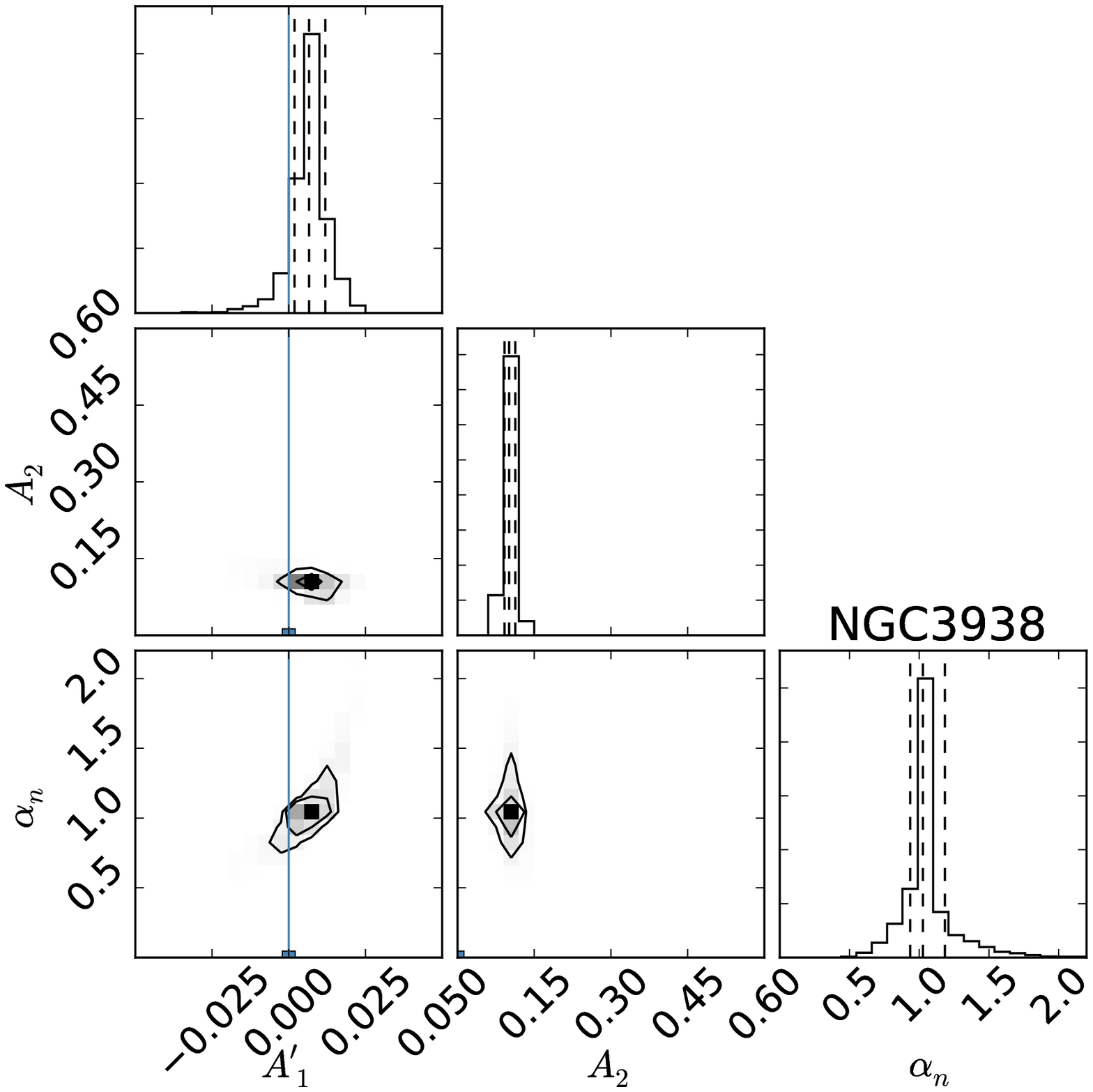}}
\caption{Bayesian corner plots for the parameters $A'_1$, $A_2$, and $\alpha_{\rm nt}$ in Eq.~5  showing the posterior probability distribution function (PDF) and their 0.16, 0.5, 0.86 percentiles (dashed lines) for 9 KINGFISH galaxies. The uncertainty contours show that the posteriors have the highest probability to occur within the confidence intervals indicated. }
\label{fig:example} 
\end{figure*}
\subsection{Modeling the radio SED}
The radio continuum (RC) spectrum is often taken as power law
\begin{equation}
S_{\nu} = A\,\nu^{-\alpha},   
\end{equation}
where $\alpha$ is the power-law index, $\nu$ the frequency, and $A$ a constant factor. However, at frequencies 1\,$<\nu\,<\,$10\,GHz, the RC emission is mainly due to two different mechanisms, the free-free emission from thermal electrons and the nonthermal emission from relativistic electrons.  In terms of these mechanisms, and assuming the optically thin condition for the thermal emission\footnote{The thermal term in this expression is equivalent to the Planck function for an optically thin ionized gas which is usually valid in the ISM and in star forming regions on $\geq$sub-kpc scales. }, the RC spectrum can be expressed as
\begin{equation}
S_{\nu} = S^{\rm th}_{\nu} + S^{\rm nt}_{\nu}= A_1\,\nu^{-0.1} + A_2\,\nu^{-\alpha_{\rm nt}},   
\end{equation}
where $\alpha_{\rm nt}$ is the nonthermal spectral index and $A_1$ and $A_2$ are constant scaling factors. We note that, globally, $\alpha_{\rm nt}$ represents the dominant energy loss mechanism experienced by the CRE population after injection from their sources in a galaxy over the 1-10\,GHz frequency range.  To avoid dependencies on the units of the frequency space, Eq. (3) can be written as 
%
%
\begin{equation}
S_{\nu} = A_1'\,(\frac{\nu}{\nu_0})^{-0.1} + A_2\, \nu_0^{-\alpha_{\rm nt}}\,(\frac{\nu}{\nu_0})^{-\alpha_{\rm nt}},   
\end{equation}
with  $A_1'\,= \nu_0^{-0.1}\,A_1$. The thermal fraction at the reference frequency $\nu_0$ is hence given by: 
\begin{equation}
f_{\rm th}({\nu_0})\equiv S^{\rm th}_{\nu_0}/S_{\nu_0}\,=\, \frac{A_1'}{S_{\nu_0}}. 
\end{equation}

We used a Bayesian MCMC interface to fit the above model to the flux densities and derive the model parameters. This approach provides robust statistical constraints on the fit parameters as it is based on a wide library of
models encompassing all plausible parameter combinations. Given an observed galaxy,  the likelihood distribution of any physical parameter can be derived by evaluating how well each model in the library accounts for the observed properties of the galaxy. The underlying assumption is that the library of models is the distribution
from which the data were randomly drawn. Thus, the prior distribution
of models must be such that the entire observational space is
reasonably well sampled, and that no a priori implausible corner of
parameter space accounts for a large fraction of the models \citep[e.g.,][]{dacunha}. 
We built a model library by generating random combinations of the parameters. To include all possible mechanisms of generation, acceleration, and cooling of cosmic ray electrons, we take $\alpha_{\rm nt}$ to be uniformly distributed over the interval from 0 to 2.2 including injection with $\alpha_{\rm nt}\sim$0.5-0.7 \citep[e.g.][]{Longair,Berkhuijsen_86}  to synchrotron and inverse Compton cooling with $\alpha_{\rm nt}\sim$1-1.2.  {The normalization factors $A_1'$ and $A_2$ are sampled uniformly in the wide ranges $-1\,<A'_1\,<1$ and $-1\,<A_2<\,30$, leading to flux densities in Jy. The negative values are not physically motivated but are included to assess the robustness of the final results and particularly the necessity for the thermal term. } 
\begin{figure}
\begin{center}
\resizebox{\hsize}{!}{\includegraphics*{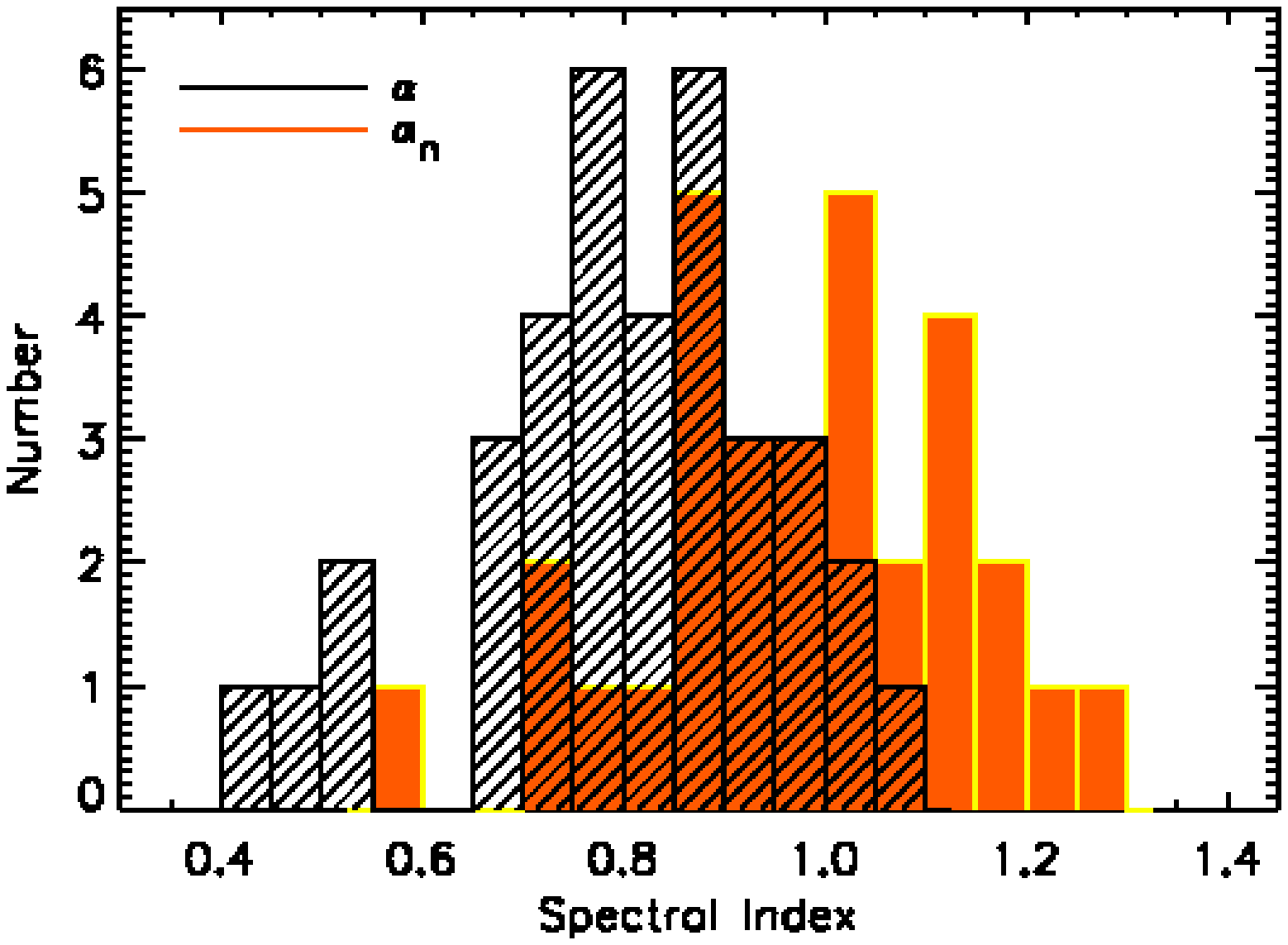}}
\resizebox{\hsize}{!}{\includegraphics*{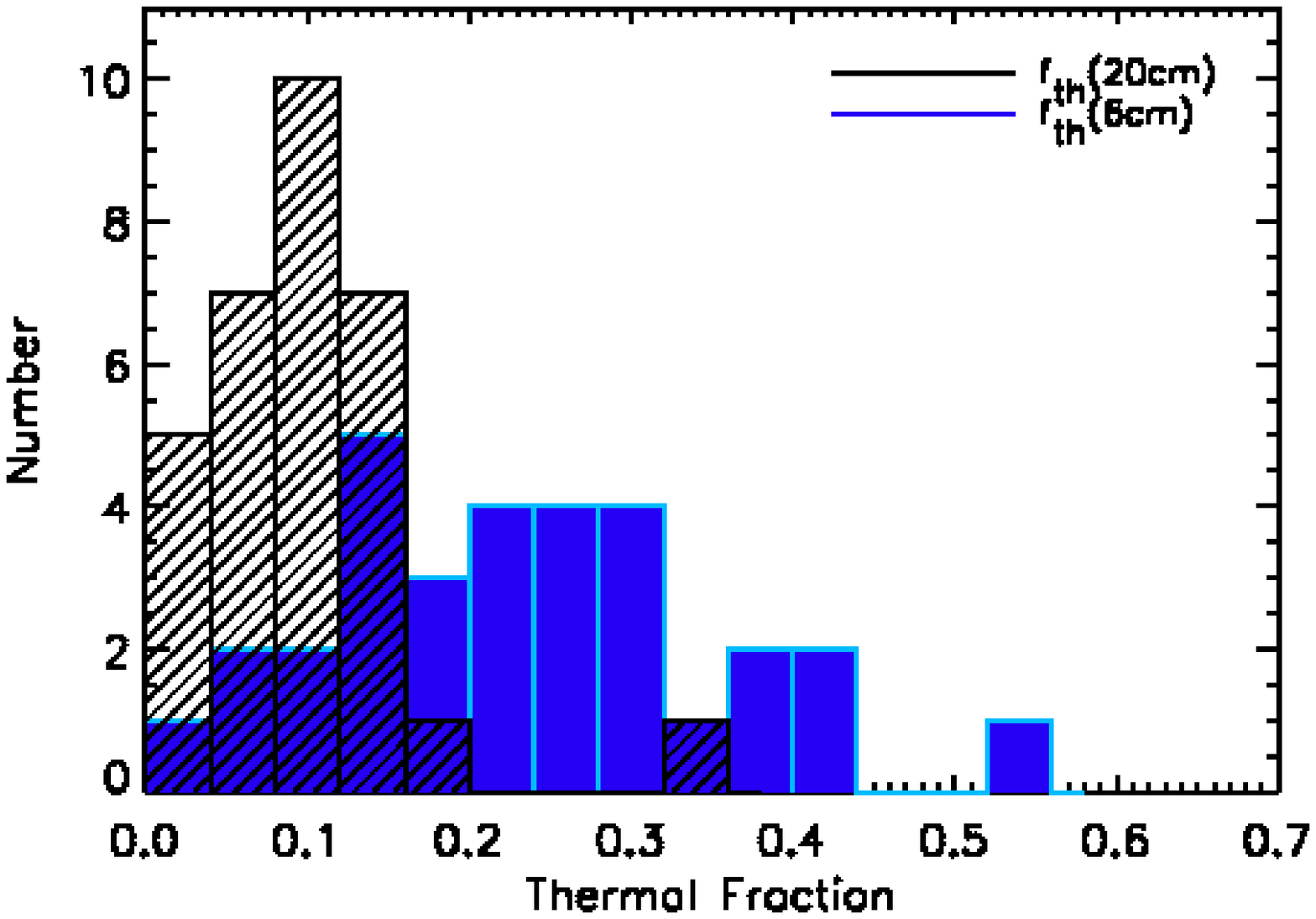}}
\resizebox{\hsize}{!}{\includegraphics*{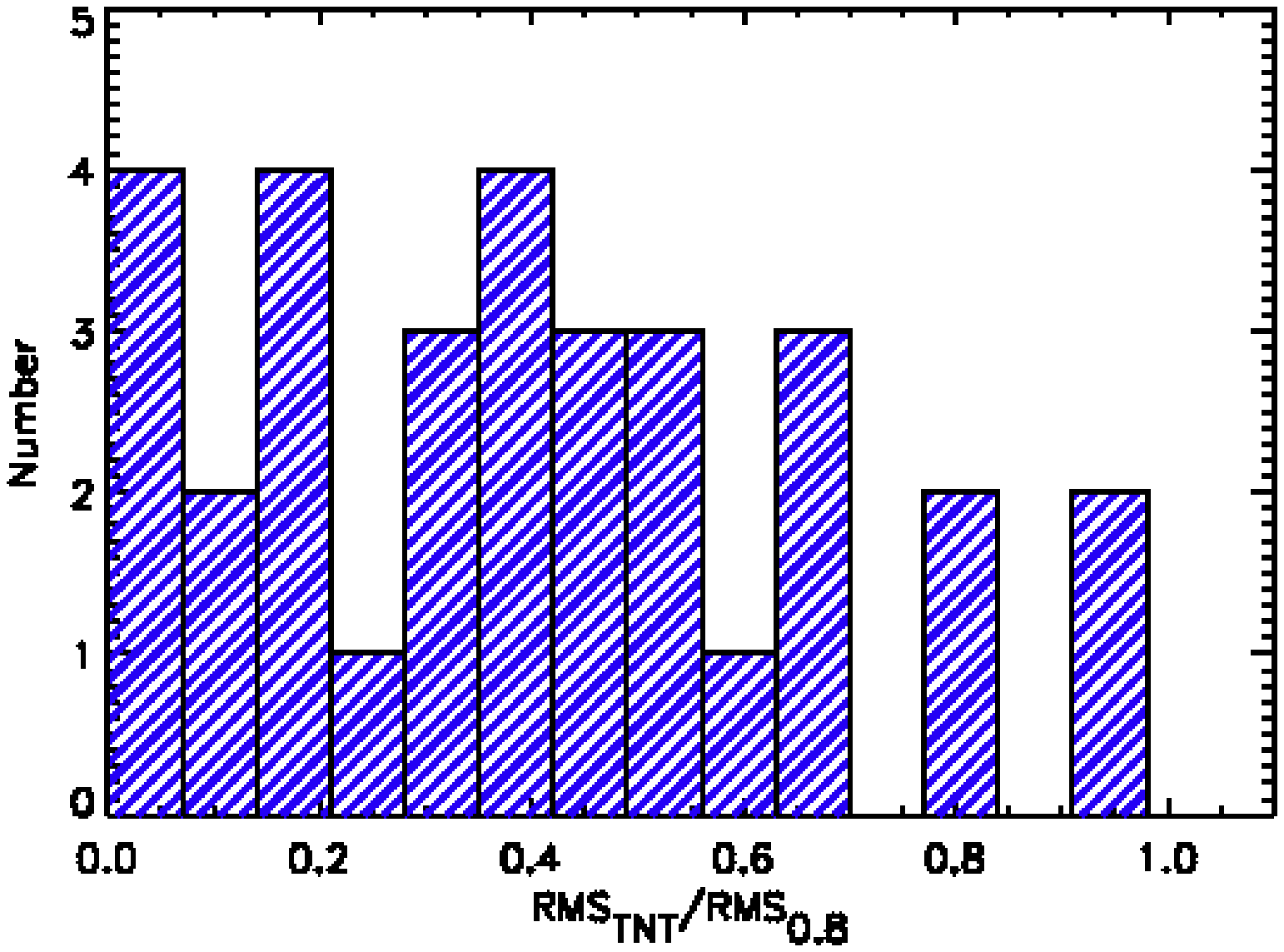}}
\caption{{\it Top}: histogram of the spectral index of the total radio continuum emission, $\alpha$, and its nonthermal component, $\alpha_{\rm nt}$,  of the KINGFISH sample. {\it Middle}: histogram of the thermal fractions at 6\,cm and 20\,cm. {\it Bottom}: {the root mean square deviation of the thermal\,+\,nonthermal model from the observation, RMS$_{\rm TNT}$, divided by the RMS assuming a single power-law model with fixed spectral index of 0.8 (RMS$_{0.8}$). The first model leads to smaller deviations and hence it is more realistic (the median RMS$_{\rm TNT}$/RMS$_{0.8}$\,$\simeq\,0.4$).}    }
\label{fig:hist}
\end{center}
\end{figure}

Using the {\it emcee} code \citep{Foreman}, we obtained the range of probable values (posteriors) for each parameter. The median of the posterior probability distribution function (PDF) is then used as the reported result. 
The uncertainties were then taken as 
the median percentile\,$\pm$\,34\% (or 16\%, 84\%, equal-tailed interval). Figure~\ref{fig:example} shows the posterior PDFs of $A_1'$, $A_2$, and $\alpha_{\rm nt}$ {for 9 representative galaxies.} The scatter plots between each posterior pair are also shown in the same figure. To have more constrained outputs, we applied this method to galaxies with $\geq$ 3 data points{\footnote{We note that, unlike the $\chi^2$ method, the Bayesian MCMC method  is not limited by the number of data points/degree of freedom as it looks for ranges of probable answers. Although the more number of data points with smaller errors leads to more localized PDFs or smaller ranges of uncertainty.}}. Hence, the galaxies with not enough detections/data points were excluded (DDO053, DDO154, DDO165, HoI, M81DwB, NGC~0584, NGC~0855, NGC~0925, NGC~1377, NGC~3198, NGC~3351, NGC~3773, NGC4625, NGC5474). %

\begin{table*}
\begin{center}
\caption{Radio monochromatic flux densities and the MRC luminosities.}
\begin{tabular}{ l l l l l l l l l l l} 
\hline
Galaxy  &S$_{\rm 2.8cm}^{\rm 10.7GHz}$ & S$_{\rm 3.6cm}^{\rm 8.4GHz}$ &  S$_{\rm 6cm}^{\rm 5GHz}$ &S$_{\rm 6.2cm}^{\rm 4.8GHz}$&  S$_{\rm 8.1cm}^{\rm 3.7GHz}$& S$_{\rm 11.1cm}^{\rm 2.7GHz}$ &S$_{\rm 20cm}^{\rm 1.4GHz}$ & S$_{\rm 22cm}^{\rm 1.36GHz}$ & MRC & B \\
Name &  [mJy] &  [mJy] & [mJy] & [mJy] & [mJy]& [mJy] &[mJy] &[mJy] &log\,[$L_{\sun}$]& [$\mu$G] \\ \hline
DDO053   & ...  & ... & ...  & 0.8\,$\pm$\,0.2$^{a}$  & ...  &...  &...  &... & ...& ... \\ 
DDO154   &...   &...  & ...  &$<$0.45$^{a}$ &...&... & $<$1.5$^d$& ...  & ...& ...\\ 
DDO165   &...   &...  & ... &$<$0.43$^{a}$  &...& ...  & $<$1.5$^d$& ...  & ...& ...\\ 
HoI      &  ... & ... & ...& 1.1\,$\pm$\,0.5$^{a}$ &...&...  & $<$1.5$^d$& ... &  ... & ... \\ 
IC0342   & ...   & 430\,$\pm$\,110$^b$  &...  & 860\,$\pm\,$160$^b$& ... &...& 1800\,$\pm$\,300$^b$& ... & 4.36 & ...\\
IC2574   &  ...  &  8.3\,$\pm$\,1.3$^{a}$ &... & 10\,$\pm$\,1$^c$&... &...&   19\,$\pm$\,8$^c$ & ... &2.63& $4.0^{0.9}_{0.5}$  \\
M81DwB   & ...    &      ...   &... &  $<$0.46$^{a}$ &...& ... & ...&... & ...& ...\\ 
NGC~0337  & ...  &  15\,$\pm$\,1$^{a}$   & ... &32\,$\pm$\,2$^{a}$ &...&...& 110\,$\pm$\,4$^d$&...   & 4.56& $14.3^{1.8}_{1.3}$ \\
NGC~0584  & ... & ... & ...  & 1.5\,$\pm$\,0.4$^{a}$ &...  & ...  & $<$1.5$^d$&...  & ... & ...  \\ 
NGC~0628  & 46\,$\pm$\,6$^e$    & 52\,$\pm$\,5$^{a}$ &... &  65\,$\pm$\,7$^{a}$&...&... &  200\,$\pm$\,10$^{a}$& 200\,$\pm$\,10$^f$  & 4.03 & $8.5^{1.5}_{1.3}$\\
NGC~0855  &  ... &...  &  ...  &3.2\,$\pm$\,0.7$^{a}$  &... &...&  4.5$^d$&  ...  &...& ...\\ 
NGC~0925  &  38\,$\pm$\,6$^e$  &  ...  & ... & ...   & ... & ... &  90\,$\pm$\,10$^f$& ...  &...& ...  \\
NGC~1266  &  ...  & 20\,$\pm$\,1$^{a}$   & ... &  35.0\,$\pm$\,6.0$^{a}$&...&...& 115\,$\pm$\,4$^d$& ...   & 5.0& $18.2^{4.7}_{4.2}$\\
NGC~1377  & ...  &...  & ...  & 52.5\,$\pm$\,1.2$^{a}$&...&  ... & $<$1.5$^d$&...  & ...& ... \\ 
NGC~1482  & ... &40.2\,$\pm$\,2.1$^{a}$ & ... &87.5\,$\pm$\,4.9$^{a}$ &...& ...  & 238\,$\pm$\,8$^d$&...    &5.06& ... \\
NGC~2146  &  224\,$\pm$\,6$^e$& ... &  472\,$\pm$\,25$^g$& 439\,$\pm$\,21$^{a}$  & ... & ... &1074\,$\pm$\,40$^d$& 1100\,$\pm$\,10$^f$  & 5.59 &$27.3^{7.8}_{5.2}$ \\
NGC~2798  &... &  23\,$\pm$\,1.5$^{a}$  & ...&   33.8\,$\pm$\,2.5$^{a}$   &...& ...&  82\,$\pm$\,3$^d$& ...  & 4.83 & $19.1^{5.2}_{4.2}$\\
NGC~2841  &  14\,$\pm$\,10$^e$  & ... &  34$\,\pm$\,11$^v$& 38$\,\pm$\,4$^{a}$  &...& 45\,$\pm$\,9$^g$&  ...& 100\,$\pm$\,7$^f$  & 4.30& $15.0^{2.5}_{3.1}$\\
NGC~2976  &  21\,$\pm$\,3$^e$  &... &... & 39\,$\pm$\,3$^{a}$ &...&...  &  125\,$\pm$\,10$^d$ & ... &3.18 & $6.7^{1.3}_{0.7}$ \\
NGC~3049  & ...   & ... &...  &4.8\,$\pm$\,0.4$^{a}$ &...& 8\,$\pm$\,4$^h$ & 12\,$\pm$\,2$^d$& ...  & 3.73&$8.8^{2.5}_{1.0}$  \\
NGC~3077  &  13\,$\pm$\,1$^e$  & ...& ...& 23\,$\pm$1\,$^{a}$  &... &... & 30\,$\pm$\,2$^d$& ...   & 2.88 &...\\
NGC~3184  &  16\,$\pm$\,8$^e$ &... & ...&28\,$\pm$\,3$^{a}$&...&...& 77\,$\pm$\,2$^x$ &   80\,$\pm$\,5$^f$ & 4.06 &$8.7^{3.4}_{1.7}$ \\
NGC~3190  &  15\,$\pm$\,7$^e$& ...& ...&13.5\,$\pm$\,0.5$^{a}$ &...& 22\,$\pm$3\,$^h$& 42\,$\pm$\,8$^{t}$ &  ...  &4.18 & $13.5^{2.6}_{2.1}$\\ 
NGC~3198  & $<$3$^e$ & ...   & ... &12\,$\pm$\,1$^{a}$ &...& ...&  ...   & 49\,$\pm$\,5$^f$  & ...& ...\\
NGC~3265  &... & 3.5\,$\pm$\,0.5$^{a}$ & ... &5.7\,$\pm$\,0.6$^{a}$ & ...  & ...&  10.1\,$\pm$\,0.9$^d$&  ...    & 3.72 & $8.2^{2.2}_{1.6}$\\
NGC~3351 &  14\,$\pm$\,2$^e$  &...   & ... &   ...   &...  &...   &  43\,$\pm$\,10$^d$&  ...   & ... & ... \\
NGC~3521  &  80\,$\pm$\,20$^e$ & ...   & ...  &170\,$\pm$\,14$^i$  & ...  &300\,$\pm$\,60$^j$ &  560\,$\pm$\,20$^{a}$&  ...  &4.82 & $19.6^{2.3}_{2.2}$  \\
NGC~3627  & 100\,$\pm$\,10$^e$ &... & 177\,$\pm$\,23$^v$ & 181\,$\pm$\,41$^b$   &... &... & ...   &500\,$\pm$\,10$^f$  &4.68 & $16.1^{5.4}_{4.5}$\\
NGC~3773  &   ...        &...  & ...& 2.9\,$\pm$\,0.3$^{a}$   &... &... &... & ...&...& ... \\
NGC~3938   &15\,$\pm\,$4$^e$   &...  & ... &26.3\,$\pm\,1.5^{a}$&...  &...   & ...  & 80\,$\pm$\,5$^f$  & 4.04& $9.1^{2.2}_{1.7}$ \\
NGC~4236  &9\,$\pm$\,1$^e$   &...  & ... & 23\,$\pm$\,3$^c$  & ...  & ...  &  48\,$\pm$\,6$^c$& ...   & 3.07& ... \\
NGC~4254  &  93\,$\pm$\,8$^e$ & 102$\,\pm$\,5$^k$ & 135 \,$\pm$\,19$^v$& 167\,$\pm$\,16$^k$    &... &...& 512\,$\pm$\,19$^k$&  510\,$\pm$\,10$^f$  &5.02 &  $16.5^{2.1}_{3.0}$\\
NGC~4321 & 61\,$\pm$\,5$^e$ & 66\,$\pm$\,6$^b$ & ... & 96\,$\pm$\,5$^l$  & ...&...&...    & 310\,$\pm$\,10$^f$ &4.79 & $13.3^{1.5}_{1.8}$\\
NGC~4536 &39\,$\pm$\,3$^m$  &42\,$\pm$\,4$^m$ & ... &80\,$\pm$\,2$^m$   &... & ...& 205\,$\pm$\,20$^d$ & ...     &4.69 & $17.3^{1.4}_{1.2}$\\
NGC~4559  &  18\,$\pm$\,11$^e$  & ...  & 31\,$\pm$\,11$^v$ & 38\,$\pm$\,3$^{a}$&...  &...  & 100\,$\pm$\,4$^{a}$& 110\,$\pm$\,10$^f$ & 3.68 &  $9.3^{0.8}_{0.7}$\\
NGC~4569 &30\,$\pm$\,6$^e$ &  36\,$\pm$\,10$^b$&...& 57\,$\pm$\,20$^{s}$ &...&...&  ...  &170\,$\pm$\,10$^f$   &4.13&  $11.7^{4.9}_{4.3}$\\ 
NGC~4579 &  82\,$\pm$\,4$^e$ & 60\,$\pm$\,10$^m$ & 57\,$\pm$\,17$^v$   &99\,$\pm$\,10$^m$& ...&...& 167\,$\pm$\,25$^n$&...  & 4.84& ...\\
NGC~4594 & 133\,$\pm$\,8$^e$ & ...& ...&156\,$\pm$\,13$^i$   &...&...& 94\,$\pm$\,20$^d$&... &...& ...\\
NGC~4625  &  ...    &   ...  &... & 3.1\,$\pm$\,0.3$^{a}$  &... &... &7.1\,$\pm$\,0.2$^{x}$&...&...& ...\\
\hline \hline 
 \label{tab:flux}
\end{tabular}
\end{center} 
\end{table*}

\begin{table*}
\begin{center}
\caption{Table 5 continued.}
\begin{tabular}{ l l l l l l l l l l l} 
\hline
Galaxy  &S$_{\rm 2.8cm}^{\rm 10.7GHz}$ & S$_{\rm 3.6cm}^{\rm 8.4GHz}$ & S$_{\rm 6cm}^{\rm 5GHz}$ & S$_{\rm 6.2cm}^{\rm 4.8GHz}$& S$_{\rm 8.1cm}^{\rm 3.7GHz}$& S$_{\rm 11.1cm}^{\rm 2.7GHz}$ &S$_{\rm 20cm}^{\rm 1.4GHz}$ &S$_{\rm 22cm}^{\rm 1.36GHz}$& MRC& B  \\
Name &  [mJy] &  [mJy] & [mJy] & [mJy]& [mJy] & [mJy] &[mJy]&[mJy] &log\,[$L_{\sun}$]& [$\mu$G] \\ \hline
NGC~4631 & 265\,$\pm$\,12$^e$ &310\,$\pm$\,16$^b$& ...   & 430\,$\pm$\,20$^b$  &...&...&  1122\,$\pm$\,50$^w$& ...   &4.69&  $24.7^{3.0}_{2.5}$\\ 
NGC~4725  & ...& 19\,$\pm$1\,$^{a}$  & .. &30\,$\pm$\,2$^{a}$ &... & ... &  92\,$\pm$\,3$^{a}$&   100\,$\pm$\,10$^{f}$& 4.11& $10.2^{1.9}_{1.8}$\\
NGC~4736  &  90\,$\pm$\,18$^e$  &...  & 111\,$\pm$\,10$^{g}$& 125\,$\pm$\,10$^b$   &... & ...&  295\,$\pm$\,5$^{a}$& 320\,$\pm$\,10$^{f}$  & 3.92& $8.9^{1.5}_{1.9}$  \\
NGC~4826  & 29\,$\pm$\,16$^e$ &... & 58\,$\pm$\,12$^v$& 54\,$\pm$\,4$^{a}$   &...&... & 126\,$\pm$\,2$^{a}$&...  & 3.63&   $8.7^{2.2}_{1.7}$ \\
NGC~5055  &  97\,$\pm$\,8$^e$  & ... & 116\,$\pm$\,21$^v$& 167\,$\pm$\,8$^{a}$    & 254\,$\pm$\,51$^{g}$ & 260\,$\pm$\,20$^{g}$& 460\,$\pm$\,5$^{a}$& 450\,$\pm$\,10$^{f}$  & 4.49& $14.1^{2.0}_{1.0}$ \\
NGC~5457  & 152\,$\pm$\,62$^g$ & ...& ...& 310\,$\pm$\,20$^b$ &...&442\,$\pm$\,30$^g$ & 760\,$\pm$\,17$^{a}$& ...  & 4.61& $12.9^{1.2}_{1.9}$ \\
NGC~5474  &  ...     & ...&... &5.0\,$\pm$\,0.6$^{a}$ &... &... & ...&...& ...& ...\\
NGC~5713  & 41\,$\pm$\,3$^e$  &31\,$\pm$\,1$^o$ & ... & 58.8\,$\pm$\,2.7$^{a}$  &... & ... & 158\,$\pm$\,6$^d$&  ...  & 4.89 & $16.4^{3.0}_{2.7}$\\ 
NGC~5866  &... & 9.1\,$\pm$\,0.6$^{a}$  &  13\,$\pm$\,6$^v$&   12.1\,$\pm$\,0.8$^{a}$    &...&...&  22\,$\pm$\,1$^r$& ...   &3.90 & $11.1^{6.0}_{3.2}$\\
NGC~6946 & 376\,$\pm$\,18$^b$& 422\,$\pm$\,65$^p$&... & 660\,$\pm$\,50$^b$  &...&794\,$\pm$\,75$^b$ & 1440\,$\pm$\,100$^p$& ... & 4.92&$16.0^{2.4}_{3.0}$\\
NGC~7331  & 77\,$\pm$\,5$^e$  & ...  & 94\,$\pm$\,13$^v$ & 173.8\,$\pm$\,8.7$^{a}$ &...&...& 540\,$\pm$\,9$^{a}$& ... & 5.00& $23.6^{2.3}_{1.8}$\\
M51      &235\,$\pm$\,32$^q$& 306\,$\pm$\,26$^b$ & ... & 420\,$\pm$\,80$^b$   & ...     &780\,$\pm$\,50$^q$& 1400\,$\pm$\,100$^z$& ...  & 4.95& $15.5^{3.5}_{3.4}$\\
\hline \hline 
 \label{tab:flux2}
\end{tabular}
\tablecomments{Upper limits at 20cm refer to the $3\sigma$ limit of the NVSS at these positions.
$a$- This work, $b$- archival Effelsberg data (IC0342: \citet{Beck_15} NGC~4569: \citet{Chyzy_06}, NGC~4631: \citet{Mora}, {NGC~5457: \citet{Berkhuijsen_16}, NGC~6946: \citet{Ehle} \& \citet{Harnett_89}}, for the rest see \citet{Stil}), $c$-\citet{Chyzy_07}, $d$- \citet{Condon_98}, $e$-\citet{Niklas_97_1}, $f$-\citet{Braun_07}, $g$- \citet{Kleinem}, $h$- \citet{Dressel_78},  $i$-\citet{Griffith} and \citet{Griffith2}, $j$- Parkes Catalogue, 1990, Australia Telescope National Facility, $k$-\citet{Chyzy_07_2},  $l$- \citet{Wezgowiec}, $m$- \citet{Vollmer_04},  $n$-\citet{Murphy_9},  $o$- \citet{Schmitt}, $p$- \citet{Taba_13}, $q$-\citet{Klein_1984}, $r$-\citet{Brown}, $s$- average of measurements by \citet{Chyzy_06} and \citet{Vollmer_04},  $t$- \citet{Gioia_87}, $v$- \citet{Sramek}, $w$- \citet{White_92}, $x$- \citet{Condon_2}, $z$- \citet{Dumas}. The MRC luminosity is calculated using Eq.\,(6). }
\end{center} 
\end{table*}
\subsection{KINGFISH radio SED parameters}
\begin{figure}
\begin{center}
\resizebox{\hsize}{!}{\includegraphics*{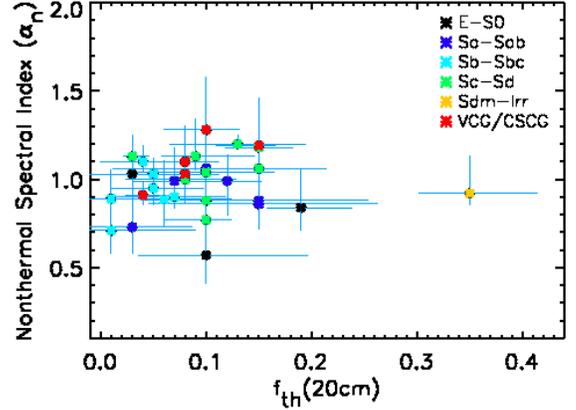}}
\caption{The nonthermal spectral index $\alpha_{\rm nt}$ against the thermal fraction at 20cm, $f_{\rm th}$(20cm), showing no correlation. }
\label{fig:anfth}
\end{center}
\end{figure}

{Figs.~\ref{fig:sed1} and \ref{fig:sed2} show the final modeled SEDs. Five galaxies, IC0342, NGC\,1482, NGC\,3077, NGC\,4236, and NGC\,4579, fit into the single-component model only. Fitting the double-component model leads to negative thermal fractions  in these galaxies which are not realistic and do not agree with other thermal-nonthermal decomposition methods (see Appendix). Inconsistent radio flux densities collected from the archive, or presence of variable radio-loud AGN \citep[as in the case of NGC\,4579 hosting a LINER, e.g.][]{Stauffer} could cause this failure.  It is also possible that $\alpha_{\rm nt}$ changes in the 1-10\,GHz frequency range for IC0342, NGC\,1482, NGC\,3077, NGC\,4236, due to the apparent curvature in their SED (Figs.~\ref{fig:sed1} and \ref{fig:sed2}). However, this cannot be judged with only 3 data points available for these galaxies.  Residuals between the thermal \& nonthermal model and the observed fluxes are less than 20\% (modeled-observed/observed) for most cases. Larger residuals are found at the high-frequency end for NGC~3190, NGC~4236, and NGC~5713}.   
The galaxy  NGC~4594 does not fit into the either double- or single-component models as it shows an inverted spectrum. This galaxy is known to host a strong radio variable source \citep[a LINER, see also][]{Hummel}. Hence this galaxy was excluded from the rest of the analysis.   The resulting $\alpha_{\rm nt}$, and the thermal fractions at 6cm, $f_{\rm th}$(6cm), and at 20cm, $f_{\rm th}$(20cm), together with their uncertainties are given in Table~\ref{tab:result}. Figure~\ref{fig:hist} illustrates the distribution of these parameters in the sample. {The nonthermal spectral index changes between  $0.57^{+0.36}_{-0.16}$ and $1.28^{+0.32}_{-0.20}$ with a mean of $\alpha_{\rm nt}\,\simeq\,0.97$ (median of 0.99) and a standard deviation of 0.16.  The mean thermal fractions are $f_{\rm th}$(6cm)$=\,(23\,\pm\,13)\%$ and $f_{\rm th}$(20cm)$=\,(10\,\pm\,9)\%$ }  over the entire sample and errors are the standard deviation. 
{The dwarf irregular (Irr) galaxy  IC~2574 shows the highest thermal fraction in the sample ($f_{\rm th}$(6cm)$\sim$55\%, $f_{\rm th}$(20cm)$\sim$35\%)}. The relatively high thermal fraction in irregular galaxies was already known from previous studies in the Magellanic clouds \citep{Loiseau,Jurusik}. 
Plotting  $\alpha_{\rm nt}$ against the thermal fractions given in Table~\ref{tab:result}, we see no obvious trend or correlation (Fig.~\ref{fig:anfth}). 
Hence, the method did not introduce a correlation between the final parameters which, in principle, could occur due to simultaneous fitting and degeneracy. 

In a separate run, we also determined the spectral index of the total continuum emission $\alpha$ following Eq.(2), for all the sample, which disregards the flattening by the thermal emission. {A uniform prior was taken for $\alpha$ in the range $0~<~\alpha ~<~2.2$ and for the normalization factor $A$ in the range $-1 < A < 30$. 
For the galaxy sample and in the 1.4-10.5\,GHz range, $\alpha$  changes from $0.40^{+0.07}_{-0.04}$ to $1.08^{+0.04}_{-0.03}$ with a mean value of $\alpha\,\simeq\,0.79$  and a standard deviation of 0.15} (Table~\ref{tab:result}). 

The average $\alpha$ and $\alpha_{\rm nt}$ are slightly higher than those reported by \citet{Israel_88}, \citet{Gioia}, \citet{Kleinem}, and \citet{Niklas} as they included frequencies lower than 1\,GHz ($\nu \sim$0.4-10.7\,GHz), i.e., the SED flattening domain. It is important to note the wide range in the parameters. Most importantly, the synchrotron spectral index is  not fixed in the sample \citep[in agreement with][]{Duric_88}.  We discuss the dependencies of $\alpha_{\rm nt}$  on star formation properties in Sect.~7.1

An almost common assumption about the radio SED is a single power-law model with a fixed spectral index of 0.8. Figure \ref{fig:hist}-bottom shows that this simple model leads, on average, to larger errors than the thermal $+$ nonthermal model.

\begin{table}
\begin{center}
\caption{Radio SED parameters of the KINGFISH sample. }
\begin{tabular}{lllll} 
\hline
Galaxy  & $\alpha_{\rm nt}$ &  $f_{\rm th}$(6cm)  & $f_{\rm th}$(20cm) &   $\alpha$\\ \hline \hline
IC0342  &  ... & ... & ...  &   $0.75^{0.14}_{0.11}$\\\vspace{.1cm}
IC2574  &   $0.92^{\mathrm{0.21}}_{\mathrm{0.07}}$  & $0.55^{0.14}_{0.12}$ &   $0.35^{0.08}_{0.06}$ & $0.50^{0.05}_{0.04}$  \\\vspace{.1cm}
NGC~0337 & $1.13^{0.12}_{0.05}$   & $0.08^{0.09}_{0.03}$ & $0.03^{0.02}_{0.01}$ &$1.08^{0.04}_{0.03}$\\\vspace{.1cm}
NGC~0628 &  $1.18^{0.17}_{0.13}$  & $0.44^{0.11}_{0.12}$ & $0.15^{0.04}_{0.04}$ & $0.84^{0.03}_{0.03}$ \\\vspace{.1cm}
NGC~1266 &  $1.03^{0.20}_{0.16}$  &  $0.08^{0.15}_{0.20}$  &$0.03^{0.07}_{0.09}$  &  $0.97^{0.03}_{0.03}$\\\vspace{.1cm}
NGC~1482 &  ... &  ...  & ...  &  $0.96^{0.03}_{0.03}$\\\vspace{.1cm}
NGC~2146  & $0.71^{0.20}_{0.13}$  &  $0.02^{0.20}_{0.25}$  &$0.01^{0.10}_{0.12}$  &  $0.68^{0.03}_{0.02}$\\  \vspace{.1cm}
NGC~2798  & $0.73^{0.19}_{0.15}$  &  $0.07^{0.10}_{0.18}$  &$0.03^{0.07}_{0.13}$  &  $0.70^{0.03}_{0.03}$\\\vspace{.1cm}
NGC~2841  & $1.06^{0.14}_{0.19}$  &  $0.22^{0.07}_{0.21}$  &$0.10^{0.04}_{0.11}$  &  $0.81^{0.09}_{0.08}$\\\vspace{.1cm}
NGC~2976 & $1.13^{0.21}_{0.08}$  &  $0.27^{0.20}_{0.14}$  &$0.09^{0.07}_{0.04}$  &  $0.93^{0.07}_{0.07}$\\ \vspace{.1cm}
NGC~3049  &$0.86^{0.24}_{0.06}$  &  $0.31^{0.27}_{0.25}$  &$0.15^{0.14}_{0.13}$  &  $0.75^{0.11}_{0.15}$\\\vspace{.1cm}
NGC~3077 &   ... &   ...  &  ... &  $0.40^{0.07}_{0.04}$\\\vspace{.1cm}
NGC~3184  &$1.06^{0.40}_{0.18}$  &  $0.39^{0.25}_{0.20}$  &$0.15^{0.08}_{0.07}$  &  $0.82^{0.17}_{0.12}$\\\vspace{.1cm}
NGC~3190  &$0.99^{0.19}_{0.15}$  &  $0.18^{0.10}_{0.11}$  & $0.07^{0.4}_{0.05}$ &  $0.89^{0.05}_{0.02}$\\\vspace{.1cm}
NGC~3265 &$0.85^{0.21}_{0.13}$   &  $0.33^{0.10}_{0.07}$  &$0.19^{0.06}_{0.04}$  &  $0.73^{0.05}_{0.02}$\\ \vspace{.1cm}
NGC~3521  &$1.04^{0.09}_{0.08}$  &  $0.15^{0.18}_{0.21}$  &$0.05^{0.04}_{0.06}$  &  $0.95^{0.08}_{0.08}$\\\vspace{.1cm}
NGC~3627  &$0.89^{0.22}_{0.15}$  &  $0.16^{0.20}_{0.24}$  &$0.06^{0.08}_{0.09}$  &  $0.79^{0.03}_{0.02}$\\\vspace{.1cm}
NGC~3938  &$1.04^{0.23}_{0.16}$  &  $0.28^{0.20}_{0.22}$  &$0.10^{0.08}_{0.09}$  &  $0.87^{0.03}_{0.02}$\\\vspace{.1cm}
NGC~4236  & ...  & ... & ...  &  $0.76^{0.02}_{0.02}$\\\vspace{.1cm}
NGC~4254  &$1.03^{0.09}_{0.16}$  &  $0.20^{0.09}_{0.14}$  &$0.07^{0.04}_{0.05}$  &  $0.88^{0.03}_{0.03}$\\\vspace{.1cm}
NGC~4321  &$1.19^{0.12}_{0.15}$  &  $0.43^{0.07}_{0.20}$  &$0.15^{0.02}_{0.10}$  &  $0.84^{0.03}_{0.04}$\\\vspace{.1cm}
NGC~4536  &$0.91^{0.07}_{0.06}$  &  $0.12^{0.06}_{0.04}$  &$0.04^{0.03}_{0.02}$  &  $0.85^{0.05}_{0.05}$\\\vspace{.1cm}
NGC~4559  &$1.20^{0.05}_{0.03}$  &  $0.31^{0.25}_{0.30}$  &$0.13^{0.10}_{0.12}$  &  $0.92^{0.14}_{0.16}$\\\vspace{.1cm}
NGC~4569  &$1.28^{0.32}_{0.20}$  &  $0.25^{0.15}_{0.18}$  &$0.10^{0.04}_{0.05}$  &  $1.01^{0.08}_{0.09}$\\\vspace{.1cm}
NGC~4579  & ...  &  ...  & ...  &  $0.50^{0.03}_{0.03}$\\\vspace{.1cm}
NGC~4631  &$0.88^{0.10}_{0.08}$  &  $0.23^{0.09}_{0.11}$  &$0.10^{0.04}_{0.05}$  &  $0.73^{0.01}_{0.01}$\\\vspace{.1cm}
NGC~4725  &$1.10^{0.20}_{0.18}$  &  $0.25^{0.13}_{0.15}$  &$0.08^{0.04}_{0.04}$  &  $0.88^{0.02}_{0.01}$\\\vspace{.1cm}
NGC~4736  &$0.99^{0.15}_{0.19}$  &  $0.25^{0.15}_{0.20}$  &$0.12^{0.04}_{0.05}$  &  $0.73^{0.01}_{0.01}$\\\vspace{.1cm}
NGC~4826  &$0.87^{0.21}_{0.16}$  &  $0.30^{0.25}_{0.27}$  &$0.15^{0.013}_{0.014}$  &  $0.68^{0.05}_{0.04}$\\\vspace{.1cm}
NGC~5055  &$0.90^{0.12}_{0.05}$  &  $0.17^{0.18}_{0.22}$  &$0.07^{0.06}_{0.08}$  &  $0.78^{0.02}_{0.03}$\\\vspace{.1cm}
NGC~5457  &$0.97^{0.07}_{0.13}$  &  $0.20^{0.13}_{0.16}$  &$0.08^{0.06}_{0.07}$  &  $0.75^{0.04}_{0.03}$\\\vspace{.1cm}
NGC~5713  &$0.89^{0.16}_{0.14}$  &  $0.04^{0.15}_{0.20}$  &$0.01^{0.08}_{0.10}$  &  $0.87^{0.02}_{0.02}$\\\vspace{.1cm}
NGC~5866 & $0.57^{0.36}_{0.16}$  &  $0.15^{0.20}_{0.15}$  &$0.10^{0.12}_{0.08}$  &  $0.48^{0.04}_{0.04}$\\\vspace{.1cm}
NGC~6946 & $0.77^{0.10}_{0.13}$  &  $0.24^{0.12}_{0.20}$  &$0.10^{0.05}_{0.08}$  &  $0.67^{0.04}_{0.05}$\\\vspace{.1cm}
NGC~7331 & $1.10^{0.09}_{0.06}$  &  $0.12^{0.15}_{0.13}$  &$0.04^{0.06}_{0.05}$  &  $1.00^{0.02}_{0.01}$\\\vspace{.1cm}
M51    & $0.95^{0.09}_{0.10}$  &  $0.15^{0.12}_{0.14}$  &$0.05^{0.04}_{0.05}$  &  $0.86^{0.02}_{0.03}$\\
\hline 
\hline
\label{tab:result}
\end{tabular}
\end{center}
\end{table}

\section{Mid-radio continuum luminosity}
\begin{figure}
\begin{center}
\resizebox{\hsize}{!}{\includegraphics*{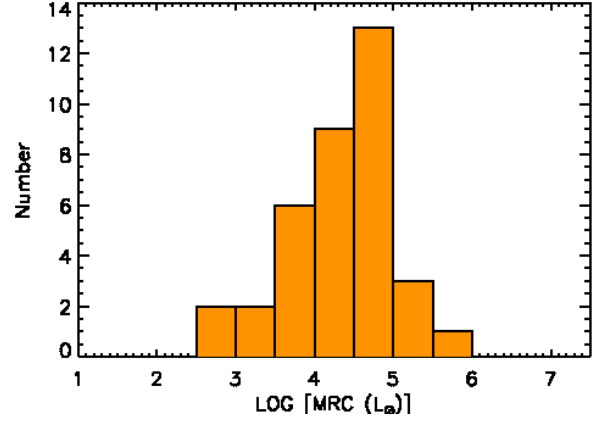}}
\caption{Distribution of the mid-radio continuum luminosity MRC of the galaxies. }
\label{fig:boloth}
\end{center}
\end{figure}

Integrating the SEDs over radio frequency intervals is needed to study the total energy output of galaxies emitted in the radio. This would provide a quantitative way to study the energy balance between the radio and non-radio domains (e.g. the IR domain) of the electromagnetic radiation emitted from galaxies. The total energy budget of the radio continuum emission in the mid-frequency range  (MRC),  is given by:

\begin{equation}
{\rm MRC} = \int_{1.4}^{10.5} L_{\nu}\,\,{\rm d}\nu,     
\end{equation} 
with $L_{\nu}=4\,\pi\,{\rm D}^2\, S_{\nu}$ and using Eq.(4) (Eq.(2) for the few cases with the single power-law model as the only possibility). The integration was performed using the Simpson's rule (see e.g.  Numerical Recipes by Press et al. 1992, 2nd edition, Section 4.2). The resulting MRC luminosities are listed in Table~\ref{tab:flux}.  {The MRC bolometric luminosity varies over $\sim$3 orders of magnitude in the sample, $4.3\times\,10^2\,L_{\sun}<$\,MRC\,$<\, 3.9\times\,10^5\,L_{\sun}$ (Fig.~\ref{fig:boloth}) with a mean luminosity of $4.8\times\,10^4\, L_{\sun}$ (median of $3.1\times\,10^4\, L_{\sun}$). The thermal MRC luminosity,
\begin{equation}
{\rm MRC_{\rm th}} = \int_{1.4}^{10.5} L_{\nu}^{\rm th}\,\,{\rm d}\nu,     
\end{equation} 
($L_{\nu}^{\rm th}=4\,\pi\,{\rm D}^2\, S_{\nu}^{\rm th}$) is about 5\% to 60\% of the MRC, depending on the galaxy. On average, the thermal emission provides about 23\% of the total energy budget emitted at 1-10\,GHz in the sample.  } 

To estimate the uncertainties in the MRC luminosities due to the uncertainties in the SED parameters $\alpha_{\rm nt}$ and $f_{\rm th}$, we first generated random datasets (100 mock datasets) assuming that they are uniformly distributed within their uncertainty intervals. Then the MRC integration (Eq.(6)) was performed for each of these mock datasets. This leads to a distribution of 100 values for the MRC luminosity. We then took the 68\% confidence interval (1 $\sigma$) as the uncertainty value.

\subsection{Contribution of the standard bands to the MRC radio energy budget}
{Taking into account the galaxy distances, the average radio SED is characterized and integrated over a slightly more extended frequency range 1-12\,GHz which covers}
all the 4 standard radio bands  L~(1-2\,GHz),  S~(2-4\,GHz), C~(4-8\,GHz), and X~(8-12\,GHz). To investigate the energetics and contributions of these standard bands to the 1-12\,GHz total energy budget, we determined the luminosity densities of the bands by integrating the average SED over the frequency width of the bands.  Table~\ref{tab:bands} shows the band-to-total ratio of the luminosity densities as well as the thermal contribution at each band.  The C band centered at 6\,cm provides the highest contribution in the total energy budget, though the band-to-band differences are not striking. Thermal sources provide {38\% of the energy emitted in the X band, highest among the bands as expected.} \citet{Condon_91} modeled  radio spectrum of a sample of compact starbursts via $$\left\langle\frac{S_{\rm nt}}{S_{\rm th}}\right\rangle\sim 10\,\left(\frac{\nu}{\rm 1\,GHz}\right)^{0.1-\alpha_{\rm nt}}.$$ {Taking the same $\alpha_{\rm nt}$ as that of the average SED ($\alpha_{\rm nt}\simeq1$), this model leads to 13\%, 21\%, 33\%, and 44\% thermal fraction at the central frequencies of the L, S, C, and X bands, respectively, which are slightly higher than the bolometric measurements in Table~\ref{tab:bands}. Instead, the  following relation:
\begin{equation}
\left\langle\frac{S_{\rm nt}}{S_{\rm th}}\right\rangle\sim 13\,\left(\frac{\nu}{\rm 1\,GHz}\right)^{0.1-\alpha_{\rm nt}},
\end{equation}
reproduces the thermal fractions at mid-radio frequencies with a higher precision for the average SED in the sample.}  
\begin{table}
\begin{center}
\caption{Relative contribution of the radio bands in 1-12\,GHz bolometric luminosity.}
\begin{tabular}{lll} 
\hline
Radio band  &   S/S$_{\rm 1-12\,GHz}$ & S$_{\rm th}$/S \\ \hline \hline
L (1-2\,GHz)& 24\%  & 10\% \\
S (2-4\,GHz)& 26\%  & 17\% \\
C (4-8\,GHz)& 30\%  & 27\% \\
X (8-12\,GHz)& 20\%  & 38\% \\ 
\hline \hline
\label{tab:bands}
\end{tabular}
\end{center}
\end{table}
\section{Radio based calibrations}
Measuring the rate at which massive stars form in galaxies is key to understand the formation and evolution of galaxies.  Various lines and continuum emission data have been used so far as SFR diagnostics, each with its advantages and shortcomings  \citep[for a review see][]{Kennicutt_12}. The most frequently used tracers, H$\alpha$ and UV (rest frame 125-250\,nm) emission, are directly related to massive star formation process, but they could be obscured or attenuated by interstellar dust.  This has motivated the use of hybrid star formation tracers combining  two or more different tracers including the IR emission to correct for the dust attenuation. The use of the IR emission itself as a SFR tracer is shadowed by a contribution from other sources/mechanisms irrelevant to massive star formation such as interstellar dust heating by solar-mass stars \citep[e.g.][]{Calzetti_10,Xu_90} and emission from the atmosphere of carbon stars \citep[mainly in mid-IR, e.g.,][]{Lu,Tabatabaei_10,Verley_09}. The radio continuum emission is an ideal SFR tracer as $a$)  it is not attenuated by dust, $b$) it emerges from different phases of  massive star formation from young stellar objects to HII regions and SNRs, and $c$) no other tracer is needed to be combined with. Even the diffuse emission, that is mainly nonthermal  \citep[e.g.][]{Tabatabaei_3_07}, also traces massive stars in normal star forming galaxies\footnote{The diffuse synchrotron emission in starburst galaxies is likely dominated by secondary CREs produced in their ISM dense gas \citep[e.g.,][]{Lacki_13}.} but those occurred in the past: The CRE lifetime is $t_{\rm syn}\simeq ~1.06 \times 10^9\,{\rm yr}\,(\frac{B}{\rm \mu G})^{-1.5}\, (\frac{\nu}{\rm GHz})^{-0.5}\sim$\,10\,Myr at  6cm ($\nu=4.85$\,GHz) where $B=13.5\,\mu$G (see Sect.~6).  
Hence, the radio SFRs must provide a more precise measure of the rate of massive star formation in a galaxy than the common non-radio SFRs. 

As follows, we calibrate the SFR, globally, using the monochromatic radio luminosities at 6\,cm and 20\,cm. 
The radio SFR tracers are further compared with the standard non-radio tracers. We also present a SFR calibration relation using the bolometric MRC luminosity.  Moreover, we construct a MRC calibration relation using the monochromatic radio luminosities at 6 and 20\,cm.
\subsection{Comparison of radio SFRs with standard SFR diagnostics}
Taking advantage of the thermal and nonthermal emission separated through the SED analysis, we can now derive the radio SFR calibration relations directly and independently from the IR SFR relations \citep[i.e., the radio-IR correlation, e.g.,][]{Condon_2}.  We further compare the radio and the commonly used SFR tracers, the 24\,$\mu$m, H$\alpha$ and FUV emission. A good correlation between those SFR tracers is the first requirement to calibrate the non-radio SFRs with the radio SFRs, particularly the thermal radio SFR as an ideal star formation diagnostic \citep{Murphy_11}.   

Assuming a solar metallicity and continuous star formation, and using a Kroupa IMF, \citet{Murphy_11} obtained a general calibration relation for the thermal radio emission:
\begin{eqnarray}
 \left(\frac{{\rm SFR_{\nu}^{\rm th}}}{M_{\sun}\,{\rm yr}^{-1}}\right)\, & =  & \,4.6\,\times\,10^{-28} \left. \left(\frac{T_e}{10^4\,{\rm K}}\right)^{-0.45} \,\, \right.
 \\
& &\left.  \times \left(\frac{\nu}{{\rm GHz}}\right)^{0.1}   \left(\frac{L_{\nu}^{\rm th}}{{\rm erg}\,{\rm s}^{-1}\,{\rm Hz}^{-1}}\right), \right.\nonumber
\end{eqnarray}
where $T_e$ is the electron temperature and $L_{\nu}^{\rm th}$ is the thermal radio luminosity.
At 6\,cm and for $T_e=~10^4$\,K, this becomes
\begin{equation}
 \left(\frac{{\rm SFR_{\rm 6cm}^{\rm th}}}{M_{\sun}\,{\rm yr}^{-1}}\right) = 1.11\,\times\,10^{-37}\left(\frac{\nu L_{\nu}^{\rm th}{\rm (6cm)}}{{\rm erg}\,{\rm s}^{-1}}\right),
\end{equation}
We note that the electron temperature could exceed the typical value of $T_e=~10^4$\,K in low-metallicity dwarf galaxies. A mean temperature of $T_e=14000$\,K has been found to be more representative in these objects \citep{Nicholls}, leading to 14\% decrease in the above calibration factor. 

Similarly, the thermal radio SFR at 20\,cm is:   
\begin{equation}
 \left(\frac{{\rm SFR_{\rm 20cm}^{\rm th}}}{M_{\sun}\,{\rm yr}^{-1}}\right) = 3.29\,\times\,10^{-37}\, \left(\frac{\nu L_{\nu}^{\rm th}{\rm (20cm)}}{{\rm erg}\,{\rm s}^{-1}}\right).
\end{equation}

Calibrating between the supernova rate and the SFR using the output of Starburst99, and using the empirical relations between supernova rate and nonthermal spectral luminosity of the Milky Way \citep{Tammann,Condon_90},  \citet{Murphy_11} found the following relation for the nonthermal synchrotron emission,
\begin{eqnarray}
\left(\frac{{\rm SFR_{\nu}^{\rm nt}}}{M_{\sun}\,{\rm yr}^{-1}}\right)\, & = & \,6.64\,\times\,10^{-29}\,\left.\left(\frac{\nu}{{\rm GHz}}\right)^{\alpha_{\rm nt}}\,\, \right.  \\
& &\left. \times \left(\frac{L_{\nu}^{\rm nt}}{{\rm erg}\,{\rm s}^{-1}\,{\rm Hz}^{-1}}\right) \right.\nonumber
\end{eqnarray}
At 6\,cm, one obtains
\begin{eqnarray}
\left(\frac{{\rm SFR_{\rm 6cm}^{\rm nt}}}{M_{\sun}\,{\rm yr}^{-1}}\right)\, & = & \,1.37\,\times\,10^{-38}\,\left. (4.85)^{\alpha_{\rm nt}} \right. \\ 
& & \left. \times \left(\frac{\nu L_{\nu}^{\rm nt}{\rm (6cm)}}{{\rm erg}\,{\rm s}^{-1}}\right), \right. \nonumber
\end{eqnarray}
and at 20cm, 
\begin{eqnarray}
\left(\frac{{\rm SFR_{\rm 20cm}^{\rm nt}}}{M_{\sun}\,{\rm yr}^{-1}}\right) & = & \,4.58\,\times\,10^{-38}\left.\,(1.45)^{\alpha_{\rm nt}} \right. \\
& & \left. \times \left(\frac{\nu L_{\nu}^{\rm nt}{\rm (20cm)}}{{\rm erg}\,{\rm s}^{-1}}\right). \right. \nonumber
\end{eqnarray}
The $\alpha_{\rm nt}$ determined in Sect.~3.1 (see Table~\ref{tab:result}) was used in the above relations (Eqs.~13, 14) to calculate the nonthermal radio SFRs at 6\,cm and 20\,cm.

As the total RC emission is a combination of the thermal and nonthermal emission, Eqs.(9) and (12) lead to the following general expression for the SFR based on the RC \citep{Murphy_11}:
\begin{eqnarray}
\left(\frac{{\rm SFR_{\nu}^{\rm RC}}}{M_{\sun}\,{\rm yr}^{-1}}\right) & = & \left. 10^{-27}  \,\,\,[2.18 \left(\frac{T_e}{10^4\,{\rm K}}\right)^{0.45}\times \right. \nonumber\\
& &\left. \left(\frac{\nu}{\rm GHz}\right)^{-0.1} + 15.1\left(\frac{\nu}{\rm GHz}\right)^{-\alpha_{\rm nt}}]^{-1} \right. \nonumber\\
& &\left.\times \left(\frac{L_{\nu}^{\rm }}{{\rm erg}\,{\rm s}^{-1}\,{\rm Hz}^{-1}}\right)\,\,\right. 
\end{eqnarray}
For instance, the case of $T_e=10^{4}$\,K and $\alpha_{\rm nt}=1$ leads to the following SFR calibration relations:
\begin{equation}
\left(\frac{{\rm SFR_{6cm}^{\rm RC}}}{M_{\sun}\,{\rm yr}^{-1}}\right) = 4.1 \times 10^{-38}  \left(\frac{\nu L_{\rm 6cm}}{{\rm erg}\,{\rm s}^{-1}}\right),
\end{equation}
and 
\begin{equation}
\left(\frac{{\rm SFR_{20cm}^{\rm RC}}}{M_{\sun}\,{\rm yr}^{-1}}\right) = 5.5 \times 10^{-38}  \left(\frac{\nu L_{\rm 20cm}}{{\rm erg}\,{\rm s}^{-1}}\right).
\end{equation}
\begin{table}
\begin{center}
\caption{SFR calibrations using radio continuum.}
\begin{tabular}{llllll} 
\hline
       X             &    Y     &  \,\,\,\,\,\, $b$ &  $a$ & $r$ & $\sigma$\\
\hline 
\hline
$I)$			&			&		&			&	&\\
SFR$_{\rm 6cm}^{\rm th}$ &   SFR$_{\rm H\alpha}$&  $1.13\pm0.13$ & -$0.03\pm0.06$ & 0.77& 0.32\\
SFR$_{\rm 6cm}^{\rm nt}$ &   SFR$_{\rm H\alpha}$&  $0.89\pm0.08$ & -$0.33\pm0.07$ & 0.75& 0.32\\
SFR$_{\rm 6cm}^{\rm RC  }$ &   SFR$_{\rm H\alpha}$& $0.94\pm0.08$ & -$0.27\pm0.06$ & 0.78& 0.32\\	
SFR$_{\rm 20cm}^{\rm th}$ &   SFR$_{\rm H\alpha}$& $1.12\pm0.13$  & $0.01\pm0.05$ & 0.74& 0.34\\
SFR$_{\rm 20cm}^{\rm nt}$ &   SFR$_{\rm H\alpha}$& $0.88\pm0.08$   &  -$0.32\pm0.07$ & 0.76& 0.34\\
SFR$_{\rm 20cm}^{\rm RC  }$ &  SFR$_{\rm H\alpha}$&  $0.90\pm0.08$ &  -$0.29\pm0.06$& 0.77& 0.33\\
SFR$_{\rm 6cm}^{\rm th}$ &   SFR$_{\rm FUV}$    & $1.11\pm0.11$  & $0.03\pm0.03$ & 0.86& 0.27\\
SFR$_{\rm 6cm}^{\rm nt}$ &   SFR$_{\rm FUV}$    &  $0.80\pm0.08$ & -$0.29\pm0.06$ & 0.89& 0.24\\
SFR$_{\rm 6cm}^{\rm RC  }$ &   SFR$_{\rm FUV}$  &  $0.86\pm0.07$ & -$0.23\pm0.05$& 0.91& 0.22\\
SFR$_{\rm 20cm}^{\rm th}$ &   SFR$_{\rm FUV}$    &  $1.15\pm0.12$   &   $0.07\pm0.03$  & 0.83& 0.30\\
SFR$_{\rm 20cm}^{\rm nt}$ &   SFR$_{\rm FUV}$    & $0.78\pm0.07$  & -$0.25\pm0.05$ & 0.89& 0.24\\
SFR$_{\rm 20cm}^{\rm RC  }$ &   SFR$_{\rm FUV}$   &  $0.81\pm0.07$ & -$0.23\pm0.06$ & 0.90& 0.23\\
SFR$_{\rm 6cm}^{\rm th}$ & SFR$_{24\mu{\rm m}}$ & $1.07\pm0.10$  & -$0.04\pm0.04$ & 0.84& 0.28\\
SFR$_{\rm 6cm}^{\rm nt}$ & SFR$_{24\mu{\rm m}}$ & $0.76\pm0.04$  & -$0.33\pm0.04$ & 0.94& 0.17\\
SFR$_{\rm 6cm}^{\rm RC  }$ & SFR$_{24\mu{\rm m}}$ & $0.81\pm0.04$  & -$0.28\pm0.03$ & 0.95& 0.15\\
SFR$_{\rm 20cm}^{\rm th}$ & SFR$_{24\mu{\rm m}}$ & $1.08\pm0.10$  & $0.01\pm0.04$ & 0.82& 0.30\\
SFR$_{\rm 20cm}^{\rm nt}$ & SFR$_{24\mu{\rm m}}$ & $0.74\pm0.04$  & -$0.31\pm0.04$ & 0.95& 0.16\\
SFR$_{\rm 20cm}^{\rm RC  }$ & SFR$_{24\mu{\rm m}}$ & $0.77\pm0.04$  & -$0.29\pm0.04$ & 0.95& 0.16\\
$II)$			&			&		&		&	&	\\
SFR$_{\rm 6cm}^{\rm RC}$ &   SFR$_{\rm H\alpha}$ & $0.80\pm0.07$ & -$0.17\pm0.05$ & 0.79& 0.33\\
SFR$_{\rm 20cm}^{\rm RC  }$ &  SFR$_{\rm H\alpha}$&  $0.79\pm0.06$ &  -$0.22\pm0.05$& 0.81& 0.31\\
SFR$_{\rm MRC}$            &  SFR$_{\rm H\alpha}$& $1.05\pm0.07$   & $0.11\pm0.02$ & 0.85 & 0.28\\
SFR$_{\rm 6cm}^{\rm RC  }$ &   SFR$_{\rm FUV}$  &  $0.82\pm0.08$ & -$0.19\pm0.06$& 0.88& 0.26\\
SFR$_{\rm 20cm}^{\rm RC  }$ &   SFR$_{\rm FUV}$  &  $0.78\pm0.07$ & -$0.22\pm0.05$& 0.90& 0.24\\
SFR$_{\rm MRC}$            &    SFR$_{\rm FUV}$  & $1.03\pm0.08$& $0.10\pm0.02$& 0.91&0.23\\
SFR$_{\rm 6cm}^{\rm RC  }$ & SFR$_{24\mu{\rm m}}$ & $0.76\pm0.05$  & -$0.23\pm0.04$ & 0.93& 0.20\\
SFR$_{\rm 20cm}^{\rm RC  }$ & SFR$_{24\mu{\rm m}}$ & $0.75\pm0.04$  & -$0.26\pm0.03$ & 0.95& 0.18\\
SFR$_{\rm MRC}$            &  SFR$_{24\mu{\rm m}}$ & $1.00\pm0.04$& $0.04\pm0.02$& 0.96& 0.15\\
\hline \hline
\label{tab:SFRcal}
\end{tabular}
\tablecomments{The linear fits in logarithmic scales (log\,Y\,~=~b\,\,\,log\,X + a) obtained  using the bisector least square fit \citep{Isobe} with $\sigma$ the scatter around the fit for $I)$ the galaxies with both the thermal and nonthermal components (Fig.~\ref{fig:monoLsfr}) and $II)$ the entire sample (Fig.~\ref{fig:TRC-SFR}).   }
\end{center}
\end{table}

As non-radio extinction-corrected SFR tracers, the 24$\mu$m emission as well as the hybrid diagnostics H$\alpha$\,+\,24$\mu$m and FUV\,+\,24$\mu$m were used.  The hybrid diagnostics could be expressed as the H$\alpha$ and FUV emission corrected for extinction. The observed H$\alpha$ luminosity is corrected following \citet{Kennicutt_09}:
$$L_{\rm H\alpha_{\rm corr}}=  L_{\rm H\alpha_{\rm obs}} + 0.02\, { L}_{24\mu m}.$$
We corrected the FUV emission for obscursion by dust using the \citet{Hao_11} calibration relation given for galaxy luminosities:  
 $$ L_{\rm FUV_{\rm corr}}=  L_{\rm FUV_{\rm obs}}+ 3.89\, {L}_{24\mu m}.$$ We note that, in this relation, the calibration factor of the 24$\mu$m term could change galaxy-by-galaxy depending on their stellar population and their contribution in the interstellar radiation field as shown by \citep{Boquien_16} for few KINGFISH galaxies. 

The SFR can be estimated using the corrected H$\alpha$ luminosity,
\begin{equation}
\left(\frac{{\rm SFR_{\rm H\alpha_{\rm corr}}}}{M_{\sun}\,{\rm yr}^{-1}}\right) = 5.37\times 10^{-42} \left(\frac{L_{\rm H\alpha_{\rm corr}}}{{\rm erg}\,{\rm s}^{-1}}\right),
\end{equation}
which is a measure of the current star formation activity \citep[$\lesssim$\,10 Myr][]{Murphy_11}.

The FUV emission traces a wider range of stellar ages and is sensitive to recent ($\lesssim$100\,Myr) star formation activity \citep{Kennicutt_98,Calzetti_05}. As in \citet{Murphy_11}, we derived the SFR based on the corrected FUV luminosity using
\begin{equation}
\left(\frac{{\rm SFR_{\rm FUV_{\rm corr}}}}{M_{\sun}\,{\rm yr}^{-1}}\right) = 4.42\times 10^{-44} \left(\frac{L_{\rm FUV_{\rm corr}}}{{\rm erg}\,{\rm s}^{-1}}\right).
\end{equation}
The mid-IR emission at 24\,$\mu$m has been widely used as a SFR tracer as well \citep[e.g.][]{Wu,Calzetti_07,Rieke_09}. This emission also traces the star formation activity over $\lesssim$100\,Myr \citep{Kennicutt_12}. We used the relation given by  \citet{Relano_07} which was calibrated for a wide range of the 24$\mu$m luminosities ($10^{38} {\rm erg\,s^{-1}}~<~L_{\rm 24\mu m}~<~3\times\,10^{44} {\rm erg\,s^{-1}}$) using a Kroupa IMF:
\begin{equation}
\left(\frac{{\rm SFR_{\rm 24\mu m}}}{M_{\sun}\,{\rm yr}^{-1}}\right) = 5.58\times 10^{-36} \left(\frac{\nu L_{\rm 24\mu m}}{{\rm erg}\,{\rm s}^{-1}}\right)^{0.826}.
\end{equation}
 
{The monochromatic RC emission at 6\,cm and 20\,cm are well correlated with the above tracers. The Pearson correlation coefficients are $r>0.7$ between the radio and the non-radio SFRs (Table~\ref{tab:SFRcal}). The relations with the thermal radio SFRs agree within the errors and are closer to linearity compared to those with the nonthermal radio SFRs, although their scatter $\sigma$ can be larger (in case of SFR$_{\rm FUV}$ and SFR$_{\rm 24\mu m}$). This is seen better in Fig.~\ref{fig:monoLsfr} showing the non-radio SFRs vs. the thermal, nonthermal, and total RC at 6\,cm. Falling within the 95\% confidence bounds, the equality between the radio and non-radio SFRs is achieved best when using the thermal radio emission as the SFR tracer.  Fig.~\ref{fig:monoLsfr} also shows that the bisector fit (used in Table~\ref{tab:SFRcal}) is more robust to the outliers than the ordinary least square (OLS) fit, although they both agree regarding the uncertainties. The SFR is over-estimated using the nonthermal radio (between 3\% to 30\%, taking into account the errors) with respect to the non-radio SFRs. The nonthermal radio emission could, on the other hand, underestimate the local SFR in resolved studies because of diffusion of CREs \citep{Murphy_11,Berkhuijsen_13}.  The tightest correlation holds between the 24$\mu$m and the nonthermal SFR, which hints on the nonthermal origin of the radio-IR correlation caused by a coupling between the gas and magnetic fields as shown in our resolved studies \citep{Taba_13_b,Taba_13}. 

The uncertainties in the radio SFRs in Fig.~\ref{fig:monoLsfr} are calculated using error propagation technique  accounting for the SED parameter errors and including the calibration, baselevel, and map fluctuation uncertainties. A 30\% uncertainty is assumed for the non-radio SFRs. We however caution that the uncertainty in the hybrid SFRs could be even larger. Taking into account contributions to the 24$\mu$m emission not associated with massive star formation, \citet{Leroy} found that the 24$\mu$m SFR estimators are systematically uncertain by a factor
of $\sim$2 leading to a calibration error of 50\% for galaxy integrated SFRs based on the hybrid SFRs. We also note that correcting the FUV emission following \citet{Boquien_16}\footnote{The FUV correction given by \citet{Boquien_16} was not applied to all galaxies due to either lack of data or being out of the applicability bound.} and \citet{Hao_11} leads to about similar SFRs, globally, considering the uncertainties.  

\begin{figure*}
\begin{center}
\resizebox{\hsize}{!}{\includegraphics*{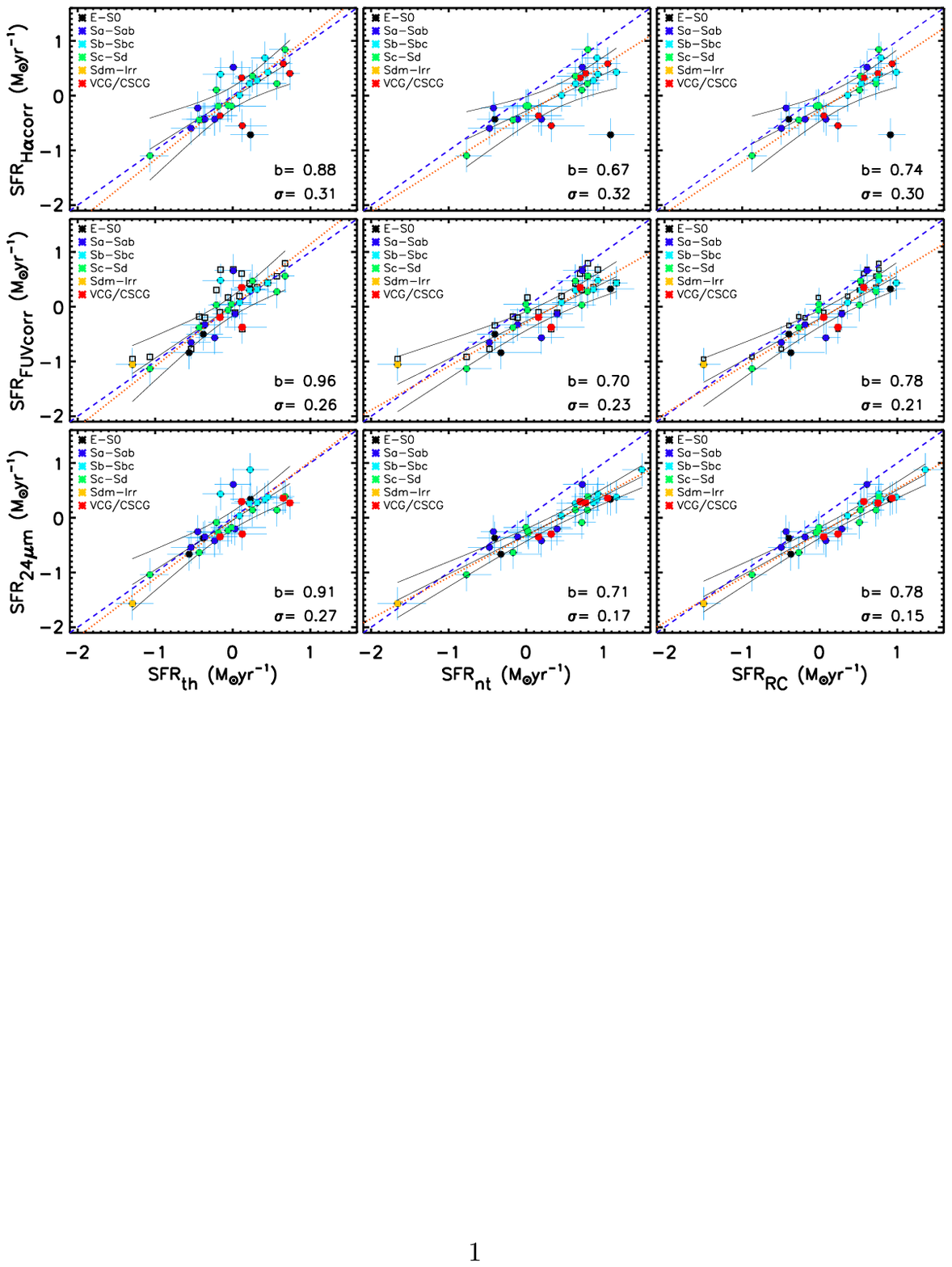}}
\caption{Comparison of the extinction-corrected SFR diagnostics,  H$\alpha$+24\,$\mu$m (H$\alpha_{\rm corr}$), FUV+ 24\,$\mu$m (FUV$_{\rm corr}$), and  24\,$\mu$m plotted from left to right against the 6\,cm radio SFRs (thermal, nonthermal, and total, respectively) in logarithmic scale. The galaxies with failed thermal/nonthermal SED fit are excluded. Also shown are the equality line (dashed), the OLS fit and its 95\% confidence bounds (solid line/curves), and the bisector fit (dotted line, see Table~\ref{tab:SFRcal}). Here, the slope b refers to the OLS fit. The squares in the second row show the FUV hybrid SFRs calibrated following \citet{Boquien_16}. An uncertainty of 30\% is assigned for the non-radio SFRs.  }
\label{fig:monoLsfr}
\end{center}
\end{figure*}
\begin{figure*}
\begin{center}
\resizebox{\hsize}{!}{\includegraphics*{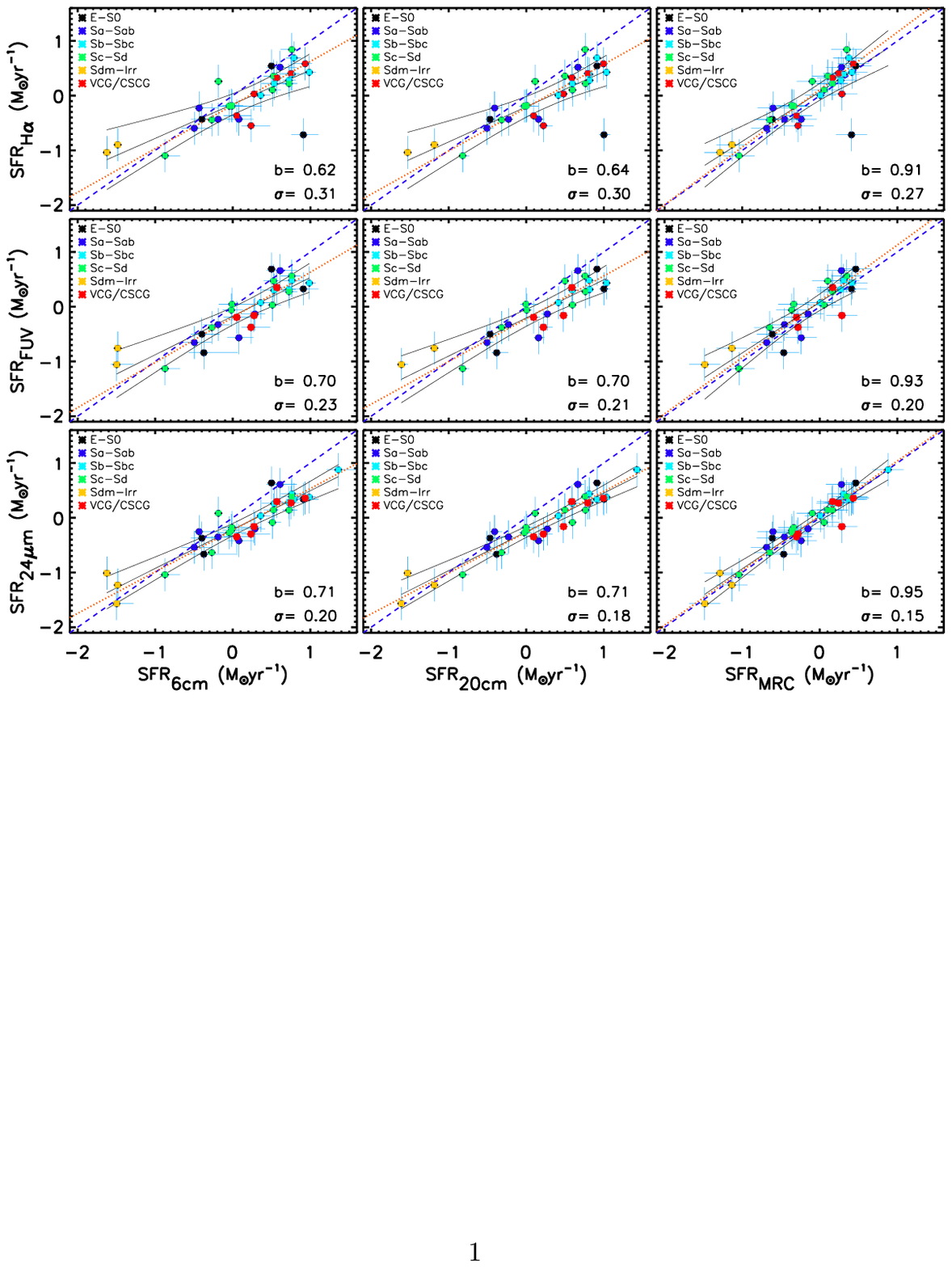}}
\caption{Comparison of the extinction-corrected SFR diagnostics,  H$\alpha$+24\,$\mu$m (H$\alpha_{\rm corr}$), FUV+ 24\,$\mu$m (FUV$_{\rm corr}$), and  24\,$\mu$m plotted from left to right against the 6\,cm, 20\,cm and MRC radio SFRs in logarithmic scale for the entire sample. Also shown are the equality line (dashed), the OLS fit and its 95\% confidence bounds (solid line/curves), and the bisector fit (dotted line, see Table~\ref{tab:SFRcal}). Here, the slope b refers to the OLS fit. An uncertainty of 30\% is assigned for the non-radio SFRs.} 
\label{fig:TRC-SFR}
\end{center}
\end{figure*}
As the next step, we investigate the use of the MRC bolometric radio luminosity, as a star formation tracer.  A tight correlation is found between the MRC and other SFR tracers ($r>0.8$) among which we select the thermal radio emission as the ideal reference SFR tracer. The following relation holds between the thermal radio luminosity at 6\,cm and the MRC,
\begin{equation}
{\rm log}\, [\nu L_{\nu}^{\rm th}{\rm (6cm)}] = (6.5\pm 1.5)\, +\, (0.80\pm0.07)\, {\rm log}\, [{\rm MRC}].
\end{equation}
The SRF calibration based on the MRC is hence derived using Eq.(10) and Eq.(21),
\begin{equation}
\left(\frac{{\rm SFR_{\rm MRC}}}{M_{\sun}\,{\rm yr}^{-1}}\right) = 3.5\,\times\,10^{-31}\left(\frac{{\rm MRC}}{{\rm erg}\,{\rm s}^{-1}}\right)^{(0.80\, \pm \,0.07)},
\end{equation}
with a dispersion of $\simeq$\,0.2\,dex. Figure~\ref{fig:TRC-SFR} shows that the non-radio SFRs agree better with SFR$_{\rm MRC}$ than with those traced monochromatically in radio (i.e., SFR$_{\rm 6cm}$ and SFR$_{\rm 20cm}$). Moreover, using the MRC as a SFR tracer reduces the scatter $\sigma$ by 5\%-30\% with respect to the monochromatic radio SFR tracers. The fitted relations are given in Table~\ref{tab:SFRcal}. 
\subsection{Calibration of MRC with monochromatic luminosities}
It would be useful to find simple relations which derive the MRC 
radio luminosity using a limited number of standard radio bands 
and applicable to a wider range of galaxy radio luminosities. This is particularly helpful when  not enough data/frequencies are available. The following combination of the 6\,cm (4.8\,GHz) and 20\,cm (1.4\,GHz) bands (C and L bands) recovers the radio 1-10\,GHz SED shapes,
\begin{equation}
{\rm MRC}= \eta_1 \nu\,L_{\nu}({\rm 20\,cm}) + \eta_2 \,\nu\, L_{\nu}({\rm 6cm}),
\end{equation}
with $\eta_1 =\,0.32 \pm 0.02$ and  $\eta_2 =\,1.68\pm 0.10$.  The
coefficients are derived from a singular value decomposition
solution to an over-determined set of linear equations \citep{press_etal_1992}. This relation reproduces the 2-component model bolometric MRC luminosities to within 1\% on average and a scatter of 8\%. For those galaxies with only single-component SED available, the model MRC and the above combination deviate by 13\%$\pm$5\%. We also emphasize that the combination given in Eq.(23) resembles the model MRC  better than a single band calibration (using either the 20\,cm or 6\,cm luminosity). 
\section{Equipartition magnetic field}
The correlation between the nonthermal radio emission and the SFR tracers could show a connection between the magnetic field and star formation activity. This is supported by the theory of amplification of magnetic fields by a small-scale turbulent dynamo (e.g. Gressel  et al. 2008) occurring in star forming regions. Assuming equipartition between cosmic rays and the magnetic field, theoretical studies suggest a relation between the magnetic field strength B and the SFR  \citep[B$\sim$ SFR$^{0.3}$, e.g., ][]{Schleicher}. We investigate this dependency in the KINGFISH sample.  

As a by-product of the SED analysis one can estimate the magnetic field strength.  In case of equipartition between the energy densities of the magnetic field and cosmic rays ($\varepsilon_{CR}\,=\,\varepsilon_{\rm B} = {\rm B}^2/8\pi$), the strength of the total magnetic field B {in Gauss} is given by
\begin{eqnarray}
{\rm B}= \big[\frac{ 4\pi (2\alpha_{\rm nt}+1)\, {\mathrm{K}}' \,I_{\rm nt} \,
E_\mathrm{p}^{1-2\alpha_{\rm nt}} \, (\frac{\nu}{2 c_1})^{\alpha_{\rm nt}}} { (2\alpha_{\rm nt}-1)\, c_2\, L\, c_3}\big]^{\frac{1}{\alpha_{\rm nt}+3}}
\end{eqnarray}
\citep{Beck_06}, where ${\mathrm{K}}'={\mathrm{K}} +1$  with ${\mathrm{K}}$ the ratio between the number densities of cosmic ray protons and electrons, $I_{\rm nt}$ is the nonthermal intensity in ${\rm erg\,s^{-1}\,cm^{-2}\,Hz^{-1}\,sr^{-1}}$,  $L$ the pathlength through the synchrotron emitting medium in cm, and $\alpha_{\rm nt}$ the mean synchrotron spectral index. $E_\mathrm{p} = 938.26$\,MeV\,$=1.50\times 10^{-3}$\,erg is the proton rest energy and
\noindent
\begin{eqnarray}
c_1 & = & \left. 3e/(4\pi {m_\mathrm{e}}^3 c^5) = 6.26428\times 10^{18}{\rm erg}^{-2}\,{\rm s^{-1}\, G^{-1}}, \right. \nonumber \\
c_2(\alpha_{\rm nt}) & = & {1\over4} c_4\,\left(\alpha_{\rm nt}+5/3\right)/(\alpha_{\rm nt}+1) \, \Gamma [(3\alpha_{\rm nt}+1)/6] \nonumber \\
& & \times \, \Gamma [(3\alpha_{\rm nt}+5)/6], \nonumber \\
c_4 & = & \sqrt{3}\,e^{3}/(4\,\pi\,m_e\,c^2) , \nonumber \\
    & = & 1.86558 \times\,10^{-23}\, {\rm erg\,G^{-1}\, sr^{-1}}, \nonumber
\end{eqnarray}
with $\Gamma$ the mathematical gamma function.
For a region where the field is completely ordered and has a constant inclination $i$ with respect to the sky plane ($i=0^o$ is the face-on view), $c_3 = [{\cos}\,(i)]^{(\alpha_{\rm nt}+1)}$.

It is usually assumed that $K\,\simeq$\,100 \citep[][]{Beck_06} and  $L\,\simeq\,1\,{\rm kpc}/ {\rm cos}\,i$. For $\alpha_{\rm nt}=1$, e.g., dominant synchrotron cooling of the CREs,  and Eq.~(22) is reduced to ${\rm B}= C\, \left(\frac {I_{\rm nt}}{{\cos}\,(i)}\right)^{1/4}$.
Since we are working with flux density and not surface brightness, a more practical expression is
\begin{equation}
{\rm B} = {\rm B_0}  \left(\frac{{\cos}\,(i)}{{\cos}\,(i_0)} \right)^{-1/4} \left(\frac{\rm S_{\rm nt}}{\rm S_{\rm nt,0}} \right)^{1/4}, 
\end{equation}
where ${\rm B_0}$ is the magnetic field strength, $i_0$ the inclination angle,  and S$_{\rm nt}$ and S$_{\rm nt,0}$ the nonthermal flux of the target and a reference galaxy. We  had determined B for one of the KINGFISH galaxies, NGC~6946, using  Eq.~(24) in \citet{Taba_13}.  This galaxy is used as the reference in Eq.~(25), i.e., B$_0$=B$_{\rm N6946}$=16\,$\mu$G and S$_{\rm nt,0}~=~[{\rm S}_{\rm 6.2cm}^{\rm 4.8GHz}~(1-~f_{\rm th(6cm)})]_{\rm N6946}$=\,0.5\,Jy (see Tables~\ref{tab:flux2} and \ref{tab:result}), and $i_0=33^{\circ}$,   to estimate B for other galaxies after correcting their fluxes for different distances. The magnetic field strength changes between $\simeq$4\,$\mu$G (IC\,2574) and 27\,$\mu$G (NGC~2146) with a mean of B~$=~13.5\,\pm\,5.5$\,$\mu$G in the KINGFISH sample (see Fig.~\ref{fig:B}-top). The B values and their uncertainties, calculated using the error propagation technique,  are listed in Tables~\ref{tab:flux} and \ref{tab:flux2}.

Fig.~\ref{fig:B}-bottom shows that B and SFR are correlated, $r=0.72\pm 0.09$\footnote{The formal error on the correlation coefficient depends on the strength of the correlation $r$ and the number of independent points n, 
$\Delta r= \sqrt{1-r^{2}}/ \sqrt{n-2}$ \citep{Edwards}.} and $r_s=0.69\pm 0.01$, as 
indicated first by the nonthermal radio--SFR correlation (Sect. 5.1).  The bisector fit shown in Fig.~\ref{fig:B} corresponds to
\begin{equation}
{\rm log B} = (0.34 \pm 0.04)\, {\rm log SFR} + (1.11 \pm 0.02), 
\end{equation}
with B in $\mu$G and SFR in $M_{\sun}$\,yr$^{-1}$. The B-SFR dependency derived agrees with the theoretical proportionality B$\sim$ SFR$^{0.3}$ due to amplification of the turbulent magnetic field in star forming regions \citep{Schleicher}. Similar relations were found in observationally resolved  studies  \citep[e.g.][]{chyzy, chyzy_11,Heesen_14}. We emphasize that the nonthermal emission traces the total magnetic field that is a combination of the turbulent and ordered fields, and dominated by the turbulent field in star forming regions. Using the radio polarization data, instead, provides a more independent probe of the ordered large-scale magnetic field in galaxies. Our recent study in a sample of non-interacting/non-cluster galaxies shows that the ordered magnetic field is closely related to the rotation and the large-scale dynamics of galaxies \citep{taba_16_1}. 
\begin{figure}
\begin{center}
\resizebox{\hsize}{!}{\includegraphics*{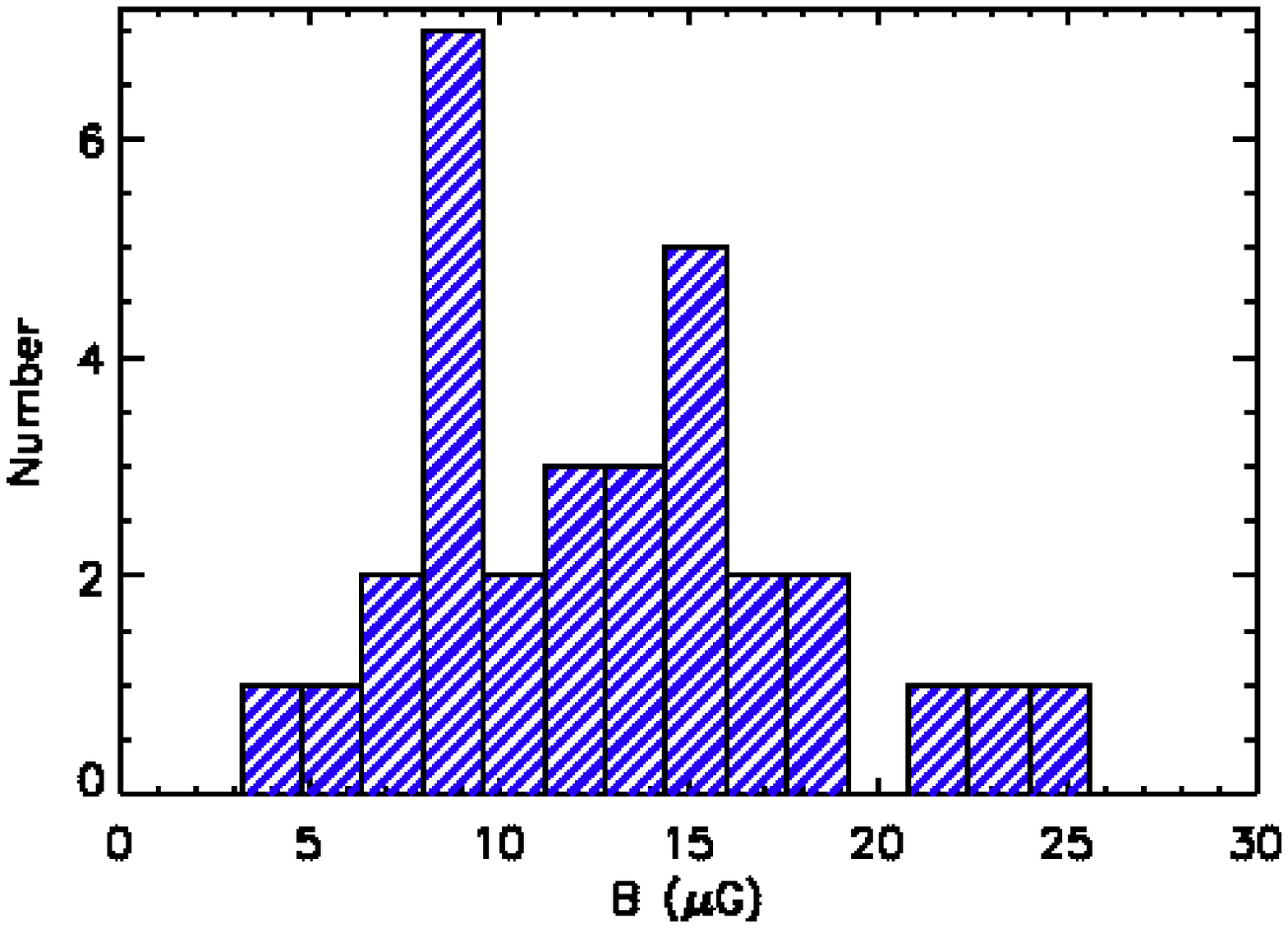}}
\resizebox{\hsize}{!}{\includegraphics*{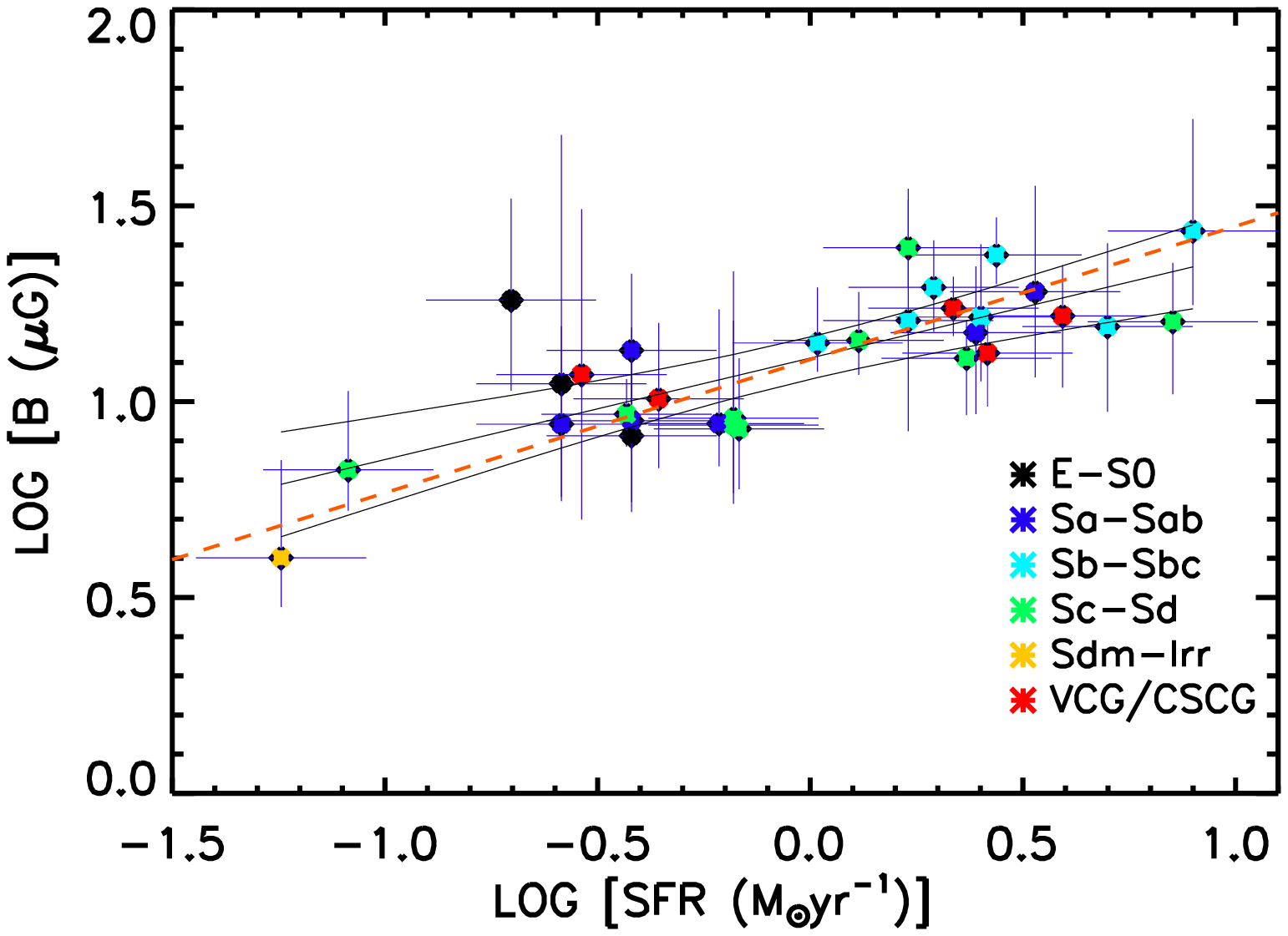}}
\caption{{\it Top}: distribution of the magnetic field strength in the KINGFISH sample. {\it Bottom}: The magnetic field strength versus the star formation rate. Also shown are the OLS fit and its 95\% confidence bounds (solid line/curves) as well as the bisector fit (dashed line).   }
\label{fig:B}
\end{center}
\end{figure}
%
%
%
\section{Further discussion}
In this section, we investigate the dependencies of the radio SED parameters $\alpha_{\rm nt}$ and $f_{\rm th}$ on star formation and equipartition magnetic field.  We also discuss the importance of this basic radio SED analysis for a better understanding of the observed IR-to-radio luminosity ratio in nearby galaxies leading to some hints for similar studies at high-z. 
\begin{figure*}
\begin{center}
\resizebox{\hsize}{!}{\includegraphics*{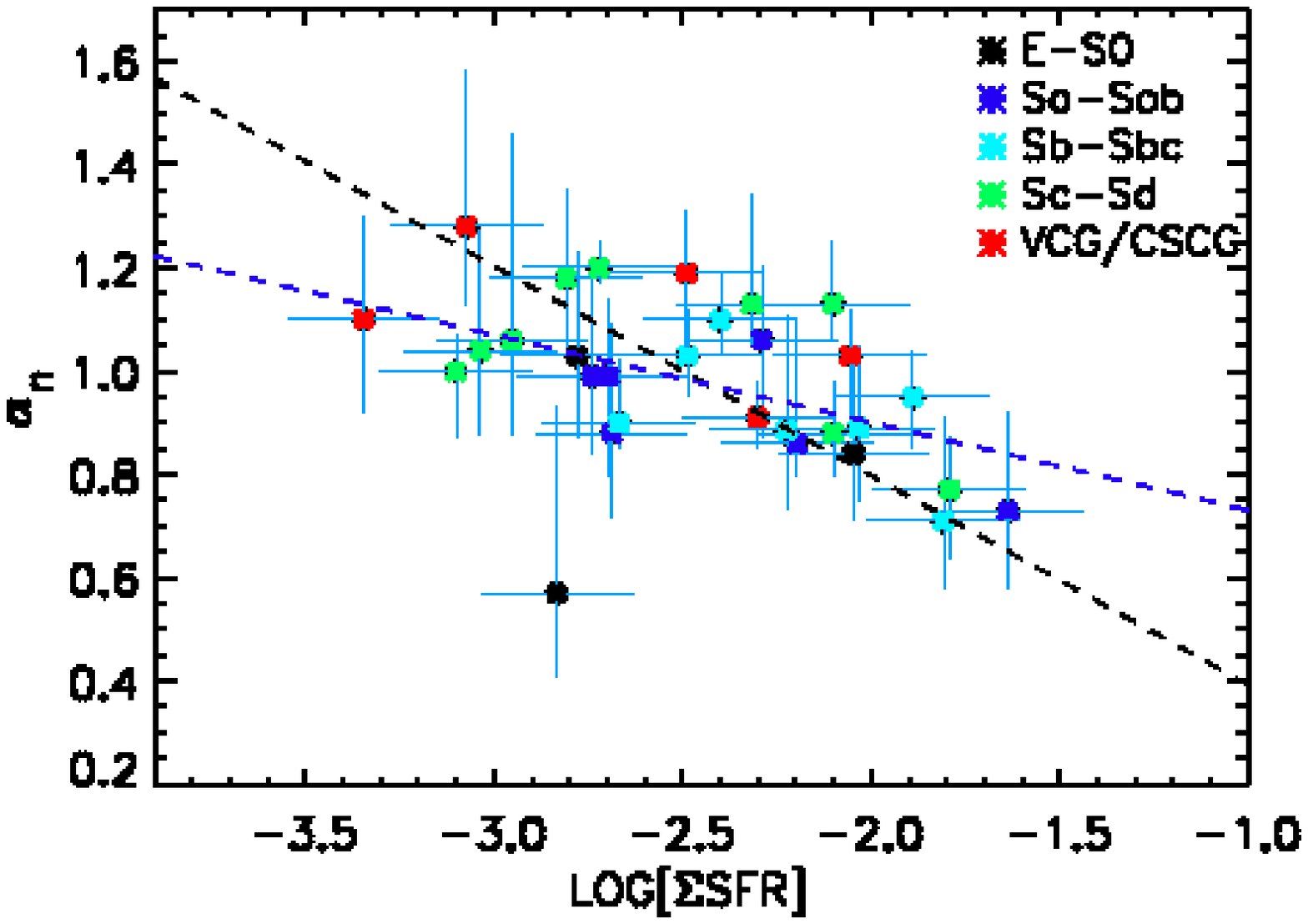}\includegraphics*{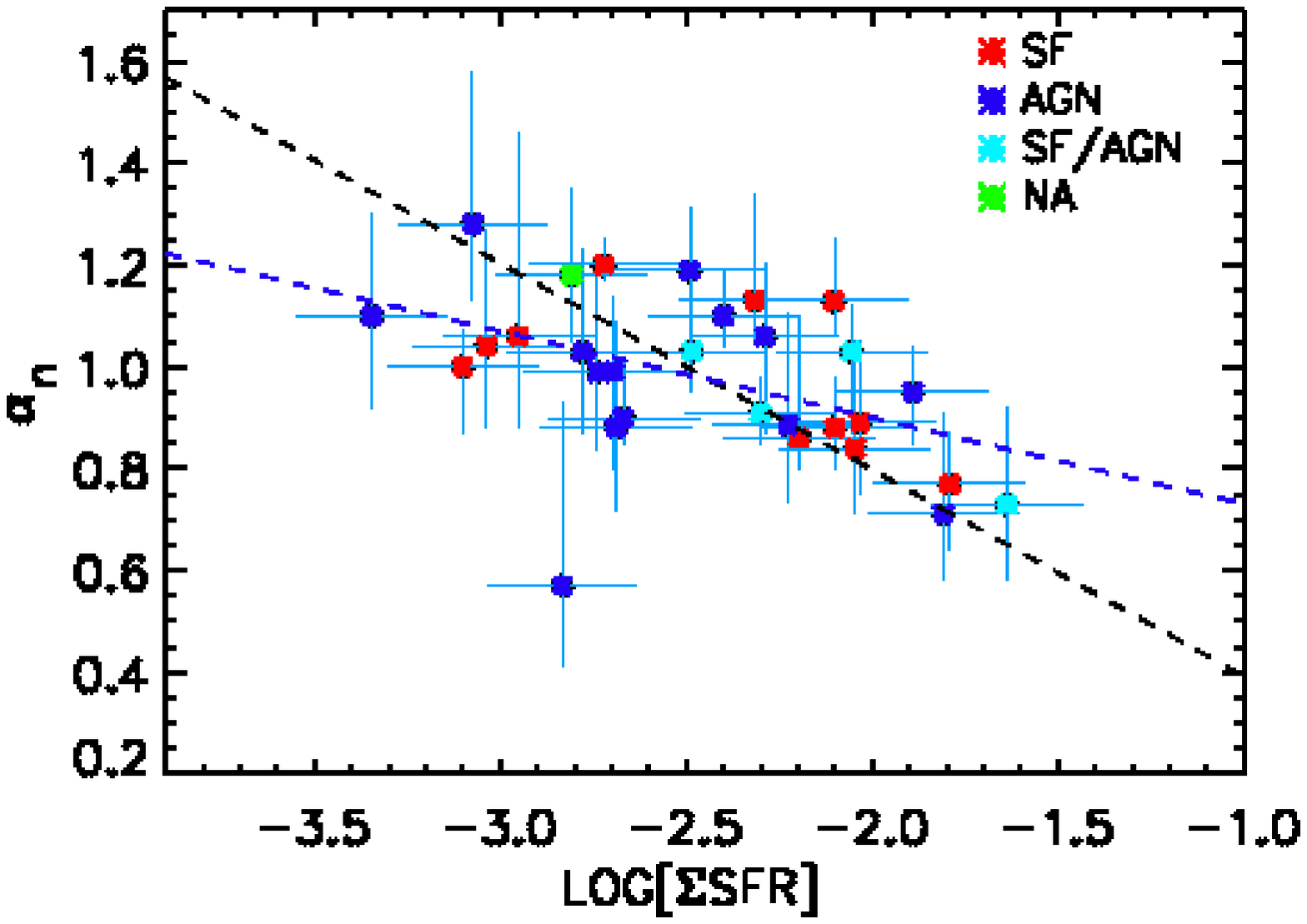}}
\caption{The nonthermal spectral index $\alpha_{\rm nt}$ vs. star formation rate surface density $\Sigma_{\rm SFR}$ for the KINGFISH sample, color coded per galaxy type (left) and nucleus type (right). Also shown are the ordinary least squares fit (blue) {with a slope of $(-0.17 \pm 0.06)$ and the bisector fit \citep{Isobe} with a slope of $(-0.41 \pm 0.15)$.} The dwarf galaxies are excluded. The decreasing trend indicates that the CRE population is younger and more energetic in galaxies with higher star formation (supernova) activity. }  
\label{fig:ansfr}
\end{center}
\end{figure*}
\subsection{The influence of star formation on the cosmic ray electron population}
After ejection from their sources in star forming regions and propagating away, young CREs lose their energy through various cooling mechanisms: synchrotron, inverse Compton, Bremsstrahlung, and ionization. These cooling mechanisms change the energy index of CREs or equivalently the spectral index of the nonthermal emission $\alpha_{\rm nt}$ in different ways.   
Hence, $\alpha_{\rm nt}$ could change from galaxy to galaxy, depending on the balance between the young particles injected in star forming regions and those cooled and aged in each galaxy \citep[also see][]{Basu_15}. 

We obtained the star formation rate surface densities $\Sigma_{\rm SFR}$ of the KINGFISH galaxies {using a non-radio SFR (the H$\alpha$ + 24\,$\mu$m hybrid SFR) to avoid possible dependencies on the radio-SED parameters}, and taking into account the optical size of the galaxies.  Fig.~\ref{fig:ansfr}-left shows a decrease in $\alpha_{\rm nt}$ with increasing $\Sigma_{\rm SFR}$ {with a Pearson correlation coefficient of $r=-0.47 \pm 0.16$ and Spearman rank of $r_s=-0.51 \pm 0.01$ for normal galaxies (log(TIR)$>$\,8.9\,$L_{\sun}$). The scatter increases when including the dwarfs and irregulars  ($r=-0.42 \pm 0.16$ and $r_s=-0.47 \pm 0.01$). NGC\,5866 appears as an outlier in Fig.~\ref{fig:ansfr} due to its flat spectrum. Excluding it, the $\alpha_{\rm nt}$--$\Sigma_{\rm SFR}$ correlation is significantly enhanced ($r=\,r_s=-0.62 \pm 0.01$). } 

Could the observed decreasing trend be partly due to AGNs? A flatter nonthermal spectrum in galaxies with higher $\Sigma_{\rm SFR}$ could occur due to the presence of flat spectrum AGNs. However, Fig.~\ref{fig:ansfr}-right shows that the AGNs could not have a direct role in the observed trend, as the galaxies with AGNs could also have steep spectrum (the 2 steepest-spectrum galaxies actually host AGNs). Hence, the observed trend is mainly due to the SFR itself and not the AGNs.

Star formation could have an important influence on the energetics of the CRE population in a galaxy by increasing the number density of young and fresh relativistic particles with a flat spectrum via supernova explosions and their strong shocks. Even in supernova remnants (SNRs) the observed nonthermal spectral index could be as flat as $\simeq$0.5-0.7 \citep[e.g.,][]{Berkhuijsen_86}.
On the other other hand, the CREs in star forming regions scatter off the very many pitch angles of the turbulent magnetic field \citep[e.g. ][]{Dorman} to the surrounding medium with a diffusion length that is smaller for smaller degree of field order \citep{Taba_13_b}. This could lead to a high concentration of high-energy particles in turbulent star forming regions causing CRE winds because of the  local pressure gradient.   They then escape with winds (see below) or are trapped in a weaker magnetic field far from star forming regions and propagate/diffuse to larger scales producing diffuse synchrotron emission. 
Hence, star formation activities/feedback could flatten the global nonthermal spectrum in galaxies by a) injecting young CREs with flat spectrum,  b) amplifying the turbulent magnetic field (Sect.~6) that helps the CREs to scatter off before they completely lose energy to synchrotron, and c) producing strong winds and outflows that increase the convective escape probability of the CREs \citep[e.g.][]{Li_beck}. In this case, the CRE escape timescale is smaller than the synchrotron cooling timescale (for CREs with an isotropic pitch angle distribution, $t_{\rm syn}=\,\frac{24.57}{\rm B^2 \gamma}$~yr with B in Gauss and $\gamma$ the Lorentz factor). Hence, the global radio spectrum of more star forming galaxies is dominated by radiation from younger CREs with flat spectrum.

A flatter nonthermal spectrum in star forming regions ($\alpha_{\rm nt}=0.5-0.7$) than in the diffuse ISM ($\alpha_{\rm nt}>0.7$) has been already found in resolved studies in M\,33 \citep{Tabatabaei_3_07} and one of the KINGFISH galaxies NGC~6946 \citep{Taba_13} for which high-resolution radio data were available. Detecting such an effect in global studies could, however, be complicated by contributions from various cooling mechanisms and inhomogenities which could induce scatter in the $\alpha_{\rm nt}$--$\Sigma_{\rm SFR}$ plane, as observed in Fig.~\ref{fig:ansfr}. 
\subsection{The influence of magnetic field on the cosmic ray electron population}
As the synchrotron emission depends on the magnetic field strength, it is also important to investigate the influence of B on the energy spectrum of the CRE population. 
Theoretically, a positive correlation is expected due to increasing synchrotron cooling, a negative correlation for a CRE escape speed increasing with B, and no correlation due to other energy losses such as the bremsstrahlung loss. The positive correlation  can be traced in the ISM far from star forming regions  where the magnetic field is more uniform/ordered. The entangled/turbulent field interrupts the continuous synchrotron cooling of the CREs and prevents further steepening of their emission spectrum by  scattering them as occurs in star forming regions (Sect.~7.1). For instance, in NGC\,6946, the nonthermal spectrum along the ordered magnetic field is steeper than in the other ISM regions particularly those with strong turbulent field \citep[][]{Taba_13}. Hence, looking for a positive $\alpha_{\rm nt}-$B correlation based on the integrated properties of the galaxies should be complicated by the presence of star forming regions having low $\alpha_{\rm nt}$ and strong B (which is mostly turbulent).

In our sample, we find a poor correlation with a rank of $r_s=\,-0.32\pm\,0.09$ at best (excluding the outliers, i.e., dwarfs and NGC\,5866).   The weakness of the correlation could be due to a combined effect from the star forming and non-star forming ISM as discussed. The negative $r_s$ indicates the large influence from star forming regions and the fact that B is dominated by the turbulent magnetic field. Other cooling/propagation effects could also cause complications in global studies. 
\subsection{The radio SED vs. the IR SED}
Bolometric luminosities are a measure of the energy budget of galaxies emitting at certain ranges of frequencies. The IR bolometric luminosities have been studied in detail at various frequency intervals,  e.g., TIR:~8-1000\,$\mu$m \citep{Sanders_96}, FIR:~42.5-122.5\,$\mu$m \citep{Rice},
FIR:~40-500\,$\mu$m \citep{Chary},  FIR/submm:~40-1000\,$\mu$m \citep{Taba_cold_13}, and TIR:~3-1100\,$\mu$m \citep{Galametz_13}. For the KINGFISH sample, \citet{Dale_12} obtained the TIR (3-1100\,$\mu$m) luminosities using the Herschel and Spitzer data (see Table~\ref{tab:list}).
To compare the emission energy budget of the KINGFISH galaxies in IR with that in radio, we must compare their IR and radio bolometric luminosities. However, to our knowledge, there is no 
definition of the radio bolometric luminosity over any frequency range in the literature apart from our current definition. Hence, Eq.(6) serves as the only available definition of the bolometric luminosity in mid-radio MRC. We compare the spectral energy distribution of the IR and radio domains by means of the ratio of their integrated luminosities in two ways, a) the TIR-to-MRC ratio:
\begin{equation}
<q>_{\rm TIR}\,\, \equiv \,\,{\rm log}\, ( \frac{\rm TIR}{\rm 10^3 \,MRC}),  
\end{equation}
and b) the FIR-to-MRC ratio:
\begin{equation}
<q>_{\rm FIR}\,\, \equiv \,\,{\rm log}\, ( \frac{\rm FIR}{\rm 10^3 \,MRC}),  
\end{equation}
with TIR, FIR, and MRC luminosities in erg\,s$^{-1}$  \citep[the MRC factor of $10^3$ in the denominator is selected arbitrarily so that $<q>$ falls in the range of the q-parameter defined traditionally using the 20\,cm radio luminosity, ][]{Helou_etal_85}. The FIR luminosities were obtained by integrating the KINGFISH SEDs \citep{Dale_12} in the frequency interval 42-122\,$\mu$m.  {In the sample,  $<q>_{\rm TIR}$ changes between 2.26 and  3.02 with a mean of $2.70\,\pm\,0.17$ (error is the scatter). The FIR-to-MRC ratio, $<q>_{\rm FIR}$, changes between 1.7 and 4.2 with a mean of $2.37\,\pm\,0.36$.}

The parameters $<q>_{\rm TIR}$ and $<q>_{\rm FIR}$ are useful to study the relative change in the IR and radio SEDs in terms of various astrophysical parameters. A first parameter is the star formation rate as an important energy source of both radio and IR emission. Figure~\ref{fig:q_SFR}-top shows a likely decreasing trend of $<q>$ vs. SFR, particularly for SFR$>1\,M_{\sun}\,{\rm yr}^{-1}$, {with a Pearson correlation coefficient $r=-0.4\pm0.1$ for both cases. The Spearman rank coefficient is $r_s=-0.45\pm0.01$ for the $<q>_{\rm TIR}$--SFR correlation, and $r_s=-0.42\pm0.02$ for the $<q>_{\rm FIR}$--SFR correlation}.  Considering the thermal and nonthermal MRC separately in Eqs.(27) and (28), a clear anti-correlation is found for $<q>$ vs. SFR when using the nonthermal emission (Fig.~\ref{fig:q_SFR}-middle). In this case, the Pearson correlation coefficient is $r=-0.5\pm0.1$ for both cases.  The $<q>$  based on the thermal emission {is not correlated} with SFR (Fig.~\ref{fig:q_SFR}-bottom). This shows that the nonthermal SED could be  more sensitive to a change in massive star formation activity than the thermal emission. One immediate cause could be the amplification of the magnetic fields in star forming regions, adding more weight to the synchrotron emission, as shown in  Sect.~4.5. As such, the observed weak anti-correlation may be due to the star formation feedback inducing the magnetic field strength in galaxies \citep[e.g.,][]{Pellegrini,taba_15}. This also explains the sublinear non-radio vs. radio SFR correlations (also the famous IR-radio correlation) shown in Sect.~5. 
\begin{figure}
\plotone{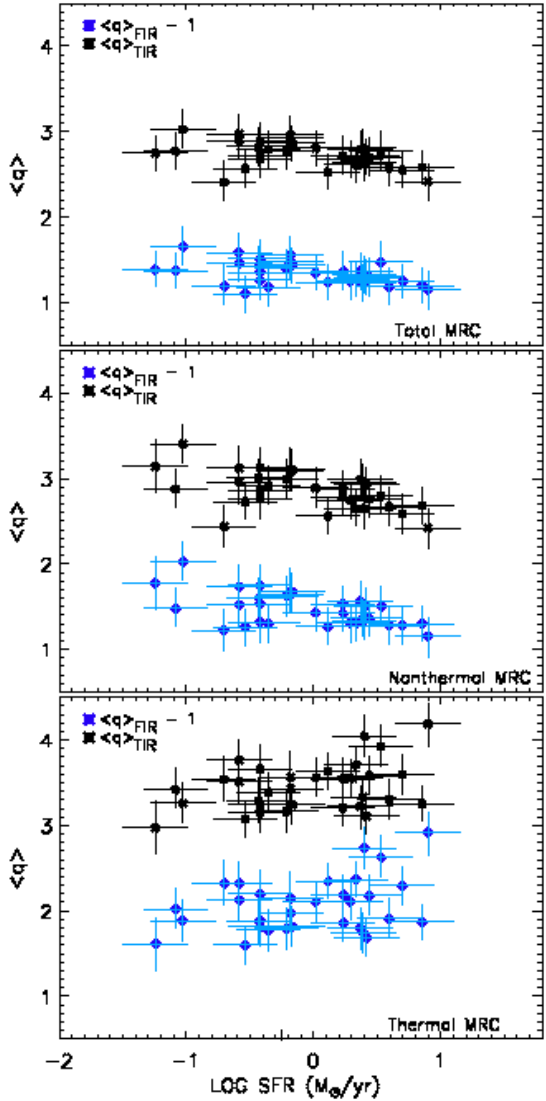}
\caption{The TIR-to-MRC ratio $<q>_{\rm TIR}$ and FIR-to-MRC ratio $<q>_{\rm FIR}$ vs. SFR for the RC (top) and its nonthermal (middle) and thermal components (bottom). A decreasing trend is indicated due to the nonthermal emission.\label{fig:q_SFR}}
\end{figure} 
\subsection{Implication for high-z studies}
As the synchrotron emission is set by the magnetic field strength, Eq.(24) implies a smaller FIR-to-nonthermal radio ratio with higher rate of star formation in galaxies, as found already in Sect.~7.3 (see Fig.~\ref{fig:q_SFR}). This also has an implication for high-z studies: we expect to see a drop in the nonthermal part of $<q>$  at high redshifts where the more luminous/higher-star-forming objects are selected. However, most of the high-z studies show either no evolution \citep[e.g.][]{Sargent,Jarvis_10} or only a tentatively slight decrease of the IR to radio ratio \citep[$q_{\rm IR}$ with z, ][]{Ivison_14,Ivison_1,Casey_12,Basu_2}. This could be of course due to the fact that no attempt is usually made to separate the thermal and nonthermal radio components when studying the IR-to-radio ratios.    

Few high-z studies have addressed variations of $q_{\rm IR}$  with dust temperature in galaxies 
leading to different results, i.e., either weak positive correlation \citep{Magnelli} or a 
negative correlation \citep{Ivison_1,smith}. As shown in Fig.~\ref{fig:q_alpha}, a 
correlation between $<q>$ and the dust temperature, derived by fitting a single 
modified black body model to the IR SEDs \citep{Dale_12}, does not occur in  nearby galaxies. 

The radio spectral index was proposed as a redshift indicator for distant galaxies \citep{Carilli}, 
but the technique was shown to have limited accuracy (50\% redshift errors) due to a change in dust 
temperatures \citep{Chapman}. This also motivated us to look for any trend between the dust temperature 
and the radio spectral index in nearby galaxies which could be used as a basic reference for high-z studies. 
Fig.~\ref{fig:vs.tdust}  shows no correlation between $\alpha$ and the dust temperature in our galaxies. On the other hand, a likely decreasing trend is found between  $\alpha_{\rm nt}$ vs. {the dust temperature ($r=-0.40 \pm 0.15$ and $r_s=-0.42 \pm 0.02$). } This can be explained by the positive correlation  between the dust temperature and the star formation surface density $\Sigma_{\rm SFR}$ with about the same quality ($r\simeq\,+0.45$), and considering that $\alpha_{\rm nt}$ decreases with $\Sigma_{\rm SFR}$  (see Sect.~4.1).
\begin{figure}
\resizebox{7.5cm}{!}{\includegraphics*{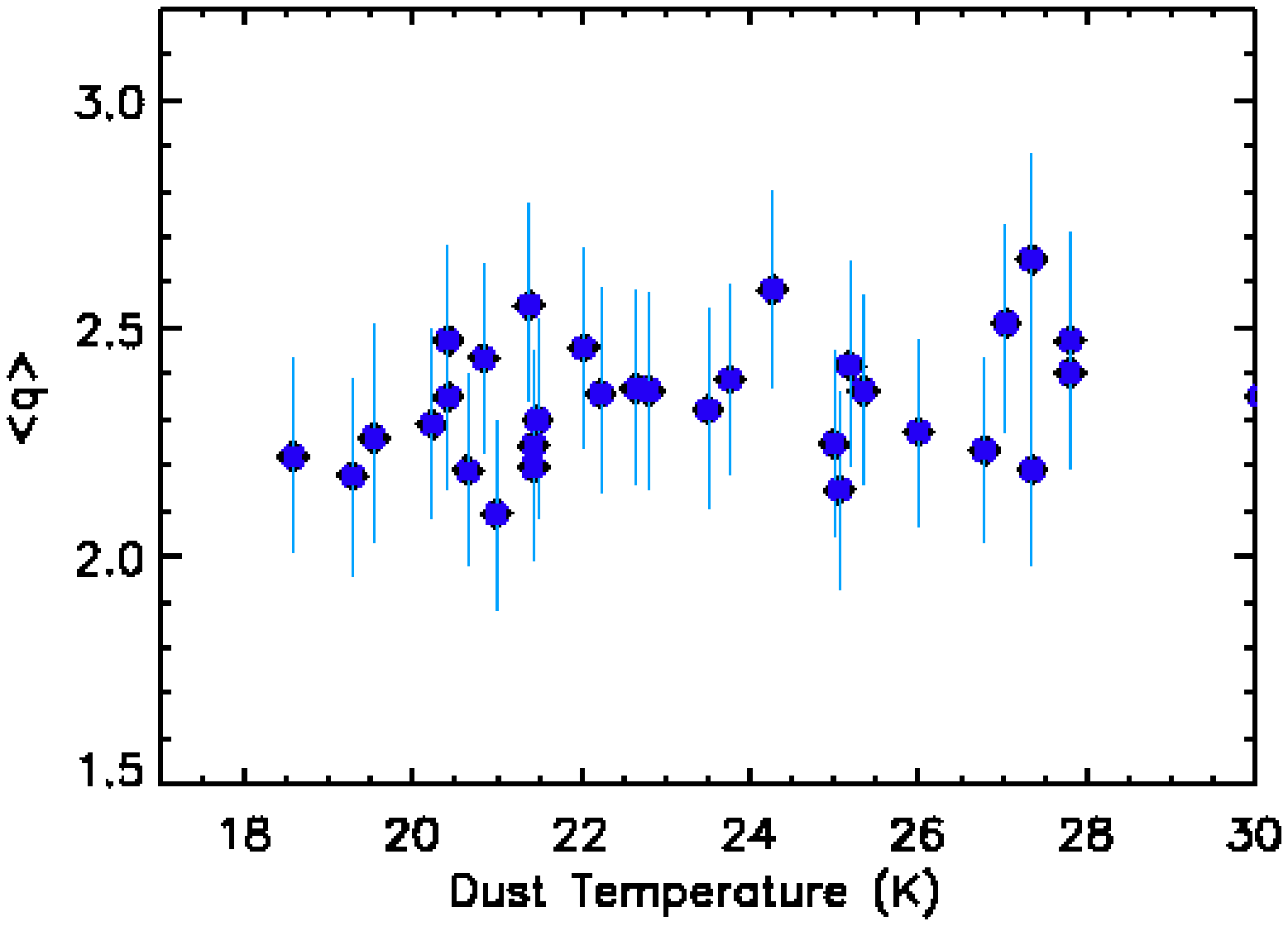}}
\caption{The FIR-to-MRC radio ratio $<q>_{\rm FIR}$ vs. dust temperature. }
\label{fig:q_alpha}
\end{figure}
\begin{figure}
\resizebox{\hsize}{!}{\includegraphics*{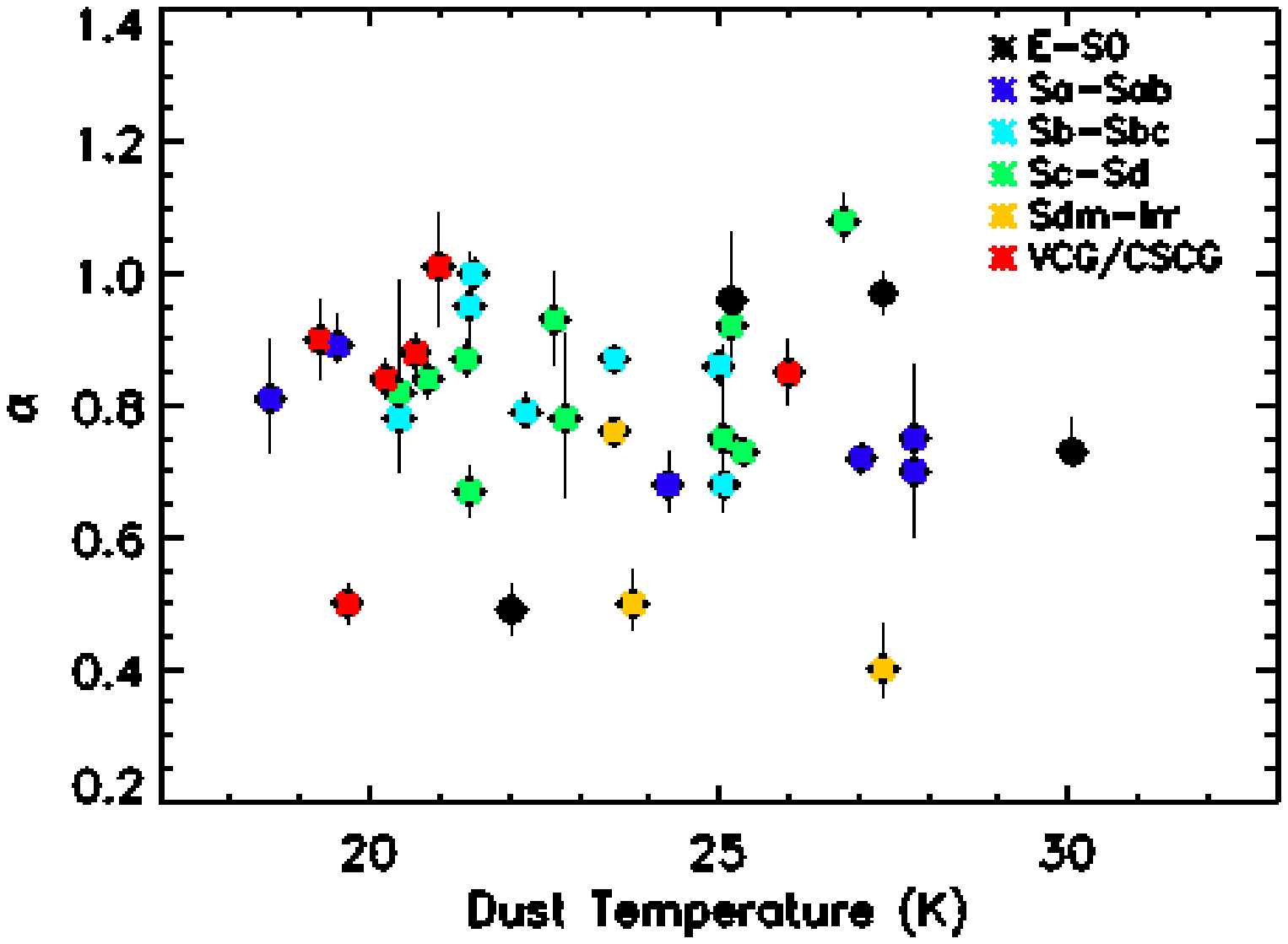}}
\resizebox{\hsize}{!}{\includegraphics*{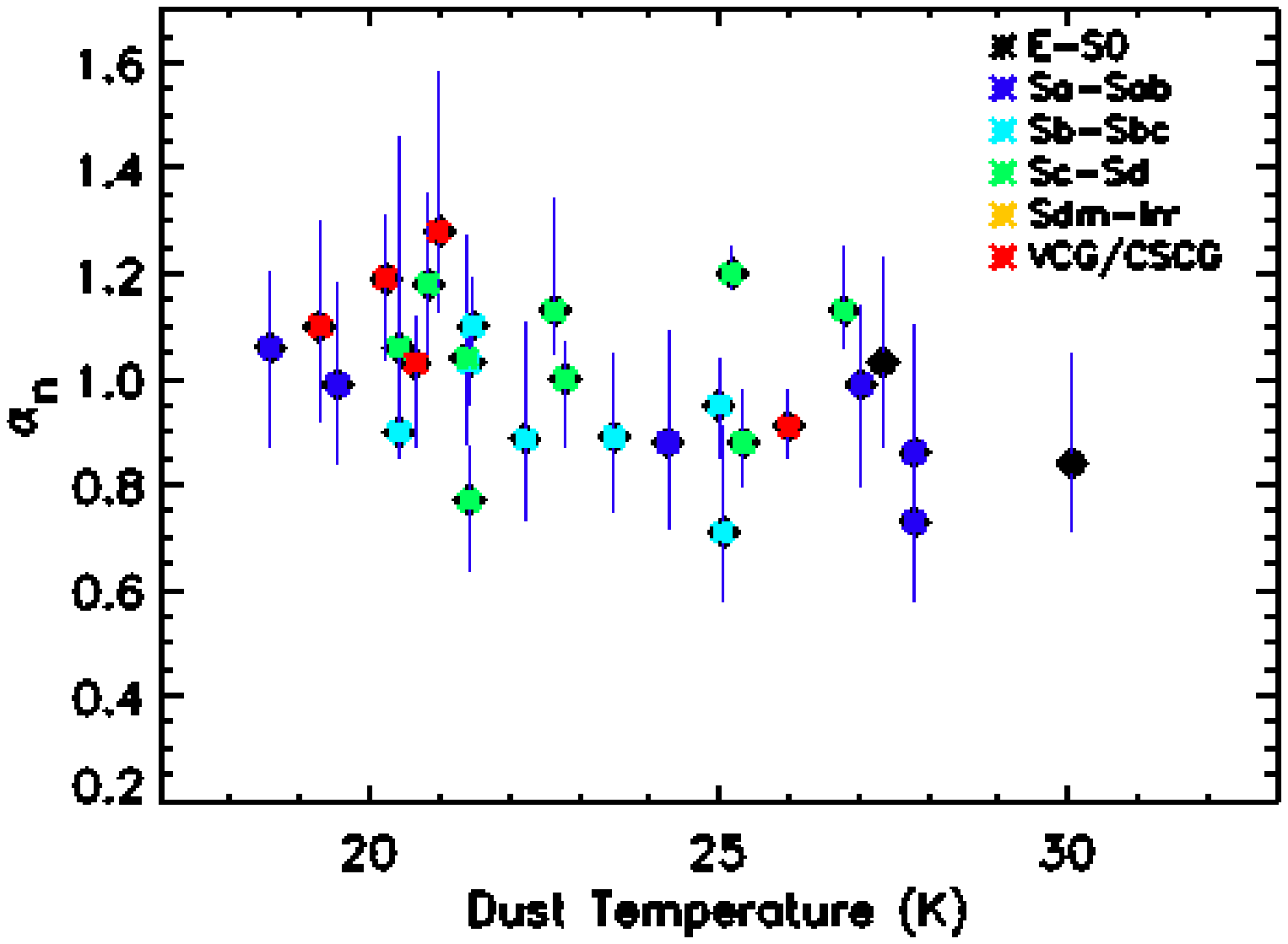}}
\caption{{\it Top:} the RC spectral index $\alpha$ vs. the dust temperature. {\it Bottom:} the nonthermal spectral index $\alpha_{\rm nt}$ exhibits a possible decreasing trend against the dust temperature excluding the flat radio sources ($\alpha< 0.6$).     }
\label{fig:vs.tdust}
\end{figure}
\section{Summary}
We compared the non-radio extinction-corrected diagnostics of star formation rates with the radio SFRs for a sample of nearby galaxies, KINGFISH, using both the MRC bolometric and monochromatic luminosities at 6\,cm and 20\,cm.
Our homogeneous and careful analysis of the 1-10\,GHz SEDs using new observations with the 100-m Effelsberg telescope allowed us to determine the MRC radio luminosities and the fractional contributions of the standard radio bands for the first time.  The 1-10\,GHz bolometric luminosity is calibrated by a linear combination of the 6 and 20\,cm bands (Eq.\,(23)). 

Unlike frequent assumptions,  the nonthermal spectral index is not fixed. {It changes over a wide range in the sample ($\sim$0.5-1.5, Table~\ref{tab:result})}, decreasing with increasing the star formation surface density of galaxies. This suggests the influence of star formation on the energetics of the CRE population, for example, by injecting high-energy cosmic rays. The average nonthermal spectral index derived for the 1-10\,GHz frequency range ($\alpha_{\rm nt}=\,0.97\pm0.16$) is slightly steeper than that derived in the {400\,MHz-10\,GHz studies ($\simeq$0.8)}, considering the uncertainties. This difference could already indicate the low-frequency flattening of the synchrotron spectrum. Neglecting the thermal component, the 1-10\,GHz radio SEDs are fitted by a single power-law model with the mean spectral index of $\alpha= 0.79\pm0.15$. 

The thermal fraction changes from zero to $\sim$60\%  with a mean of 23\% at 6\,cm, and from zero to $\sim$40\% with a mean of 10\% at 20\,cm (Table~\ref{tab:result}) and agrees with the estimates based on the H$\alpha$ methods (Table~\ref{tab:compare}). It is the highest in dwarf irregular galaxies but does not show a clear correlation with morphology, $\Sigma_{\rm SFR}$, or metallicity. 

We defined the mid-radio (1-10\,GHz) continuum bolometric luminosity, MRC, and obtained its distribution in the sample. The MRC luminosity of the KINGFISH galaxies changes over $\sim$3 orders of magnitude with a mean luminosity of $4.8\times~10^4~L_{\odot}$.  Characterizing the average radio SED, we determined the contribution of the standard radio bands (L, S, C, X) in the mid-radio luminosity. We also presented a new calibration for the simple radio model \citep{Condon_91}, although large deviations could occur in individual galaxies.

Our study of the KINGFISH sample which includes a wide range of galaxy types, shows that the MRC is an ideal star formation tracer. This is because of its good and linear correlation with other star formation tracers including the FUV and H$\alpha$ emission derived independently. {We also presented SFR calibration relations using the MRC bolometric luminosity}. 

We found that the FIR-to-MRC  luminosity ratio, $<~q~>_{\rm FIR}$, could change with star formation rate that is due to the nonthermal component and its nonlinear correlation with star formation rate.  Amplification of the equipartition turbulent magnetic fields in star forming regions could additionally strengthen the synchrotron power in galaxies with higher SFR, leading to a decrease in $<q>$. Hence, star formation feedback and magnetic fields could play a role in the balance between the radio and IR spectral energy distributions. Due to this feedback, the nonthermal radio emission overestimates the global SFR  in starbursts and galaxies with high star formation activity.

Extrapolating the SEDs beyond the 1-10\,GHz, we predicted the thermal fractions at several frequencies from 350\,MHz to 45\,GHz (Table~\ref{tab:beyond}) {based on the modeled SEDs}. Comparing to the real observations at those selected frequencies it would be possible to determine the flattening of the SED (i.e., due to the free-free absorption of the synchrotron emission)  at frequencies lower than 1\,GHz, or contribution of the spinning dust emission at frequencies higher than 10\,GHz. 
\acknowledgments

We thank the anonymous referee for helpful comments. 
FST acknowledges financial support from the Spanish Ministry of Economy and Competitiveness (MINECO) under grant number AYA2013-41243-P as well as the support by the German Research Foundation (DFG) via the grant TA 801/1-1. We thank P. Muller, A. Kraus, E. Angelakis, I. Myserlis, and the Effelsberg staff for their help and support in observations and calibration of the cross-scan observations.  RB acknowledges financial support from DFG Research Unit FOR1254. D.D.M acknowledges support from ERCStG 307215 (LODESTONE).

\bibliography{s}
%
%

\appendix

{
\section{Comparison with other thermal/nonthermal separation techniques}

A degeneracy in parameter space occurs naturally when several free parameters are fitted simultaneously using the classical $\chi^2$ method. In the Bayesian MCMC approach, the confidence intervals are the most probable posteriors taken directly from the parameter space (see Figs.~1). Hence the degeneracy is naturally included in the uncertainties (16\%-84\%, equal-tailed intervals) reported in Table~\ref{tab:result}. To check further the reliability  of the confidence intervals and the range of the uncertainties, we perform a comparison with a different thermal/nonthermal separation technique. The thermal radio emission can be optimally traced by the brightest Hydrogen recombination line, the H$\alpha$ emission, in galaxies after de-reddening \citep{Tabatabaei_3_07,Taba_13_b,Taba_13}. In global studies, combining the H$\alpha$ and the 24$\mu$m fluxes is used to de-redden the H$\alpha$ emission (Sect.~5.1). The thermal free-free emission traced based on this de-reddening could however be overestimated depending on the stellar population in a galaxy as the interstellar dust is not the only source of the 24$\mu$m emission \citep[dusty atmospheres of carbon stars also emit the IR emission at 24$\mu$m, e.g., ][]{Verley_09,Tabatabaei_10,Boquien_16}. The corrected (H$\alpha$+ 24$\mu$m) and observed H$\alpha$ fluxes can hence be used as upper-estimate and lower-estimate of the thermal radio flux, respectively. The following expression, 
$$
 {\rm S}_{\rm th}^{\nu}= 1.14\times10^{-14} \, (\frac{\rm T_e}{\rm 10^4\,K})^{0.34} \, (\frac{\nu}{\rm GHz})^{-0.1}\,{\rm S}_{\rm H\alpha}
$$
converts the H$\alpha$ flux (corrected or observed) S$_{\rm H\alpha}$ in ${\rm ergs\,s^{-1}\,cm^{-2} }$ to the thermal radio flux density S$_{\rm th}$  in ${\rm erg\,s^{-1}\,cm^{-2}\,Hz^{-1}}$ at frequency $\nu$ \citep[e.g.,][]{Deeg}. We derive the thermal fraction at 6\,cm, $f_{\rm th}$(6cm)\,=\,S$_{\rm th}^{\rm 4.8GHz}$/S$_{6cm}^{\rm 4.8GHz}$ for the KINGFISH galaxies with available H$\alpha$ flux (Tabel~\ref{tab:compare}).   The thermal fraction based on the radio method (i.e., the median of the posterior PDFs in the Bayesian approach) mostly falls in between the two H$\alpha$ estimates or is closer to the corrected H$\alpha$ estimate. In few other cases, the radio and H$\alpha$ estimates agree within the uncertainties. It is then worth noting that the Bayesian MCMC method is successful and reliable capturing the correct answer, apart from the large degeneracy caused by the large observational errors taken from the literature (particularly the 10\,GHz data)-- which in most cases widens the range of  the uncertainties in the thermal fraction to non-physical, negative values.

\begin{table}
\begin{center}
\caption{Thermal fraction $f_{\rm th}$(6cm) based on the observed H$\alpha$, H$\alpha$+ 24$\mu$m (H$\alpha_{\rm corr}$), and the radio-SED methods. }
\begin{tabular}{llll} 
\hline
Galaxy  &  H$\alpha$    &   H$\alpha_{\rm corr}$ & radio-SED \\ \hline \hline
IC0342	&    0.17    & 0.27  &  ...   \\\vspace{.1cm}
IC2574  &  0.48 & 0.54   & $0.55^{0.14}_{0.12}$ \\\vspace{.1cm}
NGC~0337 &  0.07 & 0.13  & $0.08^{0.09}_{0.03}$ \\\vspace{.1cm}
NGC~0628 &   0.30  & 0.42 & $0.44^{0.11}_{0.12}$  \\\vspace{.1cm}
NGC~1266 &  0.01  & 0.07 &  $0.08^{0.15}_{0.20}$ \\\vspace{.1cm}
NGC~1482 &     0.01   & 0.08  &  ...    \\\vspace{.1cm}
NGC~2146 &   ...     & ...   &    $0.20^{0.20}_{0.25}$        \\\vspace{.1cm}
NGC~2798  &   0.03 & 0.10 & $0.07^{0.10}_{0.18}$  \\\vspace{.1cm}
NGC~2841 & ... & ... &  $0.22^{0.07}_{0.21}$\\\vspace{.1cm}
NGC~2976  &  0.23  & 0.32 & $0.27^{0.20}_{0.14}$  \\ \vspace{.1cm}
NGC~3049  &   0.23 & 0.44 & $0.31^{0.27}_{0.25}$  \\\vspace{.1cm}
NGC~3184  & 0.27 & 0.39   & $0.39^{0.25}_{0.20}$  \\\vspace{.1cm}
NGC~3190  & 0.05  & 0.11  & $0.18^{0.10}_{0.11}$  \\\vspace{.1cm}
NGC~3265 &  0.08  & 0.29  & $0.33^{0.10}_{0.07}$  \\ \vspace{.1cm}
NGC~3521  & 0.10 & 0.18   & $0.15^{0.18}_{0.21}$  \\\vspace{.1cm}
NGC~3627  & 0.08  &  0.18 & $0.16^{0.20}_{0.24}$  \\\vspace{.1cm}
NGC~3938  &  0.24 & 0.34  & $0.28^{0.20}_{0.22}$  \\\vspace{.1cm}
NGC~4236  &   0.36     & 0.42    &  ...   \\\vspace{.1cm}
NGC~4254  &  0.10 & 0.17 & $0.20^{0.09}_{0.14}$  \\\vspace{.1cm}
NGC~4321  & 0.05 &  0.25 & $0.43^{0.07}_{0.20}$  \\\vspace{.1cm}
NGC~4536  & 0.03 & 0.14  & $0.12^{0.06}_{0.04}$ \\\vspace{.1cm}
NGC~4559  & 0.26  & 0.37 & $0.31^{0.25}_{0.30}$ \\\vspace{.1cm}
NGC~4569  & 0.03  & 0.10 & $0.25^{0.15}_{0.18}$  \\\vspace{.1cm}
NGC~4579  &   0.02    &  0.06    &   ...               \\\vspace{.1cm}
NGC~4631  &  0.07 & 0.11 & $0.23^{0.09}_{0.11}$  \\\vspace{.1cm}
NGC~4725  &  0.18 & 0.25 & $0.25^{0.13}_{0.15}$  \\\vspace{.1cm}
NGC~4736  & 0.04 & 0.21   &$0.25^{0.15}_{0.20}$  \\\vspace{.1cm}
NGC~4826  &0.04  & 0.26   & $0.30^{0.25}_{0.27}$  \\\vspace{.1cm}
NGC~5055  & 0.05 & 0.15   &$0.17^{0.18}_{0.22}$  \\\vspace{.1cm}
NGC~5457  & 0.11 & 0.25  & $0.20^{0.13}_{0.16}$  \\\vspace{.1cm}
NGC~5713  & 0.01 & 0.10  & $0.04^{0.15}_{0.20}$  \\\vspace{.1cm}
NGC~5866 &  0.09 & 0.13  & $0.15^{0.20}_{0.15}$  \\\vspace{.1cm}
NGC~6946 & 0.20  & 0.29  & $0.24^{0.12}_{0.20}$  \\\vspace{.1cm}
NGC~7331 & 0.05  & 0.15  & $0.12^{0.15}_{0.13}$  \\\vspace{.1cm}
M51      & 0.04 & 0.10   &  $0.15^{0.12}_{0.14}$  \\
\hline 
\hline
\label{tab:compare}
\end{tabular}
\end{center}
\end{table}
}

\section{Plots of the SED fitting with Bayesian MCMC}
The radio SED fits presented in Sect.~3 are shown in Figs.~\ref{fig:sed1},\ref{fig:sed2}, and \ref{fig:sed3}. 
\begin{figure*}
\begin{center}
\resizebox{\hsize}{!}
{\includegraphics*{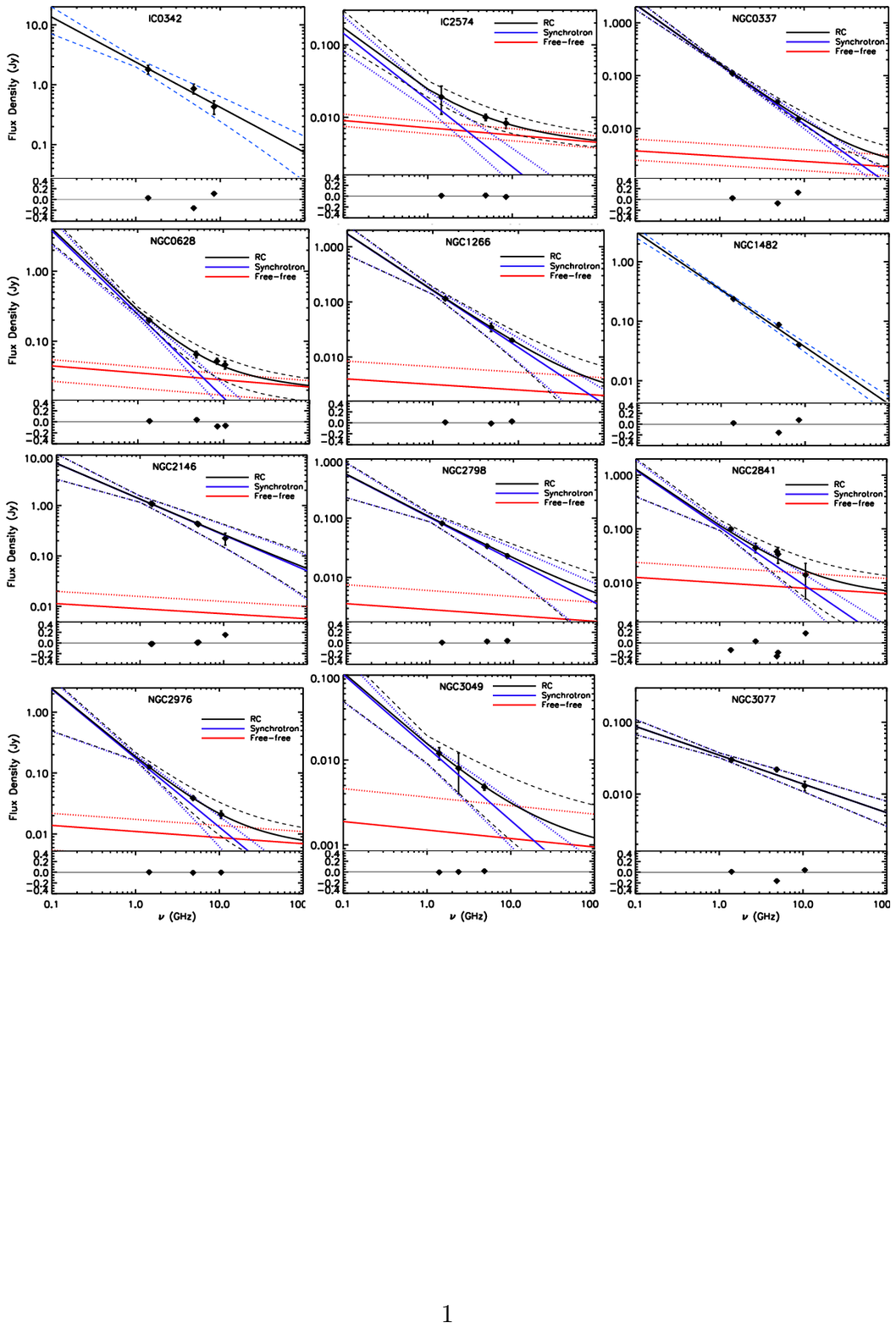}}
\caption{Radio SEDs (flux density vs. frequency) of the KINGFISH sample (solid curves) for the total RC (black) and its nonthermal (blue) and thermal (red) components as well as their uncertainty curves (dashed for RC and dotted for its components).  The points show the observed flux densities. Also shown are the relative residuals (modeled-observed/observed ratio) for each galaxy. }
\label{fig:sed1}
\end{center}
\end{figure*}
\clearpage
\begin{figure*}
\begin{center}
\resizebox{\hsize}{!}{\includegraphics*{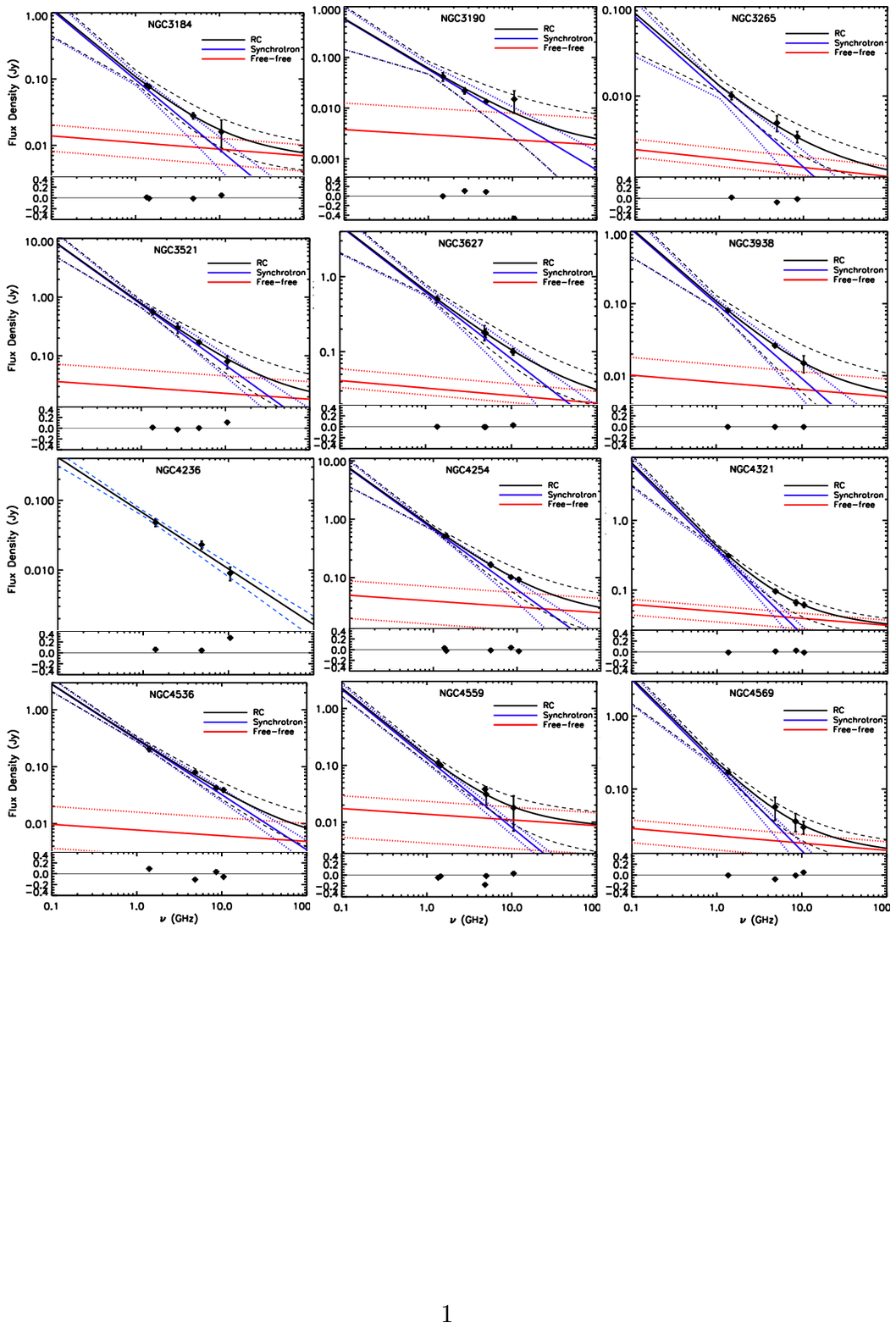}}
\caption{Same as Fig.~\ref{fig:sed1} for the rest of the sample.}
\label{fig:sed2}
\end{center}
\end{figure*}
\clearpage
\begin{figure*}
\begin{center}
\resizebox{\hsize}{!}{\includegraphics*{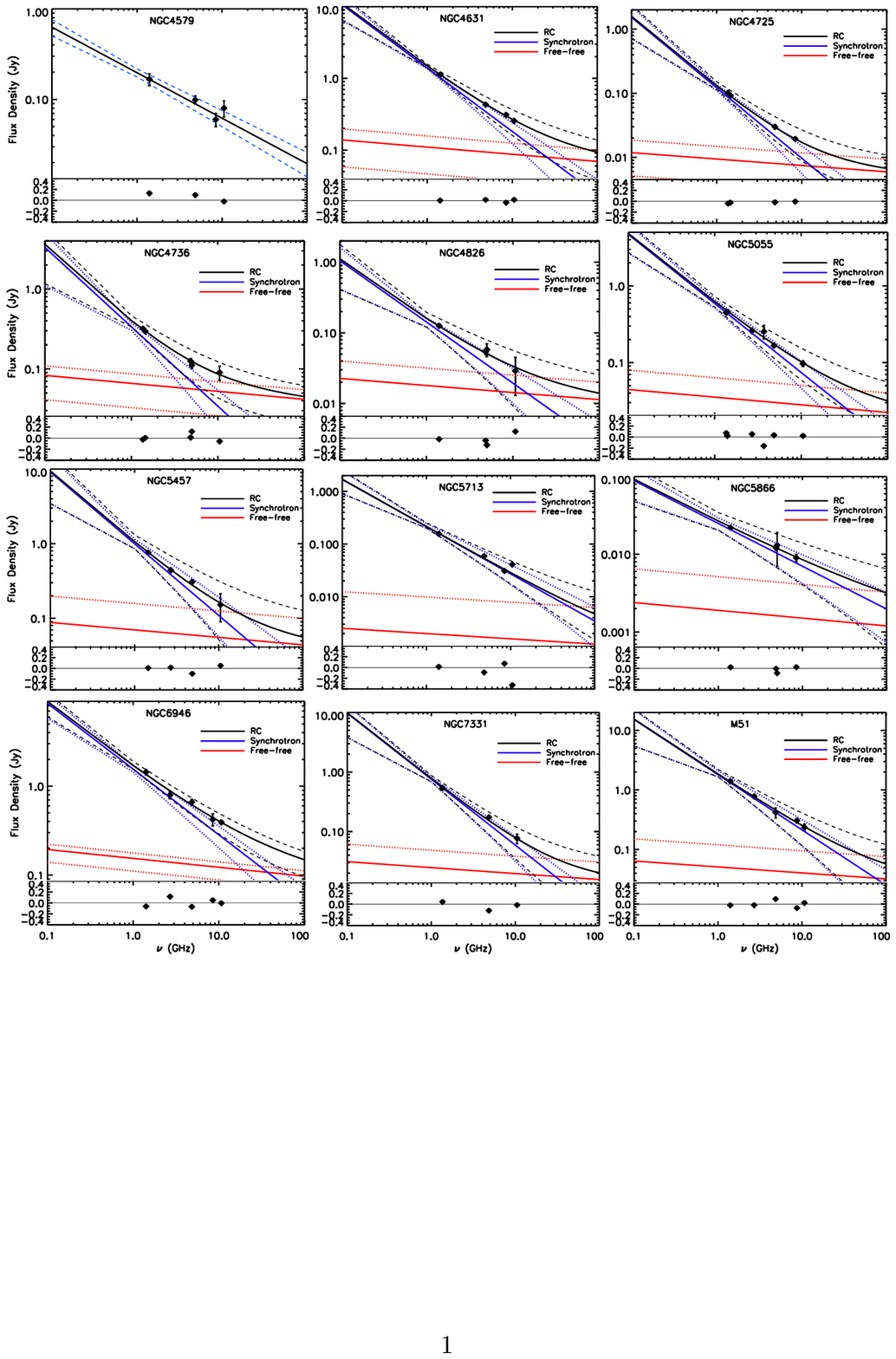}}
\caption{Same as Figs.~\ref{fig:sed1} and \ref{fig:sed2} for the rest of the sample.}
\label{fig:sed3}
\end{center}
\end{figure*}
%

%
\section{The radio SEDs beyond 1-10\,GHz}
By extrapolating the best-fit SED models, the synchrotron and free-free fluxes, and the thermal fractions can be estimated at frequencies higher and lower than 1-10\,GHz. This may not be realistic due to the curvature of the synchrotron SED and its flattening at lower frequencies. In this case, the predicted fluxes are higher than the observed fluxes and they actually provide a basis for evaluating the flattening itself. On the other extreme, the high-frequency extrapolations could result in total flux densities which are lower than the observed fluxes, due to spinning dust emission \citep{Bot} or magnetic nanoparticles \citep{Draine_12}.  Then, our predictions would help detecting such emission in the sample. Table~\ref{tab:beyond} lists the predicted thermal fractions at 350\,MHz, 15\,GHz, 22\,GHz, 33\,GHz, and 45\,GHz for each galaxy. %

\begin{table*}
\begin{center}
\caption{Thermal fractions predicted beyond the 1-10\,GHz.}
\begin{tabular}{llllll} 
\hline
Galaxy  & 350\,MHz  & 15\,GHz & 22\,GHz & 33 \,GHz & 45\,GHz  \\
\hline \hline
IC0342  & ...  & ...   & ...  & ...  & ...\\
IC2574  & 0.11  &  0.75  & 0.81  &0.86  &0.90\\
NGC~0337 & 0.01 &  0.23  & 0.31  &0.41  &0.48\\
NGC~0628 &0.05  &  0.55  & 0.60  &0.68  &0.73\\
NGC~1266 & 0.01 &  0.20  & 0.26  &0.34  &0.41\\
NGC~2146 & 0.00 & 0.03  & 0.04  &0.05  &0.06\\
NGC~2798 & 0.00 & 0.13  & 0.16  &0.20  &0.23\\
NGC~2841 & 0.03 & 0.56  & 0.64  &0.73  &0.78\\
NGC~2976 & 0.05 & 0.50  & 0.60  &0.70  &0.76\\
NGC~3049 & 0.01 & 0.45  & 0.53  &0.60  &0.65\\
NGC~3077 & ...  & ...  &  ...  & ...  & ...\\
NGC~3184 & 0.05 & 0.60  & 0.68  &0.76  &0.81\\
NGC~3190 & 0.07 &  0.37  & 0.45  &0.54  &0.61\\
NGC~3265 & 0.03 &0.57   & 0.64  &0.70  &0.75\\
NGC~3521 &0.01  &0.32   & 0.41  &0.50  &0.57\\
NGC~3627 &0.01  & 0.31  & 0.37  &0.45  &0.51\\
NGC~3938 &0.02  & 0.51  & 0.60  &0.68  &0.74\\
NGC~4236 & ... & ...   & ...  & ...  & ...\\
NGC~4254 & 0.02 & 0.42  & 0.51  &0.61  &0.67\\
NGC~4321 &0.03  &0.73   & 0.81  &0.87  &0.90\\
NGC~4536 & 0.00  &0.21   & 0.27  &0.35  &0.41\\
NGC~4559 & 0.04 &0.68   & 0.76  &0.84  &0.88\\
NGC~4569 & 0.03 &0.66   & 0.75  &0.82  &0.86\\
NGC~4579 & ...  & ...   & ...  & ...  & ...\\
NGC~4631 & 0.01 &0.40   & 0.47  &0.55  &0.61\\
NGC~4725 & 0.03 &0.53   & 0.63  &0.72  &0.78\\
NGC~4736 & 0.06 &0.65   & 0.75  &0.82  &0.85\\
NGC~4826 & 0.04 &0.50   & 0.57  &0.65  &0.70\\
NGC~5055 & 0.01 &0.34   & 0.41  &0.50  &0.56\\
NGC~5457 & 0.02 &0.42   & 0.50  &0.59  &0.65\\
NGC~5713 &0.00 &0.07   & 0.10   &0.13  &0.16\\
NGC~5866 & 0.00& 0.20   & 0.23  &0.27  &0.30\\
NGC~6946 &0.01  & 0.36   & 0.42  &0.48  &0.53\\
NGC~7331  &0.01  & 0.33   & 0.42  &0.52  &0.60\\
M51      & 0.00 & 0.20   & 0.27  &0.34  &0.40\\
\hline \hline
\label{tab:beyond}
\end{tabular}
\end{center}
\end{table*}
%

%
%
\end{document}